\documentclass[twocolumn]{aastex63}
\usepackage{hyperref}
\usepackage{verbatim}
\usepackage{tabularx}

\newcommand{\htwo}{{H\,II}\,}

\newcommand{\lya}{\mbox{Ly$\alpha$}}
\newcommand{\lyaf} {\lya\ forest}
\newcommand\LCDM{$\Lambda$CDM}

\def\D{{\Delta}}
\def\L{{\Lambda}}

\def\2pr{^{\prime \prime}}

\def\kmsmpc{{\rm km\, s^{-1} Mpc^{-1}}}

\newcommand{\oii}{[\ion{O}{2}]~} 
\newcommand{\zphot}{z_{\rm phot}}

\newcommand{\deltachisq}{{\ensuremath{\Delta \chi^2}}}
\newcommand{\snroii}{\ensuremath{{\rm SNR}(F_{\rm O\,II})}}

\begin{document}

\title{Validation of the Scientific Program for the Dark Energy Spectroscopic Instrument}


\author{DESI Collaboration: A.~G.~Adame}
\affiliation{Instituto de F\'{\i}sica Te\'{o}rica (IFT) UAM/CSIC, Universidad Aut\'{o}noma de Madrid, Cantoblanco, E-28049, Madrid, Spain}
\author{J.~Aguilar}
\affiliation{Lawrence Berkeley National Laboratory, 1 Cyclotron Road, Berkeley, CA 94720, USA}
\author[0000-0001-6098-7247]{S.~Ahlen}
\affiliation{Physics Dept., Boston University, 590 Commonwealth Avenue, Boston, MA 02215, USA}
\author[0000-0002-3757-6359]{S.~Alam}
\affiliation{Tata Institute of Fundamental Research, Homi Bhabha Road, Mumbai 400005, India}
\author{G.~Aldering}
\affiliation{Lawrence Berkeley National Laboratory, 1 Cyclotron Road, Berkeley, CA 94720, USA}
\author[0000-0002-5896-6313]{D.~M.~Alexander}
\affiliation{Centre for Extragalactic Astronomy, Department of Physics, Durham University, South Road, Durham, DH1 3LE, UK}
\affiliation{Institute for Computational Cosmology, Department of Physics, Durham University, South Road, Durham DH1 3LE, UK}
\author{R.~Alfarsy}
\affiliation{Institute of Cosmology \& Gravitation, University of Portsmouth, Dennis Sciama Building, Portsmouth, PO1 3FX, UK}
\author{C.~Allende~Prieto}
\affiliation{Departamento de Astrof\'{\i}sica, Universidad de La Laguna (ULL), E-38206, La Laguna, Tenerife, Spain}
\affiliation{Instituto de Astrof\'{i}sica de Canarias, C/ Vía L\'{a}ctea, s/n, E-38205 La Laguna, Tenerife, Spain}
\author{M.~Alvarez}
\affiliation{Lawrence Berkeley National Laboratory, 1 Cyclotron Road, Berkeley, CA 94720, USA}
\author{O.~Alves}
\affiliation{University of Michigan, Ann Arbor, MI 48109, USA}
% ...

\submitjournal{\aj}

\begin{abstract}
The Dark Energy Spectroscopic Instrument (DESI) was designed to conduct a survey covering 14,000 deg$^2$ over five years to constrain the cosmic expansion history through precise measurements of Baryon Acoustic Oscillations (BAO).
The scientific program for DESI was evaluated during a five month Survey Validation (SV) campaign before beginning full operations.
This program produced deep spectra of tens of thousands of objects from each of the stellar (MWS), bright galaxy (BGS), luminous red galaxy (LRG), emission line galaxy (ELG), and quasar target classes.
These SV spectra were used to optimize redshift distributions, characterize exposure times, determine calibration procedures, and assess observational overheads for the five-year program.
In this paper, we present the final target selection algorithms,
redshift distributions, and projected cosmology constraints resulting from those studies.
We also present a `One-Percent survey' conducted at the conclusion of Survey Validation covering 140 deg$^2$ using the final target selection algorithms with exposures
of a depth typical of the main survey.
The Survey Validation indicates that DESI will be able to complete the full 14,000 deg$^2$ program with spectroscopically-confirmed targets from the MWS, BGS, LRG, ELG, and quasar programs with total sample sizes of 7.2, 13.8, 7.46, 15.7, and 2.87 million, respectively.
These samples will allow exploration of the Milky Way halo, clustering on all scales, and BAO measurements with a statistical precision of 0.28\% over the redshift interval $z<1.1$, 0.39\% over the redshift interval $1.1<z<1.9$, and 0.46\% over the redshift interval $1.9<z<3.5$.\\
\end{abstract}




\setcounter{footnote}{0} 
\section{Introduction}
Studies of the geometry and energy content of the Universe, physics of cosmic expansion, fundamental properties of standard model particles, and growth of structure remain the key focus of cosmology studies.
Early measurements of cosmic expansion history using Type Ia supernovae
(SNe~Ia) helped to constrain the energy content, providing the first evidence for cosmic acceleration that could be explained by a form of dark energy \citep{riess98a,perlmutter99a}.
Subsequent SNe~Ia studies \citep[e.g.][]{suzuki12a,betoule14a,scolnic18a} were able to constrain the equation of state for dark energy to a precision of roughly 4\% when combined with cosmic microwave background (CMB) measurements from the {\it Planck} satellite \citep{planck11a}, consistent with a \LCDM\ model where dark energy can be explained by a cosmological constant.
Under this assumption of a flat \LCDM\ model, final CMB measurements from {\it Planck} lead to measurements of the matter density to better than 1\% precision and baryon density to better than 0.5\% precision \citep{planckcosmo}.

Wide-field, optical spectroscopy offers cosmological measurements that are complementary to SNe~Ia measurements of the distance-redshift relation and measurements of CMB anisotropy.
Spectroscopy of galaxies and quasars provides a precise, three-dimensional map of matter in the Universe in which the scale of baryon acoustic oscillations (BAO) can be measured at high precision.
As a preferred scale in the clustering of matter ($\sim$150 Mpc comoving), BAO measured in large-scale structure provide a standard ruler for observational cosmology.
Measurements from 2dFGRS \citep{colless01a} and the Sloan Digital Sky Survey \citep[SDSS;][]{york00a} 
marked the first use of BAO as a cosmological probe \citep{cole05a,eisenstein05a}, thus motivating the design of surveys dedicated to BAO and clustering measurements, such as WiggleZ \citep{blake11a,blake11b}.
The Baryon Oscillation Spectroscopic Survey \citep[BOSS;][]{dawson13a} of SDSS-III \citep{eisenstein11a}
and the extended Baryon Oscillation Spectroscopic Survey
\citep[eBOSS;][]{dawson16a} of SDSS-IV \citep{blanton17a} are the largest of those completed spectroscopic programs.
The SDSS, SDSS-II, BOSS, and eBOSS programs produced eight spectroscopic samples that led to BAO measurements spanning the redshift range $0.07<z<2.5$.
When combined with the {\it Planck} temperature and polarization data, these BAO measurements provide nearly an order of magnitude improvement on curvature constraints relative to primary CMB constraints alone.
Adding again the Pantheon SNe~Ia sample \citep{scolnic18a}, the BAO data allow constraints on the Hubble constant $H_0=67.87 \pm 0.86 \,\kmsmpc$ \citep{eboss21} under a cosmological model that allows for a time-varying equation of state for dark energy and non-zero curvature.
It has been demonstrated that this measurement of the Hubble constant is robust against both assumptions of expansion history and estimates of baryon density.
However, several local measurements of the Hubble constant find higher values \citep[e.g.][]{freedman19,wong20a,riess22a}, albeit with varying degrees of tension.

Spectroscopic samples of galaxies and quasars can also be used to probe the growth of structure through redshift-space distortions (RSD).
RSD appear in the clustering of matter due to the peculiar velocities induced by gravitational interactions, thus creating an apparent enhancement of clustering along the line of sight relative to clustering perpendicular to the line of sight \citep{kaiser87a}. 
RSD data complement recent weak lensing measurements \citep[e.g.,][]{2018PASJ...70S..25M,2019PASJ..tmp...22H,hildebrandt20a,joachimi20a,des21} by offering constraints on the gravitational infall of matter over cosmological scales.
Growth of structure measurements allow enhanced tests of the energy components, neutrino masses, and of General Relativity.
Even when assuming a cosmological model that allows for a time-varying equation of state for dark energy and non-zero curvature, percent-level constraints on $\Omega_\Lambda$, $H_0$, and $\sigma_8$ are possible when using the full sample of BAO and RSD measurements from the SDSS series of experiments \citep{ross15a,howlett15a,alam17a,LRG_corr,gil-marin19a,raichoor19a,tamone19a,demattia19a,hou19a,neveux20,2019duMasdesBourbouxH}, CMB data from {\it Planck}, SNe~Ia data from the Pantheon sample \citep{scolnic18a} and more recently \citep{scolnic22a}, and weak lensing data from the Dark Energy Survey \citep{2018PhRvD..98d3526A} and more recently \citep{DESY3}.
Under this model, the combination of BAO, RSD, CMB, SNe~Ia, and weak lensing data leads to a constraint $\Omega_k = -0.0022 \pm 0.0022$, $w_a = -0.49^{+0.35}_{-0.30}$, and
$w_p = -1.018 \pm 0.032$ at a pivot redshift $z_p=0.29$.
Here, the time-varying equation of state for dark energy is defined as $w(z) = w_p+(a_p - a)w_a$, where $a_p$, the expansion factor corresponding to the pivot redshift, is chosen to make the uncertainties on $w_p$ and $w_a$ uncorrelated.
Furthermore, the combination of samples produces tests of gravity that are consistent with General Relativity, a measurement of the clustering amplitude $\sigma_8=0.8140 \pm 0.0093$, and of the summed neutrino masses $\sum m_\nu <0.115$ eV (95\% confidence) \citep{eboss21}.
When evaluating the SDSS BAO and RSD independently from the other samples, the clustering amplitude is found to be $\sigma_8=0.85 \pm 0.03$, a measurement that does not support the somewhat low estimates of structure growth reported in recent weak lensing studies \citep[e.g.][]{DESpluskids}.

The Dark Energy Spectroscopic Instrument \citep[DESI;][]{levi13,desi16a,desi16b} was designed to advance studies of the cosmological model by large margins over previous programs through measurements of the clustering of galaxies, quasars, and the \lya\ forest.
DESI will be used to conduct a five-year survey over 14,000 deg$^2$ with a spectroscopic sample size that will be ten times that of the previous SDSS programs.
This footprint will be covered by six different classes of targets.
Following the motivation to perform BAO measurements near the cosmic-variance limit, we will use selections based on optical and infrared imaging data to identify a bright sample of low redshift galaxies (BGS; $z_{\rm median} \sim 0.2$), luminous red galaxies (LRG; $0.4<z<1.1$), emission line galaxies (ELG; $0.6<z<1.6$), quasars as direct tracers ($0.9<z<2.1$), and Lyman-$\alpha$ forest (\lyaf) quasars ($2.1 < z < 3.5$) to trace the distribution of neutral hydrogen. 
Toward this goal, data from only two months of operations has already resulted in a detection of the BAO signal in both the BGS and LRG samples \citep{1stBAO}.
The extensive program will also extract cosmological information from the derived power spectra to constrain neutrino masses, modified gravity, and the physics of inflation.
In addition, a sample of stellar targets will be observed to a high density in an overlapping Milky Way Survey \citep[MWS;]{cooper22a}.
These stellar spectra will be used to explore the stellar evolution, kinematics, and assembly history of the Milky Way.

Because the surface density and faintness of the wide-field DESI sample far exceed the capabilities of
previous spectroscopic facilities, these samples had to be extensively explored with the DESI instrument itself before the commencement of the five year program.
To do so, we conducted observations in a phase of `Survey Validation' (SV).  
These observations were used to test the quality of data against the primary BAO science requirements, optimize target selection algorithms, and inform the final DESI operational and analysis program.
The first stage of SV, the Target Selection Validation, took place from December 14, 2020 through April 2, 2021.
In the final stage of SV, we performed a pilot survey of the full DESI program that covered approximately 140 deg$^2$ (`One-Percent') using a superset of the final selection of MWS, BGS, LRG, ELG, and quasar targets.  
At least 95\% of targets were observed from each of the samples over 20 distinct fields.

In this paper, we present an overview of the DESI Target Selection and One-Percent Survey Validation programs, the results, and the implications for the five-year program.
A full description of the final target selection algorithms for the LRG, ELG, and quasar samples can be found in accompanying papers by \citet{zhou22a}, \citet{raichoor22a}, and \citet{chaussidon22a}, respectively.
The procedures for identifying all classes of targets can be found in \citet{myers22a}. The description of prioritization of targets for observation is detailed in \citet{schlafly22a}.
An overview of the observational strategy and projections for the BGS program can be found in the accompanying paper by \citet{hahn22a}, while an overview of the MWS science program can be found in \citet{cooper22a}.
Visual inspections played an essential role in verifying the performance of the instrument, the data reduction pipeline, and the target selection algorithms.
The visual inspection process and characterization of the spectroscopic performance for the galaxy samples can be found in \citet{lan22a}, while the same for quasars can be found in \citet{alexander22a}.

This paper is organized as follows.
In Section~\ref{sec:srd}, we present an overview of the initial requirements for BAO precision and the programmatic questions that Survey Validation was designed to address.
In Section~\ref{sec:observations}, we describe the Target Selection Survey Validation program, observations, and the resulting calibration procedures.
In Section~\ref{sec:ts}, we present the imaging data, target selection algorithms, and the One-Percent Survey observations that were vetted during SV and will be used for studies of clustering.
We present the exposure times, survey strategy, and redshift distributions expected for the five-year survey in Section~\ref{sec:results}.
In Section~\ref{sec:forecasts}, we present the cosmological forecasts, and in Section~\ref{sec:conclusion},
we present a summary of the plans for cosmological studies, release of data products to the broader community, and highlights of other science opportunities with the DESI data.
Throughout, we use the AB magnitude system and assume a fiducial cosmology described by the final {\it Planck} results \citep{planckcosmo}, where $\Omega_M = 0.315$, $\sigma_8=0.811$ and $h=0.674$.

\section{Survey Validation}\label{sec:srd}
The primary purpose of Survey Validation was to confirm that the survey design, instrument performance, and data quality would be sufficient to meet the top-level goals on BAO measurement precision.
Here, we present an overview of those goals, the instrument design, and the specific questions that the SV observational program was designed to address.

\subsection{DESI Science Requirements}
DESI is designed as a Stage-IV dark energy experiment as defined by the Dark Energy Task Force \citep[DETF;][]{detf}.
A Stage-IV experiment implies at least a factor of ten improvement in dark energy Figure of Merit (FoM) relative to a representative Stage-II program.
The cosmology results from the three year Supernova Legacy Survey (SNLS) were chosen as this representative program.
The SNLS sample of 472~SNe~Iae produced constraints on the time-evolving equation of state for dark energy $w_0 = -0.905 \pm 0.196$ and $w_a = -0.984^{+1.094}_{-1.097}$ under the assumption of a flat universe \citep{sullivan11}.

The detailed assumptions and forecast procedures 
using only CMB and BAO measurements are presented in the science, targeting, and survey design report \citep{desi16a}.
Briefly, we define the Figure of Merit as $[\sigma(w_p) \sigma(w_a)]^{-1}$ for dark energy with a time-evolving equation of state.
The dark energy equation of state parameters are forecast in a model where curvature is also treated as a free parameter.
A 9000 deg2 DESI survey of galaxies, quasars, and the Lya forest would achieve a DETF FoM for BAO science of 121,
whereas the FoM of the Year 3 SNLS result was found to be 11.
Doing so requires measurements of the isotropic cosmic distance scale, $R(z)$, to a precision 0.28\% over the interval $0.0 < z < 1.1$ and 0.39\% over the interval $1.1 < z < 1.9$.
Additional quasar and \lyaf\ BAO measurements of $H(z)$ are required to a precision of 1.05\% over the interval $1.9 < z < 3.7$.

These early BAO and FoM predictions were based on an assumed redshift distribution for the various target classes that had not yet been measured from imaging or spectroscopic data.
Early algorithms for selection of targets \citep{bgs20,lrg20,elg20,qso20} held promise for meeting the requirements of final spectroscopic sample size and redshift range.
The FoM will be significantly improved with a larger survey area, the addition of RSD measurements, and the inclusion of weak lensing, SNe~Ia, or other dark energy probes.
Cosmological forecasts that account for the final target selection algorithms, predicted areal coverage, and additional measurements can be found in Section~\ref{sec:forecasts}.

\subsection{DESI Instrument Design}\label{subsec:instrument}

DESI was built with the requirement of obtaining a minimum of 30 million redshifts to achieve the sub-percent precision BAO measurements described above, while providing additional margin through a 14,000 deg$^2$ footprint.
A full description of the motivation and requirements for the instrument, control system, and data management can be found in the instrument design report \citep{desi16b} and an overview of the completed instrument \citep{desi-collaboration22a}.

To enable the required performance, new corrective optics were installed at the National Optical
Astronomy Observatory’s 4-m Mayall telescope at Kitt Peak, Arizona to allow the installation of a 0.8-meter diameter focal plane \citep{miller22a}.
The field of view available to the instrument is 8.0~deg$^2$, of which 7.45~deg$^2$ is accessible for spectroscopy.
The roughly circular focal plane is divided into ten `petals' distributed over equal angles in azimuth.
The instrument design incorporates robotically-actuated fibers to minimize overhead from fiber repositioning between exposures \citep{silber22a}.
The positioners are arranged with a mean 10.525 mm pitch between
centers, each with a range of motion that covers a 12~mm (nearly three arcminute) diameter.
Each positioner hosts a fiber with a core diameter of 107~$\mu$m, corresponding to an average 1.5~arcsecond diameter projection on the sky.
A focal plane consisting of 5,020 of these fiber positioners was constructed. 5,000 fibers feed ten spectrographs that cover a wavelength range from 360~nm to 980~nm. The remaining 20 fibers feed a separate camera for independent measurements of sky background.

Each spectrograph consists of three cameras with a resolving power, $R = \lambda/\Delta \lambda$, that ranges from roughly 2000 at the shortest wavelengths to nearly 5500 at the longest wavelengths \citep{jelinsky22a}.
The focal plane is installed at prime focus, with 47.5~meter fiber runs connecting each positioner to a spectrograph in a climate controlled, enclosed environment \citep{poppett22a}.
The instrument is controlled in real-time through a series of automated data acquisition components that determine dynamic exposure times, perform data quality assessment, and convert on-sky target coordinates to fiber positions \citep{kent23}.
The automated data acquisition and rapid reconfigurability of the fiber positions enable very efficient operations with a deadtime between exposures of less than 120~seconds.

Based on experience from previous spectroscopic programs and simulated spectra for realistic target samples, the instrument was expected to complete a 14,000~deg$^2$ survey in five years.
The faint ELG targets are the most challenging spectra to classify, requiring spectra that are sufficiently deep to detect [O~\textsc{ii}] fluxes down to $8 \times 10^{-17} \, {\rm erg \, s^{-1} \, cm^{-2}}$.
Given the five-year observing window and goal for a 14,000~deg$^2$ footprint, exposure depths must be equivalent to 1000-second exposures taken at zenith through regions of sky with no Galactic extinction.
The [O~\textsc{ii}] flux requirement of the ELG sample was expected to set the observational pace for redshift completion, while the brighter LRG and quasar targets were expected to reach high completeness, even with shallower exposures.
Exposure times for BGS and MWS targets were to be tuned to balance high redshift completeness with high surface density during the times when the moon produced higher sky background levels.
The assumed redshift success rates and data quality as a function of exposure time were tested in Survey Validation with results described in Section~\ref{sec:results}.

\subsection{Questions to Inform the Survey Validation Program}

During SV, we obtained data to test the quality of spectra against the objective of completing BAO measurements to a precision required for a Stage-IV program.
These data were further used to optimize target selection algorithms and inform the final DESI operational and analysis program.
The SV observations were designed to allow us to finalize the target selection algorithms and survey strategy as follows:
\begin{itemize}
    \item By performing a selection of SV targets that exceeds the main survey target densities, various selection boundaries could be assessed so that the final algorithm could be tuned for optimal redshift distributions.
    \item By obtaining sufficiently deep spectra on SV targets to determine the parent redshift distributions with high confidence, we could thus determine the number of tracers for modified selections as a function of redshift, even with uncertainties in the data reduction pipelines. 
    \item By conducting spectroscopy over a large number of exposures, multiple data splits could be used to test repeatability and determine the statistical uncertainties on redshift estimates, completeness of spectral classification, and purity in assignment of redshifts.
    \item By assigning a sufficiently large number of standard star and white dwarf targets to each field, spectrophotometric data quality could be assessed with varying flux calibration schemes.
    \item By assigning a sufficiently large number of sky fibers to each field, sky-subtraction algorithms could be vetted to determine how many sky fibers are required to achieve nearly Poisson-limited sky subtraction.
    \item By performing observations of each field in varying conditions, exposure times as a function of sky brightness, seeing, transparency, airmass, and Galactic extinction could be computed and used to calibrate the real-time, dynamic exposure time calculator.
    \item By assessing the relationship between redshift success rate and exposure depth, exposure times for the main program could be established to optimize science return in five years of operations.
\end{itemize}

\section{Target Selection Validation}\label{sec:observations}
We conducted Target Selection Validation observations over the period December, 2020 through early April, 2021, with a few additional observations completing in May. 
We took the data to address the questions above while scheduling a program long enough to allow time to complete the studies before beginning the main survey.
These observations were divided between targets for the MWS in a dedicated stellar SV program, for the BGS on dedicated fields that also included MWS targets at lower priority, and for LRG, ELG, and quasar samples.
In all cases of Target Selection Validation observation, each field was covered by a single `tile' with one dedicated position for each of the 5,000 fibers.

\subsection{Observations}
The basic observational goal of Target Selection Validation was to obtain high quality spectra for a statistically representative sample that would include the final selection algorithm for each class of target.
The deep spectra were intended to allow tests of reliability of the redshift estimates.
The broader selection was intended to allow optimization of the sample definitions to maximize scientific yield.

To achieve roughly uniform redshift performance in the main program, exposure times will be adjusted to account for the Galactic extinction, airmass, seeing, transparency, and sky background.  
Galactic extinction and airmass can be predicted ahead of time, but seeing, transparency, and sky background are determined in real time using feedback from guide cameras and the sky monitor.
There was no calibration of the real-time exposure time estimates prior to SV, so we modified exposure times only based on extinction, airmass, the phase and position of the moon, and the seeing delivered in the previous exposure.
We assumed a power law for the relationship between exposure time and airmass, such that $t_{\rm exp}=t_0 X^{1.25}$.
Here, $X$ represents airmass and the power law form was determined empirically from BOSS/eBOSS observations.  The constant in front is a normalization factor that is defined separately for dark and bright time observations.

The LRG, ELG, and quasar programs were conducted when the sky was darkest.
Assuming median seeing of 1.1 arcseconds, photometric conditions, typical sky
in dark time, Galactic extinction $E(B-V)=0$, and observations at zenith, spectral simulations indicated that
1000-second exposures were sufficient to determine redshifts for ELG targets with [O~\textsc{ii}] line fluxes above the threshold described in Section~\ref{subsec:instrument}. 
A series of survey simulations accounting for variations in observing conditions indicated that we would complete each LRG, ELG, and quasar field with these effective exposure times in five years \citep{schlafly22a}.
Correcting only for airmass, Galactic extinction, seeing, and moon phase \& location on a field-to-field basis, we used this 1000-second effective exposure time for each epoch of LRG, ELG, and quasar observation.
Typically, four epochs were obtained for each field over four different nights.
This observing strategy provided data at varying airmass, observing conditions, and depth.

Pixel-level simulations of the spectrograph indicated that a 4000-second cumulative effective exposure time was sufficient to classify the faintest
targets in the nominal selections with a high degree of confidence.
Exposures of this depth are also sufficient to classify the majority of interlopers in the target selections
that could potentially confuse classifications in normal-depth exposures. 
Objects that could not be classified under this observation strategy will surely result in redshift failures during the shorter, main survey exposures.
An exception to the four epoch, 4000-second observing strategy was made for three fields containing only ELG targets and for three fields containing only quasar and LRG targets.
These six fields were observed to exposure times ranging from the equivalent of 6.5 to 15 epochs.
The goal for these data was to facilitate visual inspection, provide a more accurate truth table of redshift estimates, and allow multiple subsamples of the data for consistency tests.
These will be among the deepest exposures taken by DESI.

These fields were observed with various combinations of ELG, LRG, and quasar targets. 
Target acquisition efficiencies improved dramatically over these four months as a result of enhancements of the fiber assignment and focal plane control software.
Overall, the selection for all targets was designed to be a well-controlled, random sub-sampling to enable modeling of the underlying population for each tracer.

The stellar and BGS observations were primarily conducted during the times when the moon was above the horizon. 
Based on early commissioning data and simulated spectra, BGS and MWS targets can be successfully classified in 180-second exposures
under nominal conditions in dark time.  Our simplified moonlight model for most of these exposures increased their exposure time by a factor of 3.6, in addition to terms for Galactic extinction, airmass, and seeing described above.
Most fields were observed under this scheme on four different nights.
When possible, one of these observations was taken during dark time to provide high-quality reference spectra.
As with the LRG, ELG, and quasar targets, we observed several fields to a depth equivalent to ten epochs of main survey time.
These fields contained BGS targets at the highest priority and were used for tests of calibration and consistency in redshift classification.

\begin{figure*}
  \centering
	  \includegraphics[width=0.95\textwidth, angle=0]{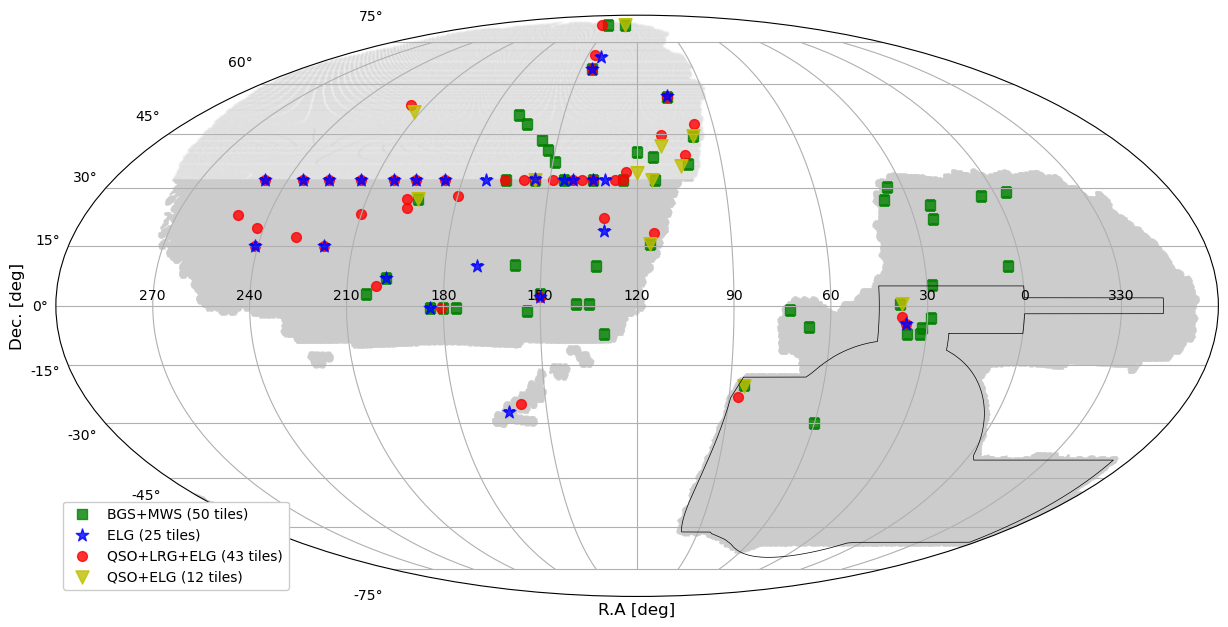}
	  \caption{The field centers for the fields designed to test MWS, BGS, LRG, ELG, and quasar selections and spectroscopic performance in the DESI Target Selection Validation program.
The light gray regions show the full imaging footprint available from Bok and Mayall imaging while the dark gray regions show the full imaging footprint available from the DECam imaging.
The black outline shows the footprint of the Dark Energy Survey (DES).
Details on the imaging can be found in Section~\ref{sec:dr9}. 
	  }
	  \label{fig:mainSV}
\end{figure*}

The footprint for all of these observations can be found in Figure~\ref{fig:mainSV}.

\subsubsection{Stellar SV Targets}
The nearest DESI targets will be Milky Way stars.
These targets will include white dwarfs, low-mass stars in the immediate solar neighborhood, rare stars, and stars in the Galactic thick disk and halo that formed more than 10 billion years ago.
In the main DESI survey, these targets will be observed concurrently with BGS targets, but at a lower priority for fiber assignment.
For validation of these targets, we designed a series of tiles with dedicated stellar targets to allow tighter control over fiber assignments in fields that were most conducive to stellar science.
The preliminary target selection algorithms are presented in \citet{allende20} while a full description of the program and results can be found in the accompanying MWS overview paper \citep{cooper22a}.

In the main survey, the bulk of the MWS sample will be magnitude-limited between
$16<r<19$ with additional proper motion and parallax criteria.
The selection in the stellar SV program was expanded by removing astrometric criteria,  
allowing fainter targets to explore the low signal-to-noise limit of the stellar pipelines, and relaxing the criteria for identifying white dwarf candidates from photometric data.
As with the main survey, high-value, sparse target classes such as Blue Horizontal Branch stars and RR Lyrae variables were prioritized for fiber assignment.
In addition, to enable comparison of derived stellar properties, priorities were adjusted to obtain spectra at high completeness from objects already observed in APOGEE \citep{majewski17a}, SEGUE \citep{yanny09a,rockosi22}, BOSS, the Gaia-ESO Spectroscopic Survey \citep{gilmore12a}, or GALAH \citep{desilva15}.

A summary of the Stellar SV observations can be found in Table~\ref{tab:SV_fields}.
Observations of these tiles were designed to address specific questions of stellar spectroscopy associated with sample selection,
performance of analysis pipelines, special field selection, and cross-calibration with previous surveys.
For this reason, field centers were chosen to sample a variety of environments.
In total, six fields were chosen to cover a range of Galactic latitudes, 10 fields to sample globular and open clusters, three fields to sample Milky Way satellite galaxies [Draco, Ursa Major II (UMaII) and Sextans], and one field was centered on a region that had a high stellar density from prior BOSS observations.

\begin{table*}[htb]
\caption{Summary statistics of Survey Validation fields.}
\begin{center}
\begin{tabular}{lccccc}
    \hline
    \hline
    Program & Number & Number & Number & Exposure & Effective\\
    & of tiles  & of nights & of exposures & time [hrs] & exposure time [hrs]\tablenotemark{\dag}\\
    \hline
    \multicolumn{6}{c}{\bf Deep fields selected for full visual inspection}\\
    \hline
    BGS & 1 & 6 & 30 & 3.0 & 0.8\\
    ELG & 3 & 9 & 51 & 12.6 & 8.7\\
    LRG \& quasar & 3 & 10 & 54 & 12.9 & 6.7\\
    \hline
    \multicolumn{6}{c}{\bf MWS, BGS, LRG, ELG, and quasar fields}\\
    \hline
    Stellar & 15 & 30 & 176 & 24.8 & 6.5\\
    BGS \& MWS & 50 & 49 & 562 & 64.6 & 15.3\\
    ELG & 22 & 26 & 157 & 36.8 & 24.3\\
    ELG \& quasar & 12 & 24 & 151 & 33.6 & 20.6\\
    LRG \& quasar & 28 & 41 & 292 & 66.5 & 39.1\\
    LRG \& quasar (updated selection) & 12 & 11 & 70 & 13.5 & 12.0\\
    \hline
    \multicolumn{6}{c}{\bf One-Percent Survey}\\
    \hline
    bright & 214 & 35 & 288 & 37.2 & 15.1\\
    dark & 239 & 33 & 374 & 96.9 & 86.4\\
    \hline
    \hline
\end{tabular}
\tablenotetext{\dag}{Effective exposure times are defined in Section~\ref{sec:cal_spectra}.}
\end{center}
\label{tab:SV_fields}
\end{table*}

\subsubsection{BGS Targets}

The lowest-redshift galaxies from DESI will come primarily from the BGS sample.
These galaxies will be observed during the time
when the Moon is significantly above the horizon, and the sky
is too bright to allow efficient observation of fainter targets.
Approximately 14 million of the brightest galaxies within the DESI
footprint will be observed over the course of the survey,
sampling galaxies at a high density with a median redshift of approximately $z=0.2$.
This sample alone will be ten times larger than the SDSS-I and SDSS-II
`main sample' that was observed from 1999-2008.
A summary of the final selection can be found in Section~\ref{sec:ts_bgs} while full description of the program and results can be found in the accompanying BGS overview paper \citep{hahn22a}.

The nominal BGS selection for the main survey is designed to rely on an $r$-band magnitude limit (BGS Bright).
One goal of SV was to test the redshift success rate as a function of exposure time and magnitude, thus providing guidance on the nominal exposure times for this sample.
Another goal of SV was to establish a selection that would prioritize galaxies over stars. 
By comparing the $G$-band magnitude from
{\it Gaia} \citep{gaia} to the  $r$-band total magnitude, we can separate stars and galaxies in the main BGS program. 
No color selection was used in the Target Selection phase of SV so that the final selection could be optimized based on $G - r$ star-galaxy separation.  
A third goal was to find a selection that maximizes completeness in the galaxy population while minimizing spurious targets from deblending and other photometric artifacts. 
To explore the signatures of spurious signal, the selection did not apply masks around large galaxies and included a subset of objects that were selected without
the quality cuts defined for the nominal BGS target selection algorithm.

A second sample of fainter BGS targets (BGS Faint) was observed at a slightly lower density than the bright targets.
The selection algorithms were extended 
to investigate whether a subsample of color-selected galaxies not in the BGS Bright sample can be spectroscopically classified at high completeness.
Fainter objects also allowed us to explore the dependence of redshift success rates on total magnitude and on an aperture magnitude matched to the DESI fiber radius.

A summary of the BGS Target Selection Validation observations can be found in Table~\ref{tab:SV_fields}.
In total, 50 fields were observed over regions with varying galactic extinction, stellar density, and imaging quality.
Eight of these fields overlapped the footprint of the Galaxy and Mass Assembly \citep[GAMA;][]{gama09} survey which is a highly complete galaxy redshift survey to a similar depth as the BGS sample.

\subsubsection{LRG Targets}

Over the approximate redshift range $0.4<z<1.1$, DESI will use LRG targets as the primary tracer for large-scale structure.
These luminous, massive galaxies have long since ceased star formation and therefore exhibit evolved, red spectral energy distributions (SEDs).
These galaxies may be most efficiently selected by taking advantage of the prominent 1.6~\micron~(rest frame) `bump' \citep{john88,sawicki02} that causes a strong correlation between optical/near-infrared (NIR) color and redshift. 

For DESI, we therefore used an algorithm similar to that used for eBOSS LRGs \citep{prakash16} to select the LRG sample from optical and infrared catalogs. A simple cut in optical/infrared colors as a function of optical color eliminates the lowest-redshift galaxies and rejects stars in an effective manner. 

Redshift estimation is informed primarily by the 4000~\AA~break and absorption features in LRG spectra. 
Given the need to reliably estimate the continuum and model these absorption features, the LRG sample was planned to be flux-limited. 
The selection was extended toward fainter magnitudes than were expected for the main program to test the redshift success rate as a function of flux and thus set the limiting magnitudes for the sample.

The selection followed the same philosophy as planned for the main survey, but with less restrictive boundaries on all colors and magnitudes to allow fine-tuning of the redshift distribution. 
In addition, two variants on color and magnitude were explored.
An optical selection relied on a sliding cut in $r-z$ color as a function of $z$-band magnitude.
An infrared selection relied on a sliding cut in $r-W1$ color as a function of $W1$-band magnitude, where $W1$ is the 3.4 micron bandpass from the Wide-field Infrared Survey Explorer \citep[WISE;][]{wise}. 

LRG targets were all observed concurrently with quasar targets.
A summary of the observations can be found in Table~\ref{tab:SV_fields}.
In total, 43 fields containing LRG and quasar targets were observed over regions with varying galactic extinction, stellar density, and imaging quality.
In all cases, LRG targets were given fibers after all quasar targets had been assigned.
In 31 of these fields, quasar selection was broadened to explore new techniques, leading to a lower yield of LRG targets.
In these fields, observations yielded roughly 1200 LRG spectra on average. 
In 12 fields, the quasar targets were selected according to an algorithm that more closely represented that of the main program, thus decreasing the number of quasar targets.
However, because those 12 fields were observed toward the end of Target Selection Validation, the instrument was near optimal performance and a much larger fraction of quasar targets were assigned fibers.
On average in these fields, roughly 800 LRG targets produced spectra.
An analysis of the LRG target selection algorithms can be found in \citet{zhou22a}.

\subsubsection{ELG Targets}

The majority of the spectroscopic redshift measurements for DESI will come from ELGs at redshifts $0.6<z<1.6$.  
These galaxies exhibit strong nebular emission lines originating in the ionized \htwo{} regions surrounding short-lived, but luminous, massive stars \citep[e.g.][]{moustakas06}.  
ELGs are typically late-type spiral and irregular galaxies, although any galaxy actively forming new stars at a sufficiently high rate will qualify as an
ELG.  
Because of their vigorous ongoing star formation, the integrated
rest-frame colors of ELGs are dominated by massive stars, and hence will
typically be bluer than LRG and other galaxies with evolved stellar populations.
This relatively blue continuum allows the efficient selection of ELG targets from optical $grz$-band photometry. 

Selection of ELG targets for DESI leverages the fact that the cosmic star formation rate was roughly an order
of magnitude higher at $z\sim1$ than today.
Galaxies with strong line-emission are therefore very common at the epoch where LRG targets become increasingly difficult to spectroscopically classify.
In particular, the prominent [O~\textsc{ii}] doublet in ELG spectra consists of a pair of emission lines separated in rest-frame wavelength by 2.783~\AA.  This wavelength separation of the doublet provides a unique signature, allowing definitive line identification (especially in cases where the doublet is resolved, enabled by the design for spectral resolution) and secure redshift measurements from [O~\textsc{ii}] alone for a large fraction of ELG targets.

During the target selection phase of SV, ELG targets were selected to explore the relationship between redshift, [O~\textsc{ii}] line strength, and $(g-r)/(r-z)$ color.
In addition we varied the definition of the magnitude limit (either $g$-band fiber magnitude or $g$-band total magnitude) and explored the performance of the instrument and selection for fainter objects than were expected for the main selection. 

A summary of the ELG observations can be found in Table~\ref{tab:SV_fields}.
In total, 25 fields containing ELG targets were observed while 12 fields containing both ELG and quasar targets were observed.
Not included in the table is the technical detail that ELG targets were also used as filler for remaining fibers in the 28 LRG and quasar fields.
As with the other target classes, these fields covered regions with varying galactic extinction, stellar density, and imaging quality.
The fields that contained ELG targets at the highest priority produced an average of roughly 3200 ELG spectra.
In the fields that also contained quasar targets, ELG targets were given fibers after all quasar targets had been assigned.
On average in these fields, roughly 2400 ELG targets produced spectra, although roughly 600 also satisfied the quasar selection and therefore had a higher priority in fiber assignment.
For training the ELG target selection algorithms, those 12 tiles required down-weighting of the targets with overlapping ELG and quasar selections so that they represent a fair fraction of the parent ELG target sample.
An analysis of the ELG target selection algorithms and the down-weighting scheme can be found in \citet{raichoor22a}.

\subsubsection{Quasar Targets}

The highest-redshift spectroscopic sample for DESI will consist of quasars.
We will measure large-scale structure using quasars as direct tracers of dark matter
in the redshift range $0.9<z<2.1$.
The DESI spectrographs cover the $\lambda=1216$~\AA\ \lya\ transition for objects with redshift above $z\sim2.0$. 
At redshifts $z=2.1$ and higher, we will use the foreground neutral-hydrogen \lya\ forest absorption observed in quasar spectra to measure large-scale structure.
In the main DESI survey, we will obtain additional exposures on confirmed \lya\ quasars to measure the \lyaf\ at the highest signal-to-noise ratio allowed under the observational constraints.

Quasars are fueled by gravitational accretion onto supermassive black holes,
leading to emission that can outshine the host galaxy.
These are the brightest population of non-transient $z>1$ targets that have a density high enough to use as tracers of large-scale structure \citep[e.g.][]{Palanque2016}. 
Even in the nearest quasars, the emitting regions are too small to be resolved, so these targets will generally appear in images as point sources that are easily confused with stars.  
Quasars are $\sim 2$ mag brighter in the near-infrared compared to stars of
similar optical magnitude and color, so we use optical photometry combined with {\it WISE} infrared photometry
in the $W1$ and $W2$ bands to discriminate against contaminating stars.

During the target selection phase of SV, we tested two different methods for identifying quasar targets.
The first was based on color cuts and the second was based on Random Forest algorithms trained to select quasars from photometric catalogs.
We also explored extensions of the initial set of photometric cuts: a relaxed definition of stellar morphology and an extension of the $r$-band magnitude limit to test the redshift distribution and population of fainter objects.
We also tested alternative methods to the color and Random Forest selections:  a selection based on variability in the \textit{WISE} light curves and a selection of high-redshift quasars based on $g$-band and $r$-band dropout techniques.

Because quasars appear at a lower density than either ELG or LRG targets, they were typically assigned fibers at the highest priority during Target Selection Validation.
As described above, several combinations of fields contained quasars.
Those 12 fields with ELG targets produced quasar spectra at an average of 1300 per field.
The first 31 quasar and LRG fields produced an average of roughly 1800 quasar spectra while the last 12 quasar and LRG fields produced an average of roughly 2200 quasar spectra.
For further details on the target selection algorithms, see~\cite{chaussidon22a}.

\subsection{Data Reduction}
\label{sec:datareduction}

The survey validation data were processed with a new spectroscopic pipeline developed specifically for DESI.
A detailed description of this pipeline can be found in \citet{guy22a}; we provide here a brief overview.

The DESI pipeline inherits much of the philosophy from SDSS, but was fully rewritten.
The most significant difference from SDSS is the spectral extraction technique.
We use here a full forward model of the CCD image, based on a precise two-dimensional model of the point spread function (PSF) in each camera.
This method, proposed by \citet{bolton10}, is more complex than the row by row extraction used in the past as it involves solving a large linear system and requires a post-processing method to provide uncorrelated spectral fluxes. The advantages are an improved statistical precision, uncorrelated fluxes on a unique wavelength grid, and a resolution matrix that provides a well-defined framework to account for the spatially varying spectrograph resolution when analyzing the spectra.

A first version of the software had been developed and tested on image simulations and spectrograph test data before the full instrument installation at Kitt Peak. It was further improved during the commissioning and the SV periods. Spectroscopic data automatically transfered to the National Energy Research Scientific Computing Center (NERSC) were processed on a daily basis with the most recent version of the software. During the SV period and the following months, several re-processing runs of the SV data were made available to the collaboration.
In each case, the internal release was accompanied by a uniform and documented software version.

Calibrations using a dome screen illuminated with a range of lamps were obtained during the afternoon prior to the nightly operations. This data set was used to determine precisely the coordinates of spectral traces, the wavelength calibration, and the PSF in each of the 30 CCDs for the upcoming night. Flat field data were also acquired to correct for the non-uniformities of the fiber transmission.

Exposures taken during each night were first pre-processed to convert ADC counts to electrons per pixel and perform bias and dark current subtraction, pixel flat-fielding, electronic cross-talk corrections, and assignment of bad pixel maps. 
A notable difference with other pipelines was the use of a model of the CCD image to estimate the Poisson noise in the pixels. 
The spectral extraction used the afternoon calibrations, but with the trace coordinates adjusted and the wavelength calibration refined using sky lines.
The output of the sequence of algorithms are uncorrelated spectral fluxes for all targets on the same wavelength grid, their variance, and a sparse resolution matrix to convert any spectral model to the resolution of the spectrograph.
The next steps of the processing comprised flat-fielding, sky subtraction, and spectrophotometric calibration.
The resolution matrix was used at each step;
for instance, a high-resolution, deconvolved sky model was derived from the sky fiber data and then re-convolved to the resolution of each fiber before being subtracted.

During SV we characterized the number of sky fibers and standard stars needed to achieve an accurate sky subtraction and flux calibration. For the former, the driver was the redshift efficiency and purity of ELGs for which sky line residuals can introduce confusion. We found that we could observe with fewer than 40 sky fibers per petal before significant degradation of the ELG redshift efficiency.
However, we conservatively maintained at least 40 sky fibers as a requirement for each petal.
This choice has minimal impact on the rate of fiber assignment for science targets because we include most non-moving or disabled fiber positioners in the list of sky fibers.
Similarly, we found that we can achieve an excellent flux calibration precision with 10 standard stars per petal and maintained this number as a goal for fiber assignment.

The data taken for the Survey Validation program were most recently processed with the {\tt fuji} version\footnote{\url{https://github.com/desihub/desispec/releases/tag/0.51.13}} of the data reduction pipeline. 
All SV data will be released publicly in 2023 using this version of the pipeline \citep{edr23a}.

\subsection{Calibration of Exposure Times}
\label{sec:cal_spectra}

Exposure times in the main survey are tuned to achieve a relatively uniform spectral data quality over all fields, in order to optimize the survey efficiency. This tuning relies on an online exposure-time calculator \citep[ETC;][]{kirkby22a} that determines when to end each exposure based on real-time monitoring of observing conditions.  Specifically, the ETC monitors the atmospheric transparency, the fraction of light entering fibers (degraded by atmospheric seeing) and sky brightness, all using dedicated instrumentation, and estimates the accumulated signal-to-noise ratio for different target categories.  The ETC also splits exposures that are predicted to be long into shorter exposures to facilitate cosmic-ray rejection.  The desired signal-to-noise ratio takes into account the observing airmass, the average galactic extinction of the field, and the type of targets being observed.

One goal of survey validation was to calibrate the ETC and characterize its performance, using the natural variations in observing conditions during SV. Calibration is needed to determine zeropoints for the transparency and sky brightness measurements, and to establish empirical scalings of spectroscopic signal-to-noise to exposure time for different target types and under varying observing conditions. 
The scatter in online ETC estimates of overall throughput (combining the effects of varying transparency and fraction of light entering a fiber) compared with offline spectrograph estimates was measured to be less than 5\%.  
The corresponding scatter in ETC estimated sky background level compared with offline reductions of sky spectra was also less than 5\%.

After calibration of the throughput and background predictions, an effective exposure time was defined to determine the completion status of each field, and corresponds to an equivalent real exposure time at airmass 1, zero galactic extinction, 1.1 arcsecond FWHM seeing, and zenith dark sky.  Table~\ref{tab:SV_fields} lists the cumulative effective exposure times achieved during SV for different programs.

Due to chromatic effects in airmass, galactic extinction, and other observational effects, it is not possible to determine an effective exposure time that is consistent across all wavelengths.
Likewise, given the varying location of spectroscopic features in observer-frame wavelength, it is not possible to determine an effective exposure time that is appropriate to all spectroscopic targets.
Instead, we designated a fiducial target profile on the sky for each observation, selected from point-like, ELG-like or BGS-like.  The target profile contributes to the fraction of light entering a fiber, along with the atmospheric and instrumental PSFs, with a larger profile reducing the fraction but also reducing the sensitivity to atmospheric seeing. We also accounted for the different expected contributions of sensor read noise from faint and bright sources.

In addition to estimating effective exposure times, the ETC monitors the `survey speed' as an indicator of the rate at which effective time is being accumulated.  
Survey speed is the instantaneous measure of the signal accumulation in the sky-noise limit. This ignores read noise, Poisson noise, and the effects of atmospheric absorption beyond airmass$=1$ or Galactic extinction. 
This is a measure of the sky conditions, normalized such that a value of one represents typical transparency, seeing, and background, while smaller values represent observations in degraded conditions.
The survey speed informs automated decisions about when to switch from BGS and MWS targets, observed mostly during bright moon conditions, to LRG, ELG and quasar targets, observed mostly during dark sky conditions.

\section{Target Selection and One-Percent Survey}\label{sec:ts}
The spectroscopic footprint for the full DESI program will cover two large footprints, one in each hemisphere of the Galactic sky. 
Over that 14,900~deg$^2$ footprint, we will observe MWS, BGS, LRG, ELG, and quasar targets using selection algorithms informed by the SV data.
After considering edge effects, these targets will be observed to high completeness in the central 14,000~deg$^2$, as intended in the instrument and survey design.

\begin{figure*}[h]
  \centering
	  \includegraphics[width=0.45\textwidth, angle=0]{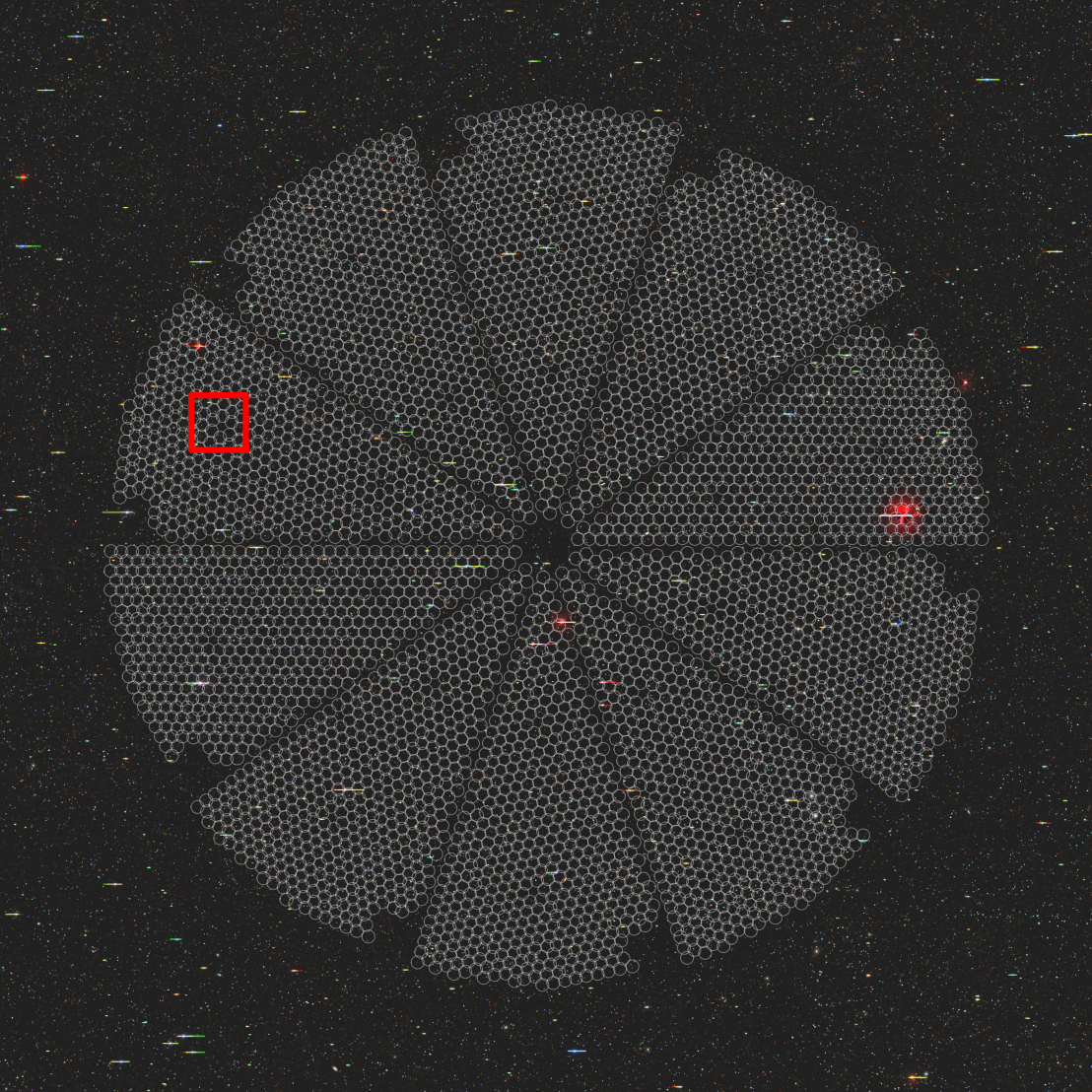}
	  \includegraphics[width=0.45\textwidth, angle=0]{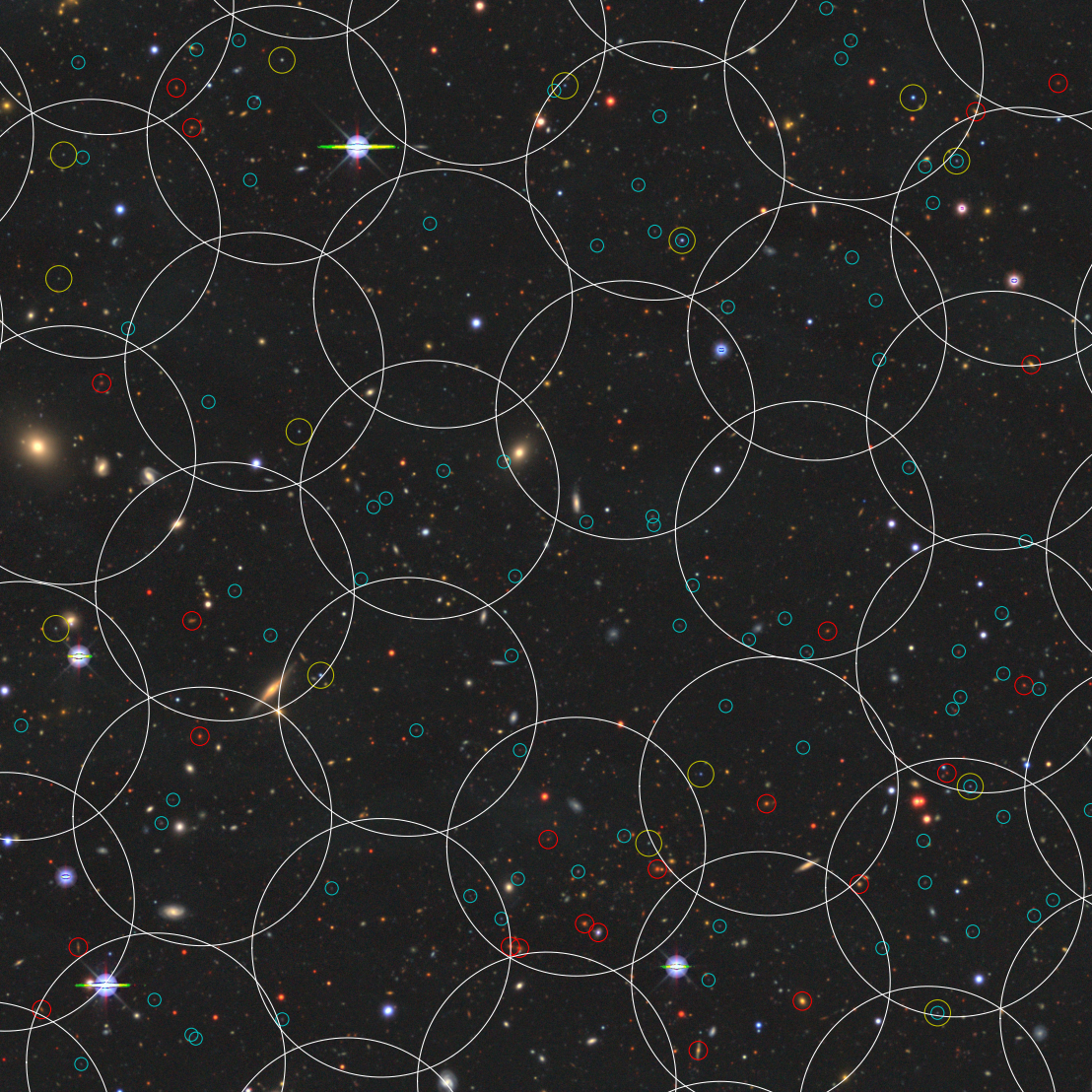}
   \caption{Example of the sky coverage of one DESI tile centered at ($\alpha$,  $\delta$)  = (0, 0).  
   The white circles display the individual fiber patrol regions.  
   {\bf Left:} an image that spans four degrees on a side, illustrating the entire DESI focal plane.
   {\bf Right:}  the smaller region identified by the red square in the left panel.
    DESI Main Survey Dark Targets are circled (LRGs in red,  ELGs in cyan,  and quasars in yellow).  
    The background imaging is a $grz$-band composite image from the DESI Legacy Surveys.
    The region was chosen to demonstrate that some fibers have limited number of targets accessible while others can be heavily oversubscribed.
    For example, even though this region will be covered by five or more tiles, the six ELG targets accessible to the central fiber close to the bottom of the image are unlikely to be observed because there is a quasar and seven LRG targets also within reach.
    There are also regions that are not accessible to fibers at all.  For example, the hole near the middle of the right panel is the location of a fiducial that is used to calibrate the focal plane coordinate system.
	  }
	  \label{fig:fiberassign}
\end{figure*}

As demonstrated in Figure~\ref{fig:fiberassign}, there are typically multiple targets accessible to any one fiber in a DESI pointing.
In the main dark time survey, quasar targets will be assigned fibers at the highest priority, followed by LRG and then ELG targets.
Because the sky is covered with multiple layers of tiles, each coordinate on the sky has roughly five opportunities of fiber assignment, allowing high completeness even for the ELG targets.
During the times when BGS and MWS targets are observed, BGS targets will be given the higher priority assignment.
The algorithms have been finalized for prioritizing targets in the assignment of fibers.
Those algorithms, including a full description of the priority scheme, are found in \citet{raichoor22b}.

A preliminary run has been performed for all targets over the full footprint, leading to estimates of the fraction of targets from each sample that will be assigned a fiber.
The efficiency of the fiber assignment algorithm for each target class depends on density, clustering, and priority.
However, for targets that are assigned a fiber, roughly 1\% of objects targeted as a galaxy will produce `failed spectra' that are not appropriate for spectral classification.
Those spectra may have had improper positioner placement, significant contamination from cosmetic defects on the detector, or other concerns.
Quasar targets are not expected to suffer from this source of incompleteness because they are given highest priority in fiber assignment and are typically re-observed when such an issue arises \citep{schlafly22a}.
In this section, we present the imaging data, the final target selection algorithms, and the fraction of targets within the central 14,000~deg$^2$ that are expected to get a meaningful spectrum.\footnote{The reported area does not correct for regions lost to masking, which account for roughly 1\% of the footprint.}

Immediately following the completion of Target Selection Validation observations, we conducted a survey of 140~deg$^2$ in all five target classes.
Titled the `One-Percent Survey', these observations produced spectroscopic samples to significantly higher fiber assignment completeness
and redshift success completeness than will be obtained in the full program.
We conclude this section with an overview of the One-Percent Survey.

\subsection{Imaging Data}
\label{sec:dr9}

The photometric catalogs used for identifying DESI targets are derived from three optical imaging surveys in combination with data from {\it WISE} and {\it Gaia}.
The three optical surveys were designed for DESI target selection in the $grz$ bands at depths that were appropriate for the selection of $z>1$ ELG targets.
Additional photometry from {\it WISE} allows selection of BGS, LRG, and QSO targets while the small PSF of the {\it Gaia} $G$-band data allows $G-r$ to be used as an accurate star-galaxy separator in the BGS sample and the {\it Gaia} astrometric data facilitates stellar selections.

The Dark Energy Camera Legacy Survey (DECaLS) was the largest of the three dedicated surveys.
DECaLS imaging in the $grz$ filters was collected in a three-pass observing strategy using the Dark Energy Camera \citep[DECam;][]{decam} at the 4-m Blanco telescope.
DECaLS made use of existing $grz$ DECam data from other programs where available, most notably the Dark Energy Survey \citep[DES;][]{DES_overview}.
DECaLS is the sole source of $grz$ imaging used for selecting targets in the South Galactic Cap (SGC).
The North Galactic Cap (NGC) coverage from DECaLS was limited to $\delta<34$ deg due to the constraints of observing at a high airmass.

In the NGC, at declination $\delta > 32.375$ deg, imaging from two coordinated programs from the Kitt Peak National Observatory is used for selecting targets.
Imaging was performed using the $g$-band and $r$-band filters in the Beijing-Arizona Sky Survey \citep[BASS;][]{bass_overview} using the 90Prime camera at the 2.3-m Bok telescope.
Imaging in the $z$-band filter was performed in the Mosaic $z$-band Legacy Survey (MzLS) using the 4-m Mayall telescope.
An upgraded camera \citep[Mosaic-3;][]{dey16} with 4k$\times$4k, thick, deep-depletion CCDs was installed at prime focus specifically to obtain better quantum efficiency at red wavelengths before starting this program.

The imaging area above $\delta>-30$ deg and median $5\sigma$ point source magnitudes for each of the three $grz$-band imaging surveys is found in Table~\ref{tab:imaging}.
This imaging area defines the region accessible to DESI at a reasonable airmass.

The full processing for all imaging data follows an approach similar to that in \citet{decals_dr8}, with several improvements in data reduction algorithms.
The $grz$ photometric catalogs were created using common positions and profiles to model the flux for each source across all images following the `Tractor' \citep{tractor} methodology.
These models were also used to compute new photometric measurements in all four of the {\it WISE} bandpasses, as in \citet{lang14,lang16,meisner17}.
The exposure times for the 3.4 and 4.6 micron bands are seven times longer than those used in the original WISE all-sky survey catalogs \citep{allwise}.
In addition, the application of Tractor to determine {\it WISE} photometry results in less confusion in extended sources due to the higher resolution optical references used to derive source models.
For targets in SV, the forced photometry on the $grz$ and {\it WISE} images was complemented by the photometric and astrometric data from the 2nd public data release from {\it Gaia} \citep{gaiadr2}.
The Legacy Survey images and photometric catalogs can be found in the 9th public data release of the DESI imaging surveys \citep{schlegel22a}.\footnote{\url{https://www.legacysurvey.org/dr9/}}
These catalogs were used to identify targets for the main survey, with one exception \cite[][Section 4.1.4]{myers22a}:
the astrometric data for all MWS targets are taken from the {\it Gaia} Early Data Release 3 catalogs \citep{gaiaedr3}.
All catalog fluxes are reported without applying Galactic extinction corrections, but target selections are extinction-corrected \citep{schlegel98} except where noted.

\begin{table*}[htb]
\caption{Imaging Statistics for selection of targets in Survey Validation and the main DESI survey}
\begin{center}
\begin{tabular}{lcc}
\hline
\hline
    & BASS+MzLS & DECaLS \\
\hline
\hline
Area [deg$^2$] & 5,170 & 11,717 \\
Median $g$ depth [mag] & 24.29 & 24.81 \\
Median $r$ depth [mag] & 23.72 & 24.24 \\
Median $z$ depth [mag] & 23.33 & 23.34 \\
Median W1 depth [mag] & 21.59 & 21.37 \\
Median W2 depth [mag] & 21.02 & 20.71 \\
Median $g$ seeing [arcsec] & 1.90 & 1.49 \\
Median $r$ seeing [arcsec] & 1.68 & 1.36 \\
Median $z$ seeing [arcsec] & 1.24 & 1.28 \\
\hline
\end{tabular}
\end{center}
\tablecomments{Areal coverage and other parameters for the BASS, MzLS and DECaLS imaging surveys limited to the sky area at $\delta>-30$ deg with coverage in all of the bands.  Median depths are for point sources detected at $5\sigma$.  Median seeing values are computed with a depth-weighted average at each location on the sky.  There are 370 deg$^2$ with imaging in all three filters in the overlap area between DECaLS and BASS+MzLS.
}
\label{tab:imaging}
\end{table*}

\subsection{Milky Way Survey (MWS)}
\label{sec:ts_mws}

\begin{figure}
 \centering
    \includegraphics[width=\columnwidth]{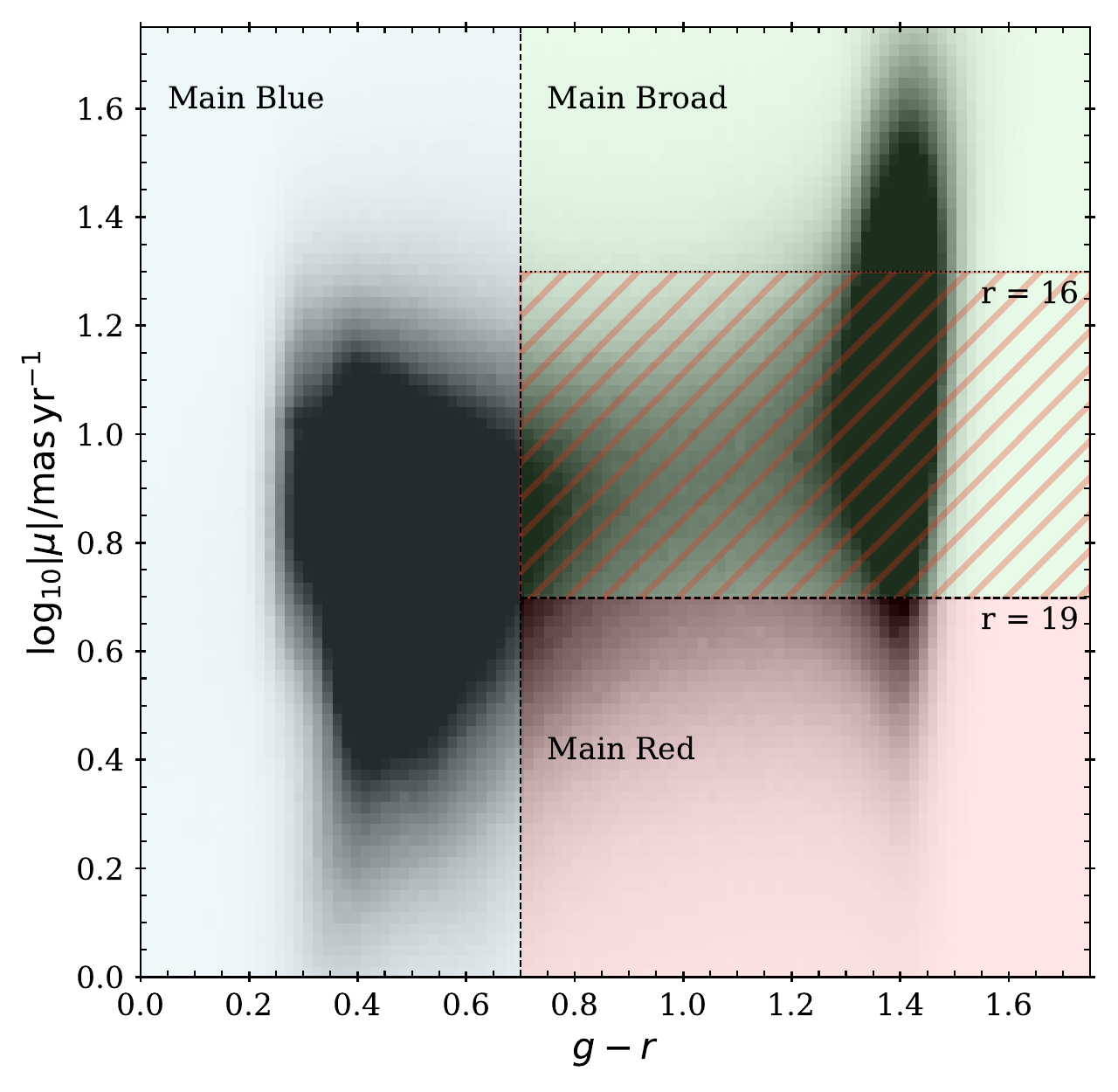}
    \caption{The distribution of stellar targets for the Milky Way Survey program as a function of color and proper motion. The two density peaks correspond to the thin disk (redder colors, higher proper motions) and the metal-poor halo and thick disk (bluer colors, lower proper motions). The blue-, red- and green-shaded regions indicate the three primary MWS target classes. All stars in the magnitude range $16 < r < 19$ are selected in one of these three categories. We do not apply any proper motion selection to stars bluer than $g-r < 0.7$ (MWS Main Blue, blue region). We divide redder stars ($g-r>0.7$) into MWS Main Broad (higher proper motion, green region) and MWS Main Red (lower proper motion, red region) using a magnitude-dependent threshold, shown by the hatched region. We give MWS Main Red stars the same fiber assignment priority as those in the MWS Main Blue sample, because they are more likely to be distant giant stars in the stellar halo. Conversely, we give MS Main Broad stars a lower fiber assignment priority. We use a more stringent proper motion threshold for fainter stars because true giants at larger distances have lower proper motions: the fiber assignment priority of a larger fraction of nearby disk stars can then be reduced without introducing a bias against high velocity giants.}
    \label{fig:main_mws_selection}
\end{figure}

Details of the scientific motivation and selection criteria for the MWS are presented in \citet{cooper22a}. 
We summarize the target selection strategy here.

As shown in Figure~\ref{fig:main_mws_selection}, the MWS is defined by three main target classes comprising an essentially magnitude limited sample of stars in the range $16 < r < 19$. 
We apply simple star-galaxy separation\footnote{The MWS star galaxy separation is different to that used for the Bright Galaxy Survey, described below.} based on {\it Gaia} EDR3 (\texttt{astrometric\_excess noise} $<3$) to Legacy Survey PSF sources and exclude sources with missing or contaminated photometry in $g$ and $r$. 
No further selection is applied to sources bluer than $g-r = 0.7$ (MWS Main Blue).
Redder sources with small {\it Gaia} parallax and proper motion (MWS Main Red), consistent with giant stars in the Milky Way's stellar halo, are separated from those more likely to be dwarf stars in the Milky Way disk (MWS Main Broad).
Sources lacking {\it Gaia} astrometry are assigned to Main Broad. 
The proper motion separation between Main Red and Main Broad increases as the square root of flux from $5\,\mathrm{mas\,yr^{-1}}$ at $r=19$ to $20\,\mathrm{mas\,yr^{-1}}$ at $r=16$ (shown by the hatched region, $0.7 <\log_{10} |\mu|/\,\mathrm{mas\,yr^{-1}} <1.3$, in Figure~\ref{fig:main_mws_selection}). 
The separation in parallax is $3\sigma_{\pi} + 0.3\,\mathrm{mas}$ where $\sigma_{\pi}$ is the parallax uncertainty reported by {\it Gaia}. 
We impose an additional cut on uncorrected flux at $r_{\mathrm{obs}}<20$.

Main Red and Main Blue targets will be given equal fiber assignment priority in the main survey program, while Main Broad targets will have lower priority.
After fiber assignment, the final dataset is expected to be approximately 33\% complete for Main Red and Main Blue targets, with slightly lower completeness for Main Broad. By default, these MWS targets will be observed at most once during the main survey.

In addition, MWS defines several classes of targets with very low surface density and high scientific value. 
These comprise white dwarf stars (selected by {\it Gaia} photometry and astrometry), stars within  100~pc of the Sun (selected by {\it Gaia} parallax), Blue Horizontal Branch (BHB) stars (selected by Legacy Survey and {\it WISE} colors) and RR Lyrae variables (selected from the {\it Gaia} variability catalog). 
In the immediate vicinity of several dwarf galaxies, globular clusters and open clusters in the DESI footprint, higher priority will be given to stars most likely to be associated with those objects, based on {\it Gaia} astrometry. These additional target categories are prioritized above the three main MWS target classes but have lower priority than BGS targets. 
An exception is made for white dwarfs, which are given higher priority than BGS galaxies because they are especially valuable both scientifically and as an additional test of flux calibration for all DESI observations. We refer to \citet{cooper22a} for details of the selection and relative prioritization of these sparse targets.

There is significant overlap between MWS Main Blue targets and the selection of metal-poor, F-type spectrophotometric standards used for all DESI survey observations.
Since standard stars will be observed in dark and bright conditions, their completeness (and the number of repeat observations) will be significantly higher than for the Main Blue sample as a whole.

The fiber assignment efficiency on MWS targets described above is expected to be 28\% from a sample with average density 1637 deg$^{-2}$ over the 14,000 deg$^2$ footprint.
Stellar densities are significantly higher towards the edges of the DESI footprint, which probe lower galactic latitudes.
To make use of any otherwise unallocated fibers in bright time observations, MWS further defines Faint Blue and Faint Red selections, separated at $g-r=0.7$ as for Main Blue and Main Red, but with magnitudes $19 < r < 20$. A weak selection for giant stars on {\it Gaia} astrometry is applied to Faint Red.
The completeness of these samples, which have the lowest priority of all DESI targets, is expected to be $\lesssim 5\%$ after fiber assignment.  
Adding these sources and including the entire footprint brings the total spectroscopic sample to 7.2 million stars, as reported in the abstract of this paper and in \citet{cooper22a}.


\subsection{Bright Galaxy Survey (BGS)}
\label{sec:ts_bgs}

\begin{figure*}
\begin{center}
    \includegraphics[width=0.95\textwidth]{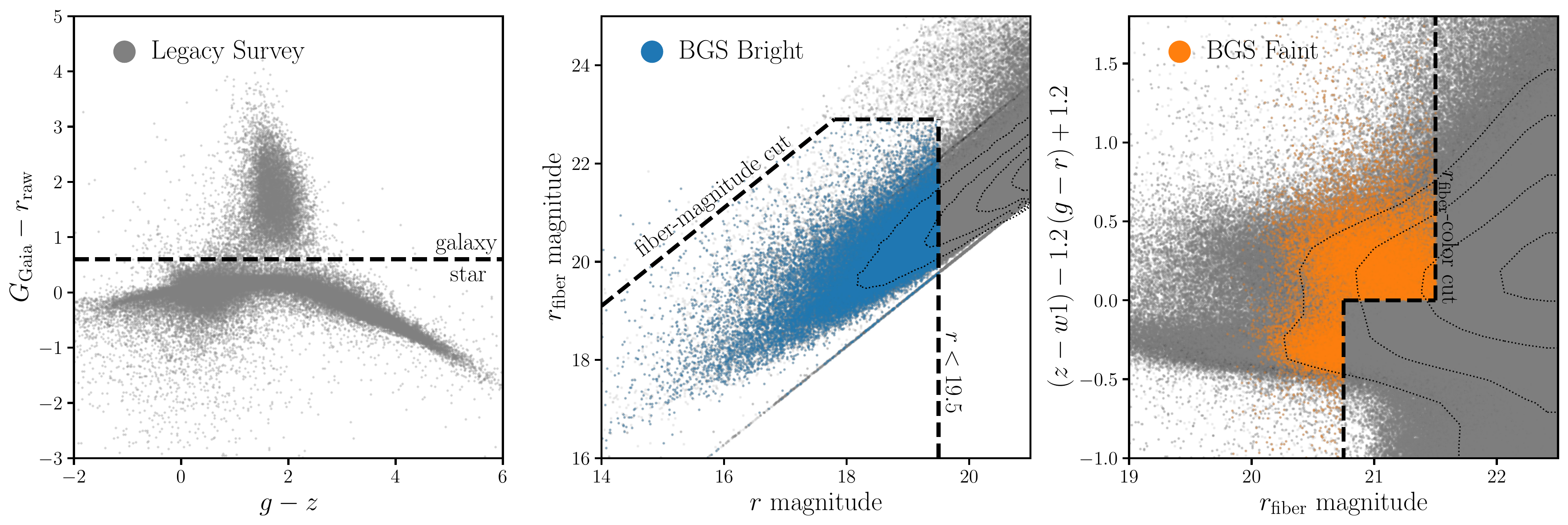}
    \caption{ \label{fig:stargalaxy}
    Representation of the target selection algorithm for the BGS program.
    {\bf Left panel:} Star-galaxy separation is performed using a $G_{Gaia} - r> 0.6$ cut (black dashed line) using $Gaia$ and Legacy Survey photometry. 
    {\bf Middle panel:} 
    The BGS Bright sample (blue) is identified using the boundaries shown by the dashed lines in the $r$ and $r_{\rm  fiber}$ magnitudes.
    No object fainter than $r_{\rm  fiber} = 22.9$ is included in the BGS Bright sample.
    {\bf Right panel:} The BGS Faint sample (orange) includes objects fainter than BGS Bright,  $19.5 < r < 20.175$, passing the ($r_{\rm fiber}$ and $(z - W1) - 1.2 (g - r) + 1.2$) cuts, illustrated by the black dashed lines.}
\end{center}
\end{figure*}

The details of the final BGS selection are presented in \citet{hahn22a}.
The BGS sample is a two tier, flux-limited, $r$-band selected sample of galaxies.
The first tier (BGS Bright) is defined by the magnitude limit $r \le 19.5$ in the DECaLs imaging areas, with a limit $r \le 19.54$ in the BASS and MzLS imaging areas to produce the same surface density of targets.
This target selection is inspired by the target selection algorithm for the SDSS main galaxy sample \citep[MGS;][]{MGS} and was chosen so that the sample includes a broad range of galaxy populations. The second tier (BGS Faint) extends the sample $r \le 20.175$.

To achieve a high completeness in galaxy targets while reducing contamination from stellar targets, we apply a star-galaxy separation for both BGS samples.
Stars with apparent magnitude $r\lesssim 20$ are sufficiently bright that they are present in the {\it Gaia} DR2 catalog.
Lower surface brightness targets with $r\lesssim 20$ that are not in the $Gaia$ catalog are, therefore, likely to be galaxies and are included in the BGS sample. 
For the objects that are in the {\it Gaia} catalogs, we compare the $G$-band magnitude from {\it Gaia} ($G_{Gaia}$) and $r$ band magnitude from DR9 Legacy imaging.
Given the similarity of the $G_{Gaia}$ bandpass to that of the $r$-band filter, the quantity $G_{Gaia}-r$ represents the difference between a PSF-fitted magnitude and a total magnitude.
As shown in left panel of Figure~\ref{fig:stargalaxy}, extended galaxies have a large $G_{Gaia}-r$ color while the locus of stars appears near $(G_{Gaia}-r)=0$,  with only a weak color dependence.
For both BGS samples, we include objects with $(G_{Gaia}-r)>0.6$ as galaxy targets.

Several additional cuts were applied to the BGS samples to reduce contamination from spurious targets that do not typically produce a valuable spectrum. 
First, we mask regions of the sky surrounding bright stars and globular clusters since these regions are typically contaminated by features such as extended halos, bleed trails, and diffraction spikes.
As an additional quality cut to remove spurious signal, we discard targets for which there are no data in one of the three $grz$ optical bands. 
We use a fiber-magnitude cut to suppress spurious signal that typically arises from imaging artifacts or fragments of `shredded' galaxies (see middle panel of Figure~\ref{fig:stargalaxy}). 
Finally, we also remove spurious objects with extreme colors from the BGS targets and very bright objects ($r>12$ and $r_{\rm  fiber}<15$) which may pollute neighboring faint fibers during DESI observations. 

Without additional filters, the BGS Faint selection would include many faint galaxies that would significantly reduce the redshift success rate of the sample.
To preferentially sample star-forming galaxies with strong emission lines \citep{AGES} and thus maintain a high redshift efficiency, we require a selection based on an $r$-band fiber aperture magnitude and $(z - W1) - 1.2 (g - r) + 1.2$ color as shown in the right panel of Figure~\ref{fig:stargalaxy}.

In addition to the BGS Bright and BGS Faint samples, BGS includes a supplementary selection to recover Active Galactic Nuclei (AGN) host galaxies that are rejected by the $G_{Gaia} - r > 0.6$ star-galaxy separation cut, but would otherwise pass BGS selection criteria.  
The presence of AGN is inferred from optical and infrared colors that trace the signatures of hot, AGN-heated dust in the spectral energy distribution (see \cite{hahn22a} for more details). 
The target density of this AGN sample is only $\sim3-4~{\rm deg}^{-2}$. 

The target densities of the BGS Bright and Faint samples are 854 deg$^{-2}$ and 526 deg$^{-2}$, respectively. 
In the main survey, BGS Bright targets will be given a higher priority when assigning fibers to ensure high completeness for this primary sample, with a typical fiber assignment efficiency of 80\% after failed spectra are taken into account.
To facilitate corrections of the BGS Faint targets for fiber assignment incompleteness, the priority of $\sim 20\%$ BGS Faint targets is randomly raised to higher priority. 
Including these upweighted targets and the effects of failed spectra, the BGS Faint sample is predicted to have $\sim 60\%$ fiber assignment efficiency, providing sufficient completeness for a range of cosmological analyses. 


\subsection{Luminous Red Galaxies (LRG)}
\label{sec:ts_lrg}

\begin{figure*}
 \centering
    \includegraphics[width=1.95\columnwidth]{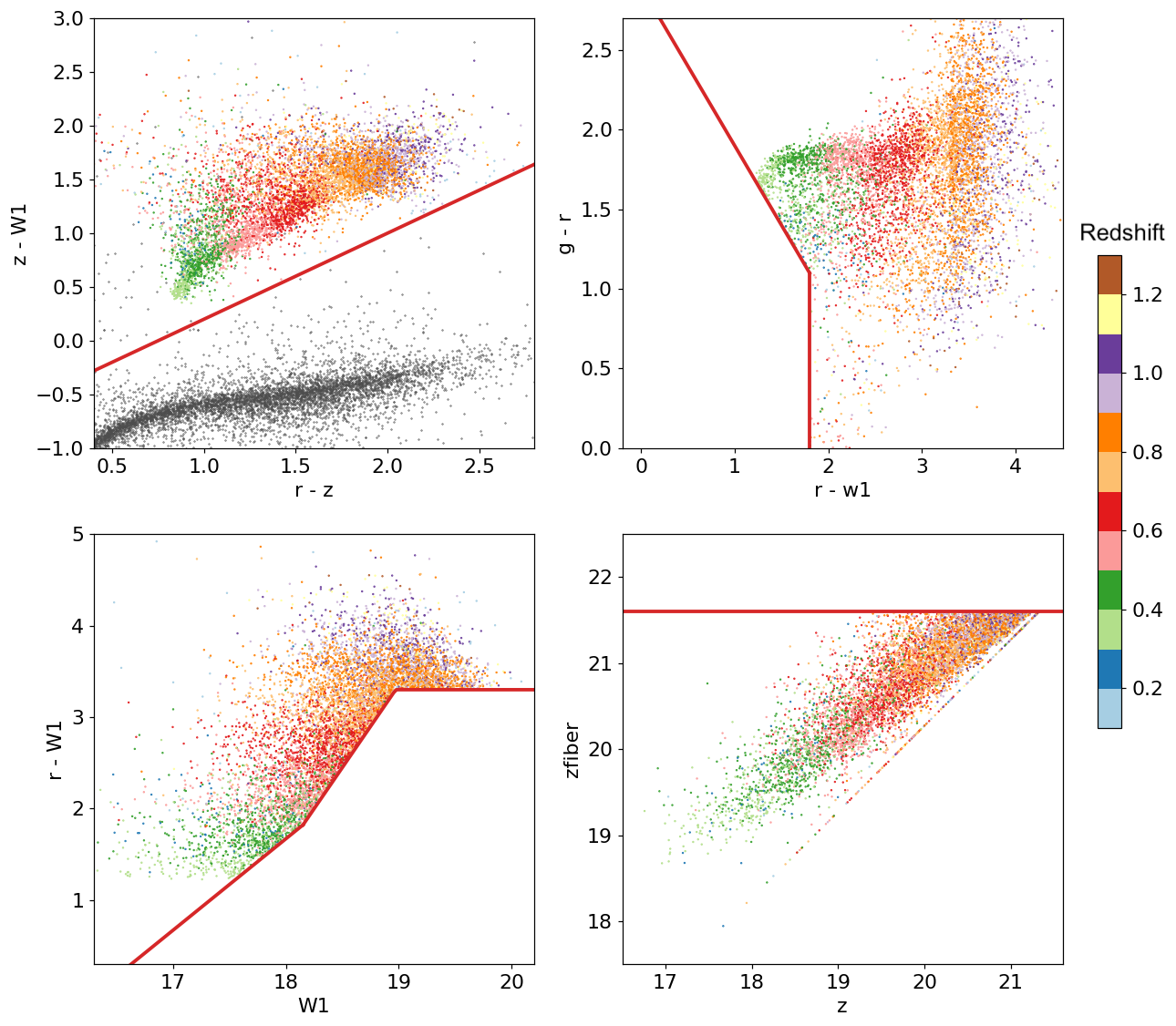}
    \caption{Selection boundaries for the LRG targets in the DECaLS footprint.  
    Redshifts are color-coded using the DESI spectroscopic redshifts. 
    The upper left panel shows the stellar rejection cut, with point sources (almost all of which are stars) in gray. 
    The upper right panel shows the cuts that remove lower-redshift galaxies and bluer galaxies.
    The lower left panel shows the color-magnitude cut that shapes the redshift distribution. 
    The lower right panel shows the magnitude limit in $z$-band fiber magnitude that ensures sufficient signal-to-noise for DESI spectroscopic observations. The objects along the diagonal line are classified as point sources in the imaging and have fixed fiber-flux to total flux ratio.}
    \label{fig:main_lrg_selection}
\end{figure*}

The details of the final LRG selection algorithm are provided in \citet{zhou22a}.
The LRG selection is tuned independently for DECaLS and MzLS/BASS catalogs to obtain roughly uniform comoving number density over the interval $0.4<z<0.8$.
With a total of 624 targets deg$^{-2}$, the surface density is approximately double that from the BOSS/eBOSS programs.
The selection maintains a high number density to higher redshifts, monotonically decreasing beyond $z=0.8$ before reaching a comoving density of $6\times10^{-5}\ h^3\mathrm{Mpc}^{-3}$ at $z=1.1$.

In the Target Selection Validation program, the {\it WISE} photometry was shown to be an effective veto against stars, with stellar contamination of less than 1\% with the selection shown in the upper left panel of Figure~\ref{fig:main_lrg_selection}.
No filters based on morphology are used to identify LRG targets, although Gaia photometry is used to remove bright stars.
 
The upper right panel of Figure~\ref{fig:main_lrg_selection} shows the criteria used to eliminate low-redshift and bluer objects. 
By requiring $g-W1>2.9$ (corresponding to the diagonal boundary line in the $g-r$ versus $r-W1$ plane), we remove galaxies with redshift $z<0.3$.  
For higher-$z$ objects, $g$-band photometry becomes less reliable, so we impose no limits on $g-W1$ for $r-W1>1.8$ galaxies. 

As shown in the lower left panel of Fig.~\ref{fig:main_lrg_selection}, the $r-W1$ color is a good proxy for redshift.
By imposing a sliding cut in $r-W1$ as a function of the $W1$ magnitude, we select the most luminous (in observed $W1$) galaxies at any redshift. 
The exact slope is trained to produce a nearly constant comoving number density of targets over $0.4<z<0.8$.
The highest redshift LRGs in our sample typically appear with the reddest colors.
These are also the faintest LRGs in our sample.
Regardless of $W1$ magnitude, all objects redder than $r-W1>3.3$ are included in order to boost the number density of the highest-redshift LRGs. 

Finally, the lower right panel of Figure~\ref{fig:main_lrg_selection} shows the faint limit used to ensure high quality spectra.
The limiting flux is defined using a magnitude measured over an aperture matched to the DESI fiber diameter ($z_{\rm fiber}$).
This fiber magnitude is more strongly correlated with the spectroscopic signal-to-noise ratio and is thus a better predictor for obtaining a successful redshift classification than the total magnitude.

LRG targets will be given a higher priority in fiber assignment than ELG targets but a lower priority than quasar targets.
In total, the LRG sample will achieve a high completeness, with a typical fiber assignment efficiency of 89\% after accounting for failed spectra.


\subsection{Emission Line Galaxies (ELG)}
\label{sec:ts_elg}

\begin{figure}
	\begin{center}
		\includegraphics[width=0.95\columnwidth]{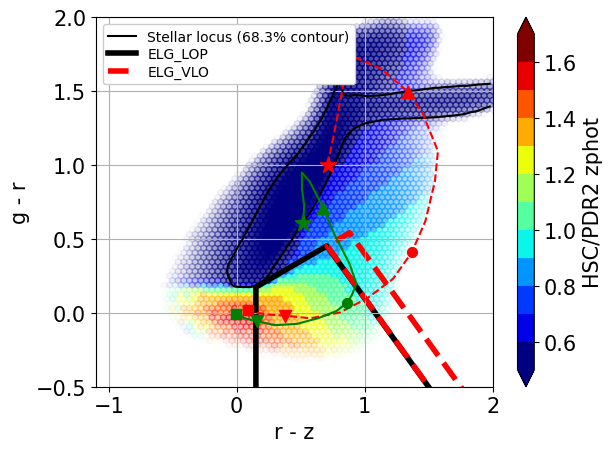}
		
			\caption{The density (illustrated by varying transparency) of $g_{\rm fiber}<24.1$ objects using Legacy imaging photometry.
			Each point is color-coded by the mean photometric redshift ($\zphot$) from the HSC/DR2 \citep{aihara19a}.
			The selection boundaries of the {\tt ELG\_LOP} sample are illustrated by the black lines while the extended selection associated with the {\tt ELG\_VLO} sample is represented by the red dashed lines.
			The green and red tracks demonstrate the redshift evolution \citep{bruzual03a} of a star-forming and passive galaxy, respectively.
			The symbols represent different epochs in that evolution: $z= 2$ (square), $z=1.6$ (downward facing triangle), $z=1.1$ (circle), $z=0.6$ (upward facing triangle), and $z=0.1$ (star).
		}
		\label{fig:maingrz}
	\end{center}
\end{figure}

The DESI ELG sample is designed to cover the redshift range $0.6 < z < 1.6$, with one of two samples selected to emphasize redshifts over the interval $1.1 < z < 1.6$.
The details of the final ELG selection algorithm are provided in \citet{raichoor22a}.
The targets selected by the algorithm optimized for the higher redshift range are labeled as the {\tt ELG\_LOP} sample, with an average target density of 1941 deg$^{-2}$ and assigned fibers at a lower priority than the quasar and LRG targets.
The ELGs in this higher redshift range will provide distinct clustering measurements at earlier epochs than can be explored with the LRG sample.
These {\tt ELG\_LOP} targets are assigned fibers at a lower priority than LRG or quasar targets, hence the acronym for the target class.
The second sample is defined by the {\tt ELG\_VLO} selection and tends to have lower redshifts but a higher redshift success rate.
The {\tt ELG\_VLO} sample has a density 463 deg$^{-2}$ and is given fibers at a very low (VLO) priority, below that of the higher redshift {\tt ELG\_LOP} sample.

In identifying ELG targets, we first impose filters on the data quality to reduce spurious signal.
We require that there is at least one observation in each of the three $grz$-bands and that the measured flux is greater than zero in all three bands.
We also reject targets that are in regions around very bright stars, large galaxies, or globular clusters.

The selection algorithm relies on fluxes measured in the $g$-band filter.
To avoid targets that are unlikely to be at a redshift $z > 0.6$, we remove all potential targets with magnitude $g<20$.
To increase the likelihood of obtaining good spectroscopic signal \citep{comparat16a}, we remove objects with $g_{fiber}> 24.1$, where the flux aperture is matched to the aperture of the DESI fibers.
 
Finally, the ELG selection algorithm is tuned to identify objects over the favored redshift range with strong [O~\textsc{ii}] line strength using colors in the $(g-r)$ vs. $(r-z)$ plane.
The general motivation for this selection is illustrated by the stellar evolution tracks in Figure~\ref{fig:maingrz} that show two evolution models of galaxies with different star formation histories. 
The star-forming galaxy exhibits bluer colors over $1.1<z<1.6$ than the passive galaxy, justifying the selection of objects that are fairly blue in both $(g-r)$ and $(r-z)$. 

More specifically, as illustrated in Figure~\ref{fig:maingrz}, the mean photometric redshift increases with decreasing $(r-z)$.
The boundary of the {\tt ELG\_LOP} sample is set at $(r-z) > 0.15$ to exclude $z>1.6$ objects where the [O~\textsc{ii}] emission line appears outside the DESI wavelength coverage.
The stellar locus appears clearly at colors that become redder in $(g-r)$ with increasing $(r-z)$, easily separable from the higher redshift galaxies.
A color-color cut with increasing $(g-r)$ as a function of increasing $(r-z)$ is common to both the {\tt ELG\_LOP} and {\tt ELG\_VLO} selections to avoid the stellar locus and reject $z< 0.6$ galaxies.
Finally, a second sliding cut with a negative slope ($(g - r) < -1.2 \times (r - z) + 1.3$) is applied to limit the sample size and to limit the number of low redshift galaxies.
The boundary for the {\tt ELG\_LOP} is set where the median redshift is approximately equal to $z=1.1$ so that higher redshift galaxies are favored.
This boundary is shifted to redder colors for the {\tt ELG\_VLO} sample to pick up objects around $z=1$ that were shown in Target Selection Validation to be spectroscopically-classified at high efficiency.

A large part of the ELG sample overlaps in redshift with the LRG sample.
This overlap will facilitate cross-correlation studies between the two tracers.
However, this overlap also means that ELGs clustered with LRGs will compete for the same fibers.
In the main program, we assign a higher priority to LRG targets than either the {\tt ELG\_LOP} or {\tt ELG\_VLO} samples, leading to a fairly high completeness in LRG targets at the cost of lower completeness in ELG targets. 
In order to obtain statistics on the ELG targets that typically get bumped by higher priority LRG targets, we increase the priority of 10\% of all ELG targets to be identical to the priority of LRG.
Fiber assignment between two objects of equal priority get resolved by a random number generator, thus allowing a full characterization of completeness using mock catalogs as in \citet{mohammad20}.

In the main survey, the {\tt ELG\_LOP} sample will achieve a fiber assignment efficiency of 69\% while the {\tt ELG\_VLO} sample will achieve a fiber assignment efficiency of 42\% after taking failed spectra into account.


\subsection{Quasars (QSO)}
\label{sec:ts_qso}

\begin{figure}
	\includegraphics[width=\columnwidth]{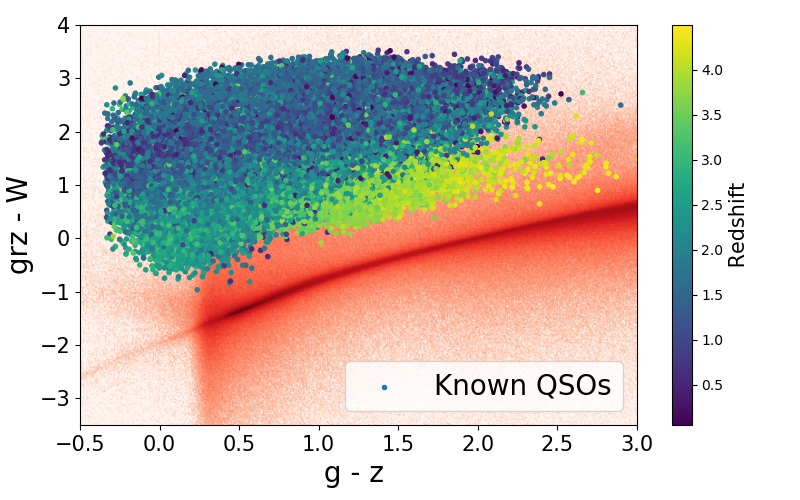}
    \caption{Colors in the optical and near-infrared  for objects photometrically classified as stars (red) and spectroscopically classified as quasars.
    Quasar redshifts are color-coded as described in the sidebar.
    Here, $grz$ is the magnitude corresponding to the weighted flux defined as flux($grz$) = [flux($g$) + 0.8$\times$flux($r$) + 0.5$\times$flux($z)$] / 2.3.
    $W$ is the magnitude corresponding to a weighted flux average defined as flux($W$)=0.75$\times$flux($W1$)+0.25$\times$flux($W2$).
    }
    \label{fig:qso_colors}
\end{figure}

The selection of quasar samples has historically relied on identification through excess UV emission \citep{Richards2002, Ross2012}. 
In DESI, we use an alternative approach that relies on flux excess in near-infrared bandpasses instead, as demonstrated in eBOSS~\citep{Myers2015}. 
We use three optical bands ($g, r, z$) combined with $W1$ and $W2$ photometry to select our primary sample of quasars.
The separation between stars and quasars allowed by optical and infrared colors is illustrated in Figure~\ref{fig:qso_colors}.
The relatively blue color of stars is due to the rapidly declining tail of the blackbody spectrum at infrared wavelengths.
The relatively red color of quasars is due to the onset of infrared emission from the dusty torus \citep{hickox18}, leading to a flatter spectral energy distribution.

For BOSS observations, an algorithm based on a neural network for selecting quasars was found to increase selection efficiency by approximately $20$\% compared to a selection based on strict boundaries in color and magnitude \citep{Yeche2010}.
To further improve the efficiency of target selection for DESI, we use a new machine-learning algorithm based on Random Forests (RF). 
At a fixed density, the observations from the Target Selection Validation program demonstrated that the random forest produced 15\% more quasars than the alternate color selection, including 21\% more quasars at $z>2.1$ for Ly-$\alpha$ forest measurements.
The RF selection was therefore chosen for the main selection of quasars.

\begin{figure*}[t!]
  \centering
	  \includegraphics[width=0.9\textwidth, angle=0]{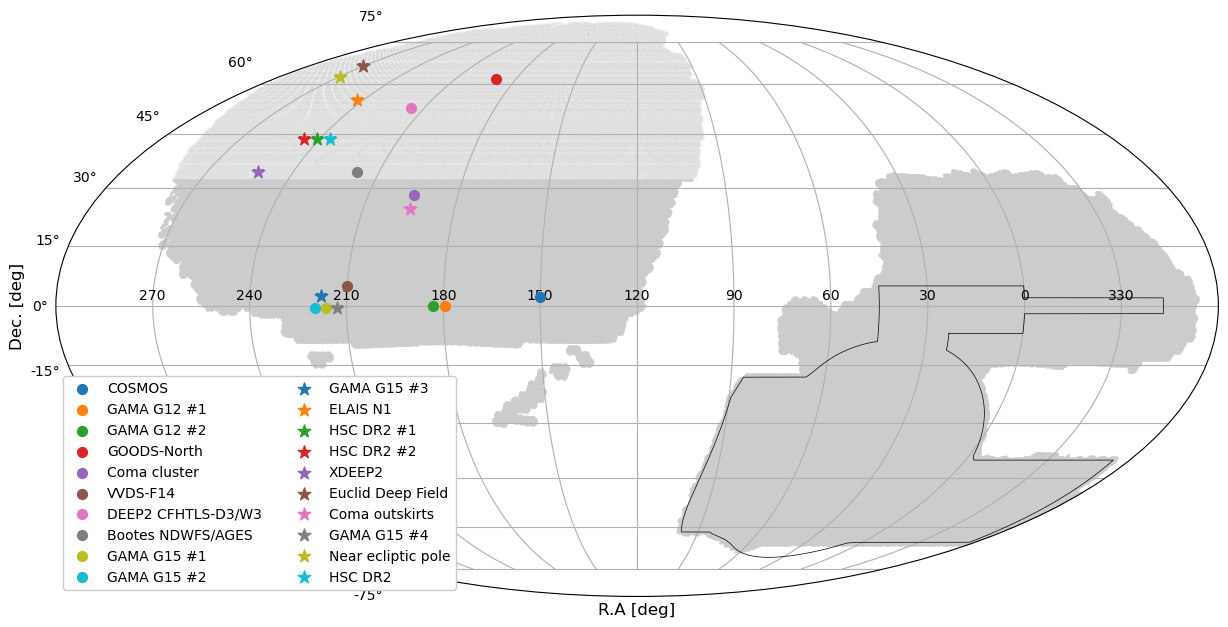}
	  \caption{The field centers for the 20 rosette pointings in the One-Percent Survey, with same shading scheme as in Figure~\ref{fig:mainSV}
	  }
	  \label{fig:1percent}
\end{figure*}

Before utilizing the RF, we only include objects that have stellar morphology (`PSF') as identified in the legacy imaging.
This rejection reduces contamination from extended source galaxies by an order of magnitude.
We also require targets to have $16.5<r_{\rm AB}<23.0$, with the bright limit set to remove residual stars and the faint limit set to ensure high quality spectral data. 
Finally, to further reject stars with low infrared flux, we require $W1<22.3$ and $W2<22.3$. 
To obtain training samples not biased by anterior color selections, we initially used quasars selected by their intrinsic time-variability~\citep{Palanque2011} in SDSS and `stars' that exhibited no significant variation in their SDSS light curves.
This selection was used for the Target Selection Validation target sample.
We then re-trained the RF selection on DESI spectra using 11 input parameters: the 10 possible colors using the five optical and NIR bands $grzW1W2$, and the $r$-band magnitude. 
By using correlated parameters in the 10 color measurements, the RF is trained to assign more or less importance to each input parameter. 

Finally, the RF probabilities were tuned to obtain quasar targets, leading to a sample of targets at $\sim 310$ deg$^{-2}$.
To ensure a uniform target density over the full DESI footprint, we apply slightly different $r$-dependent thresholds on the RF probability in the three regions (North, South (DES) and South (non-DES) \citep[for more details, see][]{chaussidon22a}. 
Because quasar targets will be given the highest priority during the five year program, fiber assignment efficiency will
be 99\%.
In addition, quasars classified at $z>2.1$ after a single epoch of observation will be prioritized for up to four observations.
Simulations of fiber assignment indicate that on average, a $z>2.1$ quasar will receive 3.4 epochs of observation, leading to increased signal-to-noise in the measurements of the Ly-$\alpha$ flux density field.

\subsection{One-Percent Survey}

Immediately after the Target Selection Validation program, the One-Percent Survey was completed to obtain a DESI-like sample across all target classes.
These data were used to determine the efficiency of automated routines for data acquisition and to create a sample that was highly complete  in both fiber assignment and redshift classification in all target classes over roughly 1\% of the final DESI footprint.
The clustering measured in these data will be used to calibrate the halo occupation statistics in the DESI mock catalogs, allow early clustering measurements, and provide a sample that is comparable in size to previous spectroscopic programs for studies of stellar, galaxy, and quasar physics, but to fainter magnitudes.
The footprint of these observations can be found in Figure~\ref{fig:1percent} and details of the targeted fields are given in Table~\ref{tab:onepercent}.

\begin{table*}[t]
\caption{List of fields included in the One-Percent Survey.}
\begin{center}
\begin{tabular}{lccccccc}
\hline
\hline
Field & \multicolumn{2}{c}{Field Center}  & MWS  & BGS  & LRG  & ELG  & Quasar \\
&  $\alpha$ (deg)  & $\delta$ (deg) & completeness & completeness & completeness & completeness & completeness \\
\hline
COSMOS & 150.10 & 2.18 & 0.96 & 0.99 & 0.99 & 0.95 & 0.99 \\
GAMA G12 \#1 & 179.60 & 0.00 & 0.97 & 0.99 & 0.99 & 0.97 & 0.99 \\
GAMA G12 \#2 & 183.10 & 0.00 & 0.98 & 1.00 & 0.99 & 0.96 & 1.00 \\
GOODS-North & 189.90 & 61.80 & 0.98 & 0.99 & 0.99 & 0.98 & 1.00 \\
Coma cluster & 194.75 & 28.20 & 0.96 & 0.99 & 0.99 & 0.97 & 1.00 \\
Coma outskirts & 194.75 & 24.70 & 0.98 & 0.99 & 0.99 & 0.95 & 1.00 \\
VVDS-F14 & 210.00 & 5.00 & 0.96 & 0.99 & 0.99 & 0.95 & 1.00 \\
GAMA G15 \#4 & 212.80 & -0.60 & 0.94 & 0.98 & 0.99 & 0.93 & 1.00 \\
DEEP2 CFHTLS-D3/W3 & 215.50 & 52.50 & 0.99 & 1.00 & 0.98 & 0.96 & 0.99 \\
GAMA G15 \#1 & 216.30 & -0.60 & 0.96 & 0.99 & 0.98 & 0.93 & 0.99 \\
Bootes NDWFS/AGES & 217.80 & 34.40 & 0.97 & 0.99 & 0.99 & 0.96 & 0.99 \\
GAMA G15 \#3 & 218.05 & 2.43 & 0.96 & 0.99 & 0.99 & 0.94 & 1.00 \\
GAMA G15 \#2 & 219.80 & -0.60 & 0.96 & 0.99 & 0.98 & 0.93 & 0.99 \\
HSC DR2 & 236.10 & 43.45 & 0.99 & 1.00 & 0.98 & 0.94 & 1.00 \\
HSC DR2 \#1 & 241.05 & 43.45 & 0.98 & 0.99 & 0.97 & 0.95 & 0.97 \\
ELAIS N1 & 242.75 & 54.98 & 0.93 & 0.98 & 0.99 & 0.97 & 1.00 \\
HSC DR2 \#2 & 245.88 & 43.45 & 0.95 & 0.99 & 0.98 & 0.95 & 1.00 \\
XDEEP2 & 252.50 & 34.50 & 0.95 & 0.99 & 0.99 & 0.97 & 1.00 \\
Euclid Deep Field & 269.73 & 66.02 & 0.96 & 0.99 & 0.99 & 0.97 & 1.00 \\
Near ecliptic pole & 269.73 & 62.52 & 0.95 & 0.99 & 0.97 & 0.95 & 1.00 \\
\hline
Average &  &  & 0.962 & 0.989 & 0.986 & 0.952 & 0.994 \\
\hline
\end{tabular}
\end{center}
\tablecomments{Completeness values correspond to the fraction of targets that received a \textbf{valid} observation within each target class.
Completeness is computed using all targets that lie a distance $0.2^\circ$ to $1.45^\circ$ from the field center.}
\label{tab:onepercent}
\end{table*}

The target selection algorithms for the One-Percent Survey were nearly the same as for the five year program, with only two minor modifications.
First, the magnitude limit in the selection algorithm for the BGS Faint sample was increased to $r < 20.3$, rather than $r < 20.175$ as in the main survey.
This led to an increase in the total surface density of the BGS Bright and the BGS Faint samples to 1,480~deg$^{-2}$ in the One-Percent Survey. 
Second, the selection algorithm for LRG targets was modified to be somewhat more inclusive in the One-Percent Survey.
The faint limit was moved 0.1 magnitude fainter in $z_{fiber}$ and the sliding cut in $r-W1$ versus $W1$ was moved to include slightly fainter galaxies.  
The color-color cut in $g-r$ versus $r-W1$ was also shifted toward slightly bluer colors to increase the number density of lower redshift galaxies.

The observations for the One-Percent Survey were conducted in nearly the same mode as expected for the five year program.
The summary of observations can be found in Table~\ref{tab:SV_fields}.

\begin{figure}[h]
  \centering
	  \includegraphics[width=0.48\textwidth, angle=0]{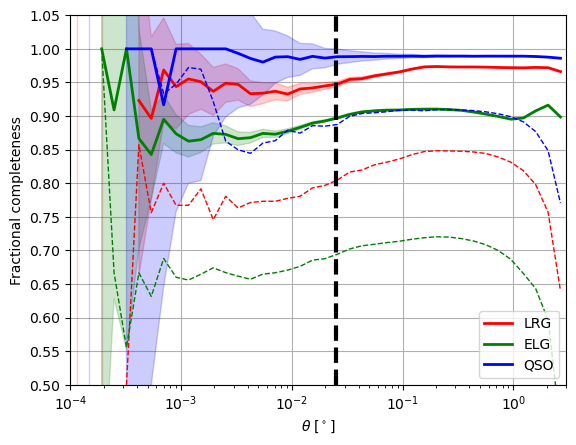}
	  \caption{The completeness for pairs of targets in the One-Percent Survey as a function of angular separation.
   Pairs found in the regions that lie between 0.2 and 1.45 degrees from each rosette field center are shown with solid lines, demonstrating the areas of highest completeness. Completeness of pairs with no cut on the distance to the rosette centers is presented in dashed lines.
   The pairwise completeness will be lower for the main survey when fewer tiles are dedicated to each coordinate on the sky.
	  }
	  \label{fig:pairwise}
\end{figure}

LRG, ELG, and quasar targets were observed during the times when sky background, seeing, and transparency were most conducive to spectroscopy of faint sources.
In most cases, these targets were observed using the automated field selection when conditions indicated at least a 40\% survey speed.
To achieve the goal of very high completeness in all target classes, the tiling pattern followed a rosette pattern over 20 unique field centers.
Each rosette consisted of at least 12 individual tiles offset by 0.12 deg from the field center. 
A minimum of 11 these tiles were observed for each rosette.
Targets within a 7.0 deg$^2$ annulus had significant coverage, while those targets over an additional 2 deg$^2$ were observed with fewer visits and lower completeness in fiber assignment.
Up to three additional visits were given to targets that produced spectra without secure redshift estimates.
These additional visits were assigned only if there were no unobserved LRG, ELG, or quasar targets competing for the fiber.
Over the regions between 0.2 and 1.45 degrees from the field center, spectra were successfully obtained for 99\% of LRG targets, 95\% of ELG targets, and nearly 100\% of quasar targets.  
The completeness statistics for each field are found in Table~\ref{tab:onepercent} and the completeness as a function of pairwise separation is presented in Figure~\ref{fig:pairwise}.
At such a high completeness in fiber assignment, the primary LRG sample for the main survey is nearly fully covered, even with the extensions described above.

BGS and MWS targets were observed during the times with slightly degraded conditions, where good signal-to-noise would still be achieved due to the higher source fluxes.
In some cases, BGS and MWS tiles were observed during conditions that would have otherwise been considered for LRG, ELG, quasar observations because we needed the tiles to fill observing queue for the entire night.
The same rosette pattern was used for these targets, but with a minimum of 10 to be observed for each field.
Spectroscopy was obtained for 99\% for BGS Bright and for BGS Faint, and 96\% for MWS targets.
As with the LRG program, the extension to fainter magnitudes had little impact on the fiber assignment completeness for the BGS Faint sample.
Following the same strategy as above, additional fibers were given to targets that did not produce robust spectral classifications, with final statistics for all fields found in Table~\ref{tab:onepercent}.

Unlike the Target Selection Validation program, the information from the guide cameras and sky monitors was used to compute the exposure times.
Following the definition of effective exposure times in Section~\ref{sec:observations}, exposures of LRG, ELG, and quasar targets were taken to an equivalent of 1200 seconds, a factor 1.2 longer than the design for the main program.
Exposure times of BGS and MWS targets were also increased by a factor of 1.2. 
The increased exposure time provided margin against lingering uncertainties in calibration and real-time estimates of accumulated signal-to-noise.

In total, the One-Percent Survey produced 939,000 secure spectral classifications in only one month.
The summary over all target classes compared to other major spectroscopic programs can be found in Table~\ref{tab:samplesizes}.
In comparing to eBOSS, we use only data from the One-Percent dark time program.
DESI observed about half of the number of targets, but required a factor of 38 less exposure time.

\begin{table}[t]
\caption{Sample sizes for each target class in the one-month DESI One-Percent Survey compared to related spectroscopic samples \citep{gama09,dawson16a}. 
We report here the number of observed, secure redshifts. For DESI, we use the criteria described in Section~\ref{sec:results}, except for MWS where we simply use ZWARN=0 and SPECTYPE=STAR.}
\begin{center}
\begin{tabular}{lccc}
    \hline
    \hline
    Target Type & GAMA & eBOSS & DESI\\
    & & & One-Percent\\
    \hline
     LRG &  & 232k & 139k\\
     ELG &  & 223k & 298k\\
     QSO &  & 545k & 38k\\
     BGS & 150k &  & 252k\\
     MWS &  &  & 212k\\
    \hline
\end{tabular}
\end{center}
\label{tab:samplesizes}
\end{table}

\section{Results}\label{sec:results}
Following the reductions of the SV data, the one-dimensional spectra were modeled as a function of redshift and spectral type.
The modeling software for this classification is named `Redrock' \citep{bailey22a}\footnote{\url{https://github.com/desihub/redrock/releases/tag/0.15.4}}
and follows a procedure similar to BOSS spectral characterization \citep{bolton12}.
Redrock is run on all spectra, but in this section we ignore statistics for those spectra that were assigned a flag corresponding to a failed spectrum in the data reductions.

The primary methodology within Redrock is a $\chi^2$ minimization computed from a linear combination of spectral templates over all trial redshifts.
A suite of stellar, galaxy, and quasar templates is fit to each spectrum over a unique redshift range appropriate to each spectral class.
The linear combination that gives the best fit to the data over all redshifts is assumed to be the best model.
The key parameters describing the best fit are the redshift, the redshift uncertainty, the spectral class (star, galaxy, or quasar), the coefficients to the spectral templates, the $\chi^2$, and the value $\Delta \chi^2$.
The parameter $\Delta \chi^2$ is defined as the difference in $\chi^2$ between the fit at the most likely redshift and the fit at the secondary minima in the $\chi^2$ function that denotes the the second best fit to the data.
$\Delta \chi^2$ therefore characterizes the likelihood that the best-fit redshift is correct, with higher values of $\Delta \chi^2$ reflecting an increased probability that the estimate is correct.

Visual inspections of galaxy spectra \citep{lan22a} and quasar spectra \citep{alexander22a} were used to provide a first estimate of Redrock performance.
These visual inspections were performed on composite spectra using all available exposures in the deepest fields in Table~\ref{tab:SV_fields}.
Because the effective exposure times for these fields were roughly ten times longer than planned in the initial survey design, visual inspectors were able to classify spectra at high confidence relative to the main survey, identify sources of spurious signal, and find common failure modes in the Redrock modeling.
The feedback on spurious signal informed improvements to the data reduction algorithms in successive internal releases.
In finding common failure modes in the Redrock modeling, the visual inspection process also allowed the DESI collaboration to customize algorithms for reliable spectral classification to each target class.

Once an algorithm was defined for determining whether a redshift was reliable, we used the spectra in the Target Selection Validation data sample to empirically determine the quality of redshift estimates.
In all cases, we subsampled the spectra to have effective exposure times of roughly 180 seconds for BGS and MWS targets and 1000 seconds for LRG, ELG, and quasar targets.
With these data splits, we quantified the following for each target sample: 
\begin{itemize}
    \item the total redshift efficiency, which is simply the fraction of spectra that produce a reliable redshift;
    \item the target redshift efficiency, which is the fraction of spectra that produce a reliable redshift in the desired redshift range; 
    \item the catastrophic failure rate, which is the fraction of targets assigned an incorrect redshift; and 
    \item the statistical precision of the redshift estimates.
    \end{itemize}
Catastrophic failure rates and redshift precision were computed using either pairs of repeated spectra, or individual exposures relative to a much deeper reference exposure.
For those spectra that are assigned a reliable redshift in the desired redshift range, the catastrophic failure rate is defined as the fraction of spectra that produce a pairwise velocity difference exceeding 1000 km s$^{-1}$ for galaxies and 3000 km s$^{-1}$ for quasars.
In this computation using pairs of repeated spectra, we implicitly assume that one redshift in the pair is correct, so the catastrophic failure rate is equal to half the fraction of pairs that produce discrepant redshifts.
When using deeper spectra as a reference, we implicitly assume that the redshift in the higher quality spectrum is correct.
Where possible, we used redshift estimates from other surveys to assess the systematic errors in these redshift estimates.

In what follows, we describe the custom algorithm for determining the redshift of each target class.
Using the Target Selection Validation sample, we report the statistics regarding total redshift efficiency, target redshift efficiency, catastrophic failure rate, and statistical redshift precision for each sample.
These statistics are reported in the other papers associated with the target selection validation, particularly those that describe the visual inspection process \citep{lan22a,alexander22a}.
Occasional differences in the reported values are due to different assumptions in the samples.
We then use the custom algorithms for each target class to summarize redshift completeness as a function of exposure time, thus setting the conditions for the exposure sequence in the main survey.
We conclude with a summary of the performance on all spectroscopic samples compared to the requirements that drove the instrument design as described in \citet{desi-collaboration22a}.
The results are found in Tables~\ref{tab:level2} and ~\ref{tab:level2_cont}.

\subsection{Redshift Determination}

\subsubsection{Milky Way Survey (MWS)}
\label{sec:res_mws}

Beyond the Redrock classification of stellar spectra and radial velocities, several MWS measurements (including radial velocities, stellar parameters and chemical abundances) will be obtained by running additional template-fitting codes specialized for stellar spectra.
As described in \citet{cooper22a}, two of these codes are developed for all stars while one is specific to white dwarfs. 
These additional algorithms will be run on all MWS targets (regardless of their classification by Redrock) and on any other sources that are classified as stars by Redrock. 
Here we show results from the RVSpecfit \citep{koposov11a,rvspecfit} code, which performs least-square fitting of stellar DESI spectra by interpolating between spectral templates to obtain radial (line-of-sight) velocities and stellar parameters. 

\begin{figure}
    \centering
    \includegraphics{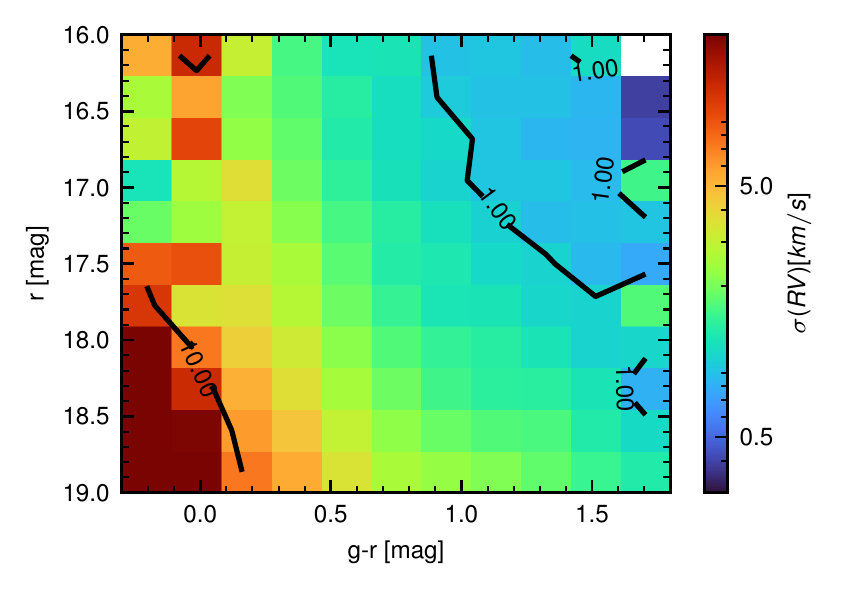}
    \caption{The precision of radial velocity estimates from the stellar radial velocity pipeline as a function of color and magnitude. The precision is measured in each color-magnitude bin by determining the 68\% confidence region in offsets between radial velocities from 100-300 second individual exposures and radial velocities measured from deep exposures.}
    \label{fig:cmd_rv_accuracy}
\end{figure}

The radial velocity precision of stars based on fits with RVSpecfit is illustrated in Figure~\ref{fig:cmd_rv_accuracy}. 
Only individual exposures are included with effective exposure times ranging from 100 to 300 seconds to sample the expected exposure times in the full MWS survey.
The mean velocity measured from deep stacked exposures with effective exposure time larger than 1000 seconds is subtracted from each individual measurement to capture the statistical errors in each measurement.
We then use the residuals in bins of color and magnitude to compute the standard deviation as estimated from the difference between the 84th and 16th percentiles. 
The RV precision determined in this way is somewhat worse than the formal uncertainty returned from spectral fitting due to additional velocity systematic errors likely associated with wavelength calibration at the level of $\sim 1\, \mathrm{km\,s}^{-1}$ \citep[see further discussion in][]{cooper22a,guy22a}.
Using measurements with radial velocity deviating from the value measured from the stack by more than $20\, \mathrm{km\,s}^{-1}$, the fraction of `catastrophic' errors is 0.6\% for stars in the color magnitude range of the MWS survey $0<g-r<2$ and $16<r<19$.
Some fraction of these may be due to stars with binary motions.

\subsubsection{Bright Galaxy Survey (BGS)}
\label{sec:res_bgs}

The BGS sample is largely magnitude-limited and thus tuned for lower redshifts than the LRG, ELG, or quasar samples.
The redshift range $0<z<0.4$ is covered at very high density with minimal overlap (roughly 5\%) with the other tracers.
These redshifts are assumed as the target range for BGS clustering studies.
Good redshifts for BGS galaxies are those that produce a $\Delta \chi^2 > 40$, a spectral classification of `galaxy', and a reported statistical redshift uncertainty less than $0.0005 (1 + z)$.
We report the performance of Redrock for both the BGS Bright and BGS Faint samples according to this definition for a reliable redshift.
We find a high redshift success rate for all magnitudes, with redshift completeness exceeding 95\% even near the magnitude limits of the BGS Bright sample.

\begin{figure*}
\begin{center}
    \includegraphics[width=0.9\textwidth]{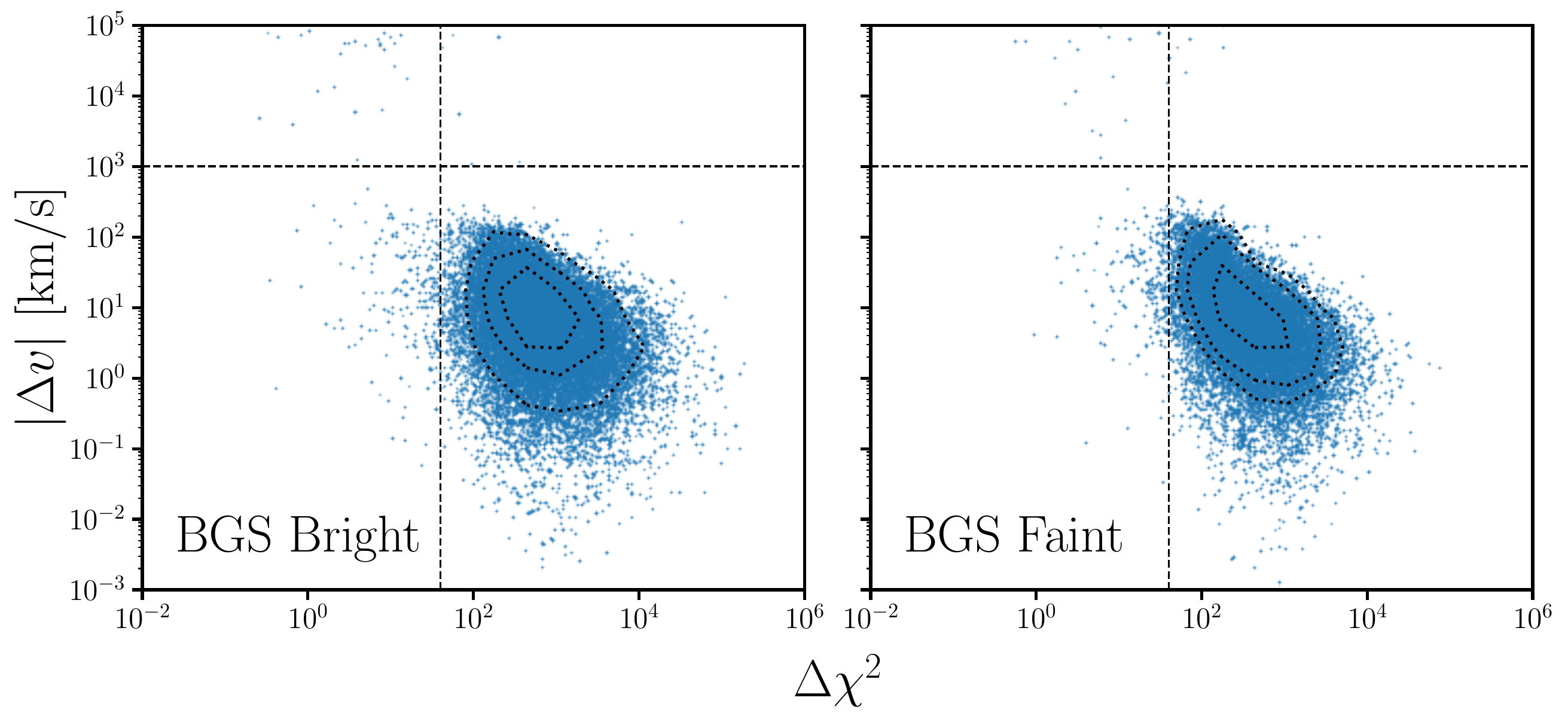}
    \caption{The difference in redshift ($\mathrm{km\,s}^{-1}$) between individual observations and deep exposures of the same BGS target.
The $\Delta \chi^2$ assigned to each data point is taken from the single epoch observation.
Spectra are only included if they were characterized as having a good redshift at $z<0.4$.
The vertical line represents the threshold for the $\Delta \chi^2$ value of a good redshift  while 
the horizontal line represents the limit at which a redshift discrepancy is considered a catastrophic failure.
The left panel shows the distribution of pairs for the BGS Bright sample while the right panel shows the distribution for the BGS Faint sample.
}
\label{fig:BGS_dv}
\end{center}
\end{figure*}

The velocity differences relative to deep coadds for BGS Bright and BGS Faint targets at redshifts $z<0.4$ are presented in the left and right panels of Figure~\ref{fig:BGS_dv}, respectively.
Because of its small size, we do not report statistics for the BGS AGN sample.
We compared the redshift estimates from DESI to those from the DEEP2 survey \citep{newman13} for objects in common between the two surveys.
We find average systematic offsets of only $6.5 \pm 1.7\, \mathrm{km\,s}^{-1}$.

\subsubsection{Luminous Red Galaxies (LRG)}
\label{sec:res_lrg}

We use the redshifts $0.4<z<1.1$ as the target range for the LRG sample.
To reject incorrect redshifts, we require that observed LRG spectra meet the following quality cuts: $\Delta \chi^2>15$, $z_\mathrm{redrock}<1.5$, and the redshift warning flag, ZWARN=0.
Here, the ZWARN flag is a bitwise output from redrock that captures the quality of the model fit.
A value of zero indicates that there is no clear evidence for a corrupted redshift estimate.
The quality cuts remove roughly 1.1\% of the observed LRG targets. 
Based on comparison with deep observations, we estimate that roughly 0.2\% of the LRGs that meet the quality cuts are catastrophic failures.

\begin{figure}
\begin{center}
    \includegraphics[width=0.45\textwidth]{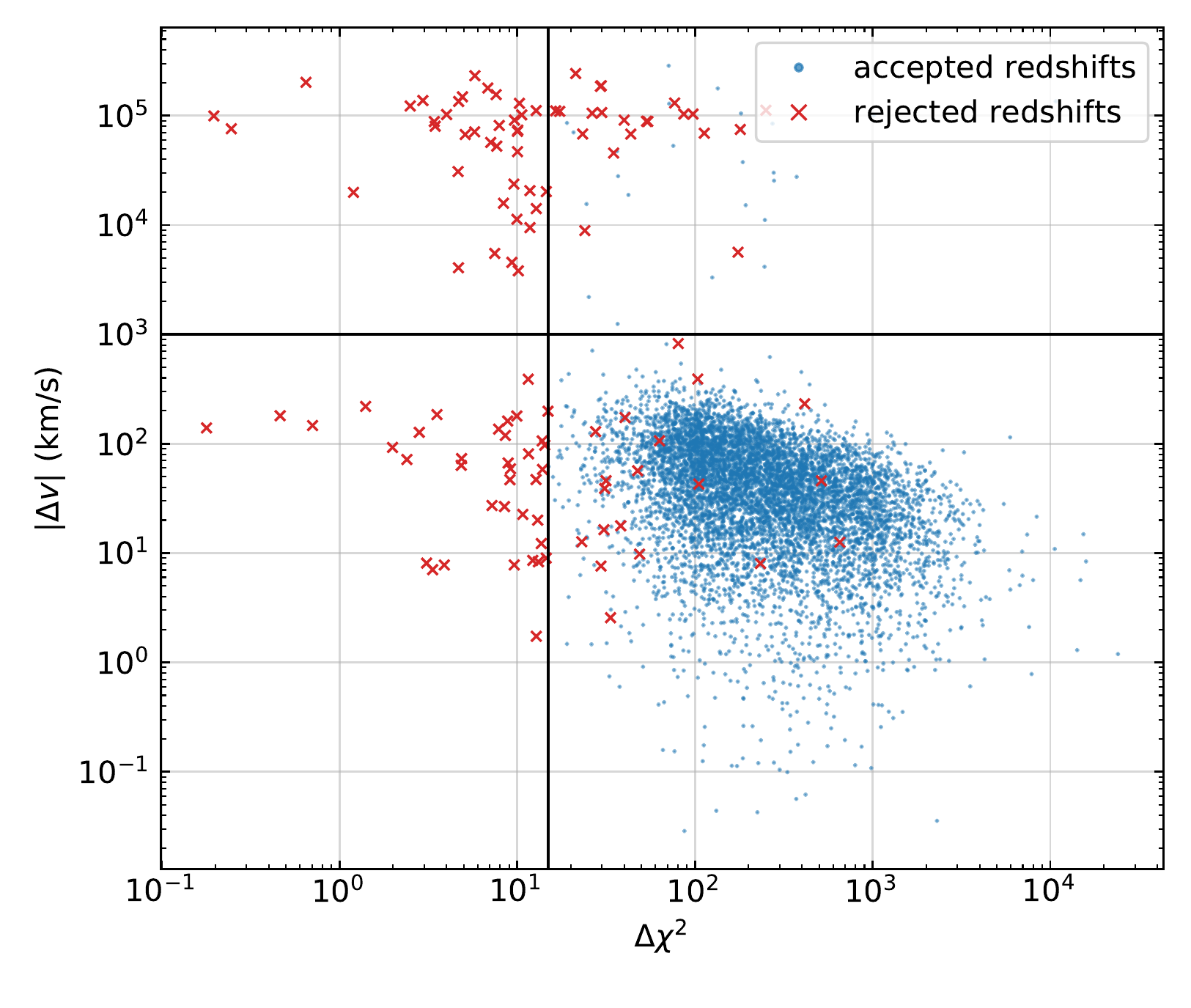}
    \caption{The difference in redshift between pairs of observations of the same LRG target.
The vertical line represents the $\Delta \chi^2>15$ threshold for an accepted redshift while 
the horizontal line represents the limit at which a redshift discrepancy is considered a catastrophic failure. The LRG redshift quality cuts also rejects any object with $z>1.5$, which causes some of the redshifts with $\Delta \chi^2>15$ to be rejected (red).}
\label{fig:LRG_dv}
\end{center}
\end{figure}

The preliminary algorithm for classifying spectra produces a total redshift efficiency of 98.9\%, and a target redshift efficiency (i.e., LRGs in $0.4<z<1.1$ with secure redshifts) of 89.4\%.
The pairwise velocity differences are presented in Figure~\ref{fig:LRG_dv}.

To estimate the systematic errors in the LRG redshifts, we compare the DESI redshifts with redshifts of the same objects from SDSS, BOSS, eBOSS \citep{ross20a} and the DEEP2 Survey \citep{newman13}. We find average offsets of less than $10\, \mathrm{km\,s}^{-1}$.

\subsubsection{Emission Line Galaxies (ELG)}
\label{sec:res_elg}

\begin{figure*}
\begin{center}
    \includegraphics[width=0.45\textwidth]{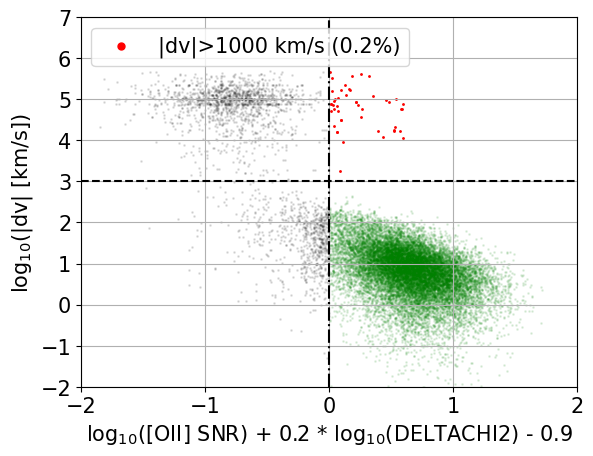}
    \includegraphics[width=0.45\textwidth]{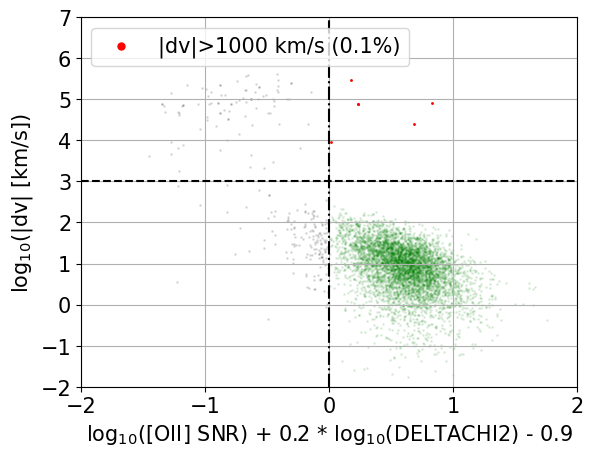}
\caption{The difference in redshift ($\mathrm{km\,s}^{-1}$) between pairs of observations taken of the
same ELG target.
A linear combination of log(\snroii) and log(\deltachisq) is used to determine whether a redshift is reliable.
The vertical line represents the threshold for the preliminary value where we assume a good redshift estimate.
The horizontal line represents the limit at which a redshift discrepancy is considered a catastrophic failure.
{\bf Left:}  The distribution of pairs for the {\tt ELG\_LOP} sample. 
{\bf Right:}  The distribution for the {\tt ELG\_VLO} sample.  
In both panels, a measurement is only included if one spectrum in the pair was characterized as having a redshift estimate in the range $0.6<z<1.6$ and if both measurements of the pair have a valid estimate of \snroii\ and \deltachisq.
}
\label{fig:ELG_dv}
\end{center}
\end{figure*}

The {\tt ELG\_LOP} targets were optimized for redshifts $1.1<z<1.6$, while the {\tt ELG\_VLO} targets were optimized to cover the full redshift range $0.6<z<1.6$.
The lower redshift, {\tt ELG\_VLO} sample produces a higher fraction of reliable classifications because there is more information from the continuum and additional emission lines.
In what follows, we present results for both the {\tt ELG\_LOP} and {\tt ELG\_VLO} targets over the interval $0.6<z<1.6$.

The same criteria are used to assign reliable redshifts to both the {\tt ELG\_LOP} and {\tt ELG\_VLO} samples.
Because the ELG spectra will be the faintest targets observed in DESI, the spectra typically yield flux measurements at a low signal-to-noise ratio over the continuum, with high signal-to-noise measurements localized to the \oii\ emission regions.
However, redshifts that are reliably estimated based on flux from the \oii\ doublet may not have a large \deltachisq\ value because of the small number of pixels and the lack of meaningful information over the continuum.
For this reason, we use the measurement \snroii\ in addition to \deltachisq\ for ELG redshift estimates reported here.
\snroii\ is a measurement of the signal-to-noise ratio of the \oii flux customized to ELG spectra.
By applying a cut in the (\snroii, \deltachisq) plane, we can reliably estimate the redshift based on the \oii doublet in cases where a low value of \deltachisq\ may otherwise indicate a poor redshift estimate.
Specifically, we adopt the preliminary selection $\rm{log}_{10} (\snroii) > 0.9 - 0.2 \times \rm{log}_{10} (\deltachisq)$ for determining whether a redshift estimate is reliable.

The preliminary algorithm for classifying spectra produces total redshift efficiencies of 70\% and 95\% for the {\tt ELG\_LOP} and {\tt ELG\_VLO} samples, respectively.
{\bf The pairwise velocity differences for {\tt ELG\_LOP} ({\tt ELG\_VLO}) targets over the target redshift range are presented in the left panel (right panel) of Figure~\ref{fig:ELG_dv}.}

To estimate the systematic errors in redshift, we compare the results from individual galaxies that DESI observed in common with eBOSS \citep{raichoor19a} or the DEEP2 survey \citep{newman13}.
In both cases, we find average offsets of only $1 \pm 0.4\, \mathrm{km\,s}^{-1}$, indicating that the systematic errors are small enough to be ignored.

\subsubsection{Quasars (QSO)}
\label{sec:res_qso}

The quasars will have the largest redshift range of all samples in DESI, reaching redshifts $z>5$.
Previous BAO studies \citep[e.g.][]{hou19a,neveux20} used quasars as discrete tracers over the redshift range $0.8<z<2.2$ while those using the Ly-$\alpha$ forest \citep{2019duMasdesBourbouxH} relied on quasars at redshifts $z>2.1$.
Although the redshift ranges for DESI cosmology studies are yet to be established, we report redshift performance statistics for quasar tracers over two independent redshift ranges for simplicity.
We assume target redshift ranges of $0.9<z<2.1$ and $z \ge 2.1$ for discrete tracer and Ly-$\alpha$ forest quasars, respectively.
Those spectra that are not classified as a quasar are not included in the statistics of redshift performance.

The visual inspection process indicated that the quasar selection produces spectra of which 71\% are classified as quasar, 16\% as galaxy, and 6\% as star.
Around 7\% of visually inspected spectra did not produce a conclusive classification or redshift.
A large number of quasar spectra with broad emission lines were misclassified as galaxies by Redrock, often at the incorrect redshift.
Based on this performance, we developed a method for automated classification based on Redrock estimates of redshift and classification with additional filtering from two customized algorithms.
The first of these algorithms provides an estimate of Mg~\textsc{ii} flux while the second relies on a machine learning classifier called {\tt QuasarNet} \citep{Busca2018,Farr2020}.

The first application of these algorithms is designed to recover quasar spectra that were misclassified by Redrock.
If a spectrum from the quasar sample is classified by Redrock as a galaxy,
we first assess the fits of the Mg~\textsc{ii} emission line.
If a line is detected with an equivalent width between 10 \AA\ and 200 \AA, a significance of at least three standard deviations, and an overall improvement to the fit of $\chi^2>16$, then we assume that the redshift estimate was correct and change the classification to that of a quasar.
If no significant Mg~\textsc{ii} flux was detected, we then assess the output of {\tt QuasarNet} to determine whether the spectrum is actually a quasar at a different redshift.
If {\tt QuasarNet} classifies the object as a quasar with a probability higher than 95\%, we compute a new redshift based on the Redrock $\chi^2$ surface evaluated only over a narrow redshift interval ($\Delta z = 0.05$) centered on the {\tt QuasarNet} redshift estimate. 

The second application of these algorithms is designed to recover quasar redshifts that were estimated incorrectly by Redrock. This occurs when both Redrock and {\tt QuasarNet} identify the object as a quasar but when the two redshifts differ by more than $0.05$. Then, as before, we recalculate a new redshift based on the Redrock $\chi^2$ surface evaluated only over a narrow redshift interval ($\Delta z = 0.05$) centered on the {\tt QuasarNet} redshift estimate.

If either of the conditions described above is satisfied, the spectrum is included in the catalog.
In cases where the classification of quasars is not based on the first estimate from Redrock, the value of \deltachisq\ for the final estimate is negative.
Likewise, because the second-best estimate from Redrock is often at the same redshift but of a different class, low values of \deltachisq\ are not always an indication of degraded confidence in the redshift estimate.
For these reasons, we do not use \deltachisq\ in the determination of redshifts.

Using the layered automated classification scheme, the target redshift efficiency for the joint tracer and Ly-$\alpha$ forest quasar samples was found to be 65\%, a bit lower than found in the visual inspection process.
The random redshift error as indicated by pairwise velocity differences was found to be $1 \pm 0.4\, \mathrm{km\,s}^{-1}$ for the tracer quasars and $1 \pm 0.4\, \mathrm{km\,s}^{-1}$ for the Ly-$\alpha$ forest quasars.
These errors do not include theoretical uncertainties due to internal kinematics, so are therefore a lower bound on the true redshift errors.  
An updated classification of errors and systematic biases in the DESI quasar redshift estimates can be found in \citet{brodzeller23}.

\subsection{Exposure Depth}

\begin{figure}
	\begin{center}
		\includegraphics[width=0.45\textwidth]{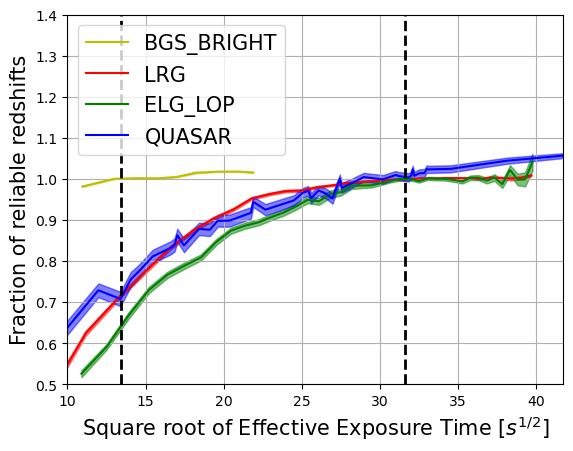}
		\caption{
			Target redshift efficiency and 68\% confidence intervals normalized to the efficiency at the fiducial exposure times (vertical dashed lines) for each of the BGS Bright, LRG, {\tt ELG\_LOP}, and quasar samples. 
            BGS Faint and {\tt ELG\_VLO} targets are not included because they overlap in redshift range with higher priority targets of a similar class.
			Effective exposure time determined for each epoch is computed as described in Section~\ref{sec:cal_spectra}.
			}
		\label{fig:efficiency_efftime}
	\end{center}
\end{figure}

In determining the performance of redshift classification for all tracers, we generally assumed the effective exposure times from the initial instrument design and pixel-level spectroscopic simulations.
For the brighter targets that will be observed in more marginal conditions, these simulations indicated that an effective exposure time of 180 seconds would be sufficient for robust classification.
The individual epochs of BGS and MWS spectra for studies described in this section were therefore tuned to this exposure time, but with significant scatter due to changing conditions.
Likewise, simulations indicated that effective exposure times of 1000 seconds were sufficient to characterize the fainter LRG, ELG, and quasar targets.
Single epochs were tuned to this exposure time for the studies presented above, but again with significant scatter.

The variations in effective exposure times allowed us to assess the redshift efficiency as a function of exposure depth (Figure~\ref{fig:efficiency_efftime}).
Here, we present the target redshift efficiency for each of the BGS, LRG, ELG, and quasar samples.
As expected, the redshift efficiency values measured above are consistent with the values evaluated at the nominal effective exposure times, with decreasing efficiency at decreasing exposure depth.

\begin{figure*}
\begin{center}
    \includegraphics[width=0.45\textwidth]{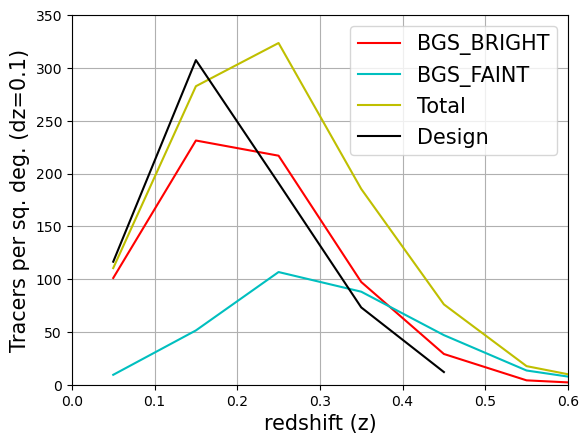}
    \includegraphics[width=0.45\textwidth]{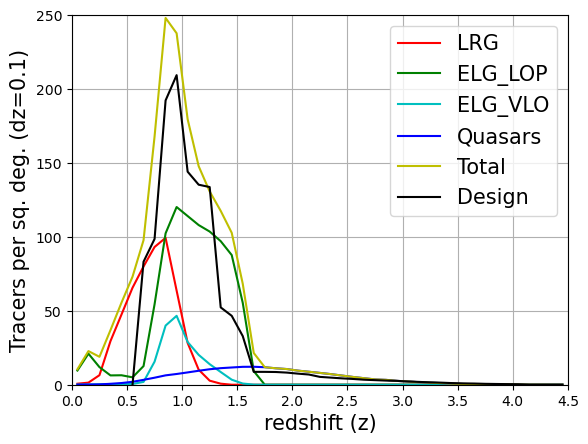}
    \caption{The surface density as a function of redshift for each extragalactic tracer in the DESI spectroscopic sample.
    Only objects with a successful fiber assignment and reliable redshift are shown and no corrections are made for incompleteness or interlopers due to catastrophic failures in redshift assignment.
    {\bf Left:  }Redshift distribution for galaxies that will be observed in bright conditions.
    The black curve corresponds to the projections from a single magnitude-limited BGS sample as assumed in the original design \citep{desi16a}.
    {\bf Right:  }Redshift distribution for galaxies and quasars that will be observed in dark conditions.
    The black curve (Design) corresponds to the predictions from an LRG sample over $0.6<z<0.8$, an ELG sample over $0.8<z<1.6$, and a quasar sample at higher redshifts as assumed in the original design \citep[FDR;][]{desi16a}.
}
\label{fig:nz}
\end{center}
\end{figure*}

The redshift efficiencies presented in Figure~\ref{fig:efficiency_efftime} allowed us to identify what integration time was best for high redshift success without overexposing and thus entering the regime of diminishing returns.
The ELG sample set the pace for integrations, where we see
a 3\% decrease in target redshift efficiency in the {\tt ELG\_LOP} sample when effective exposure times are decreased from 800 seconds to 700 seconds.
However, for the ELG samples, the slope of the efficiency curve flattens at effective exposure times larger than 800 seconds, indicating an approach to diminishing returns.
The target efficiency in the {\tt ELG\_LOP} sample only increases from 67.8\% to 68.8\% as effective exposure times are increased from 800 seconds to 1200 seconds.
The change in the LRG redshift efficiencies is even lower over this range, while the quasar redshift efficiency climbs from 91.1\% to 94.2\%.
Likewise, the change in BGS Bright target redshift efficiency only changes by 2\% as effective exposure times are increased from 120 seconds to 210 seconds.

These results were determined from the Target Selection Validation program, and we find consistent behavior in the One-Percent Survey. 
As a balance between consistent redshift efficiency and areal coverage, we therefore plan effective exposure times of 1000 seconds for the LRG, ELG, and QSO programs and 180 seconds for the BGS and MWS programs.
This stability in redshift efficiency demonstrates that the clustering analyses will be subject to peak-to-peak variations in redshift efficiency of less than 3.4\% as long as estimates with the real-time exposure time calculator are accurate to 20\%.
As described in Section~\ref{sec:cal_spectra}, the expected scatter in real-time exposure time estimates is expected to be less than 10\%.

While effective exposure times are expected to be the primary driver for redshift efficiency, it is equally important to assess whether survey speed (Section~\ref{sec:cal_spectra}) plays a role.
The nominal survey plan prescribes observations to LRG, ELG, and quasar targets when survey speeds exceed 0.4 and observations to BGS and MWS targets when survey speed is within a range of $[\frac{1}{6}, 0.4]$.
During times of slower survey speeds due to degraded observing conditions, observing will be dedicated to a backup program of bright stellar targets.
That backup program is described elsewhere \citep{cooper23}.

It is possible that the lower survey speeds will lead to reduced redshift efficiencies for BGS, LRG, ELG, or quasar targets even at a constant effective exposure time.
For example, exposures taken at lower survey speeds may be more susceptible to sky subtraction or other residuals in the limit of lower signal-to-noise.
However, studies revealed very little change in the redshift efficiency for BGS, LRG, ELG, or quasar targets with survey speeds that were below the nominal thresholds.
As with the dependence on effective exposure times, the relationship between survey speed and redshift efficiency observed in the One-Percent Survey is consistent with what was found in Target Selection Validation, thus confirming the scheme for allocating spectroscopic time between higher and lower redshift samples.

\subsection{Overall Performance}

\begin{table*}[t]
\caption{Requirements and Performance for LRG, ELG, and Quasar Spectroscopy}
\begin{center}
\begin{tabular}{l p{2.8in} | p{2.8in}}
\hline
\hline
\# & Requirement & Performance \\
\hline
L2   & {\bf Survey Data Set Requirements}  \\
L2.2 & {\bf Luminous Red Galaxies}   \\
L2.2.1 & The average density with redshift $0.4 < z < 1.0$ shall be at least 300 deg$^{-2}$. & The average density with redshift $0.4 < z < 1.1$ is 478 deg$^{-2}$.\\
L2.2.2 & The random redshift error shall be less than $\sigma_z = 0.0005 (1 + z)$. & The typical random redshift error is $\sigma_z = 0.00014 (1+z)$. \\
L2.2.3 & The systematic in the mean redshift shall be less than $\Delta z = 0.0002 (1+z)$. & The systematic error in the mean redshift is $\Delta z = 0.00001 (1+z)$. \\
L2.2.4 & The catastrophic redshift failures exceeding $1000\, \mathrm{km\,s}^{-1}$ shall be $<5$\%. & The rate of catastrophic redshift failures exceeding $1000\, \mathrm{km\,s}^{-1}$ is 0.2\%. \\
L2.2.5 & The redshift completeness shall be $>95$\% for each pointing averaged over all fibers with targets. & The fraction of targets confirmed as galaxies is 96\% over all fibers that receive targets.\\
L2.3 & {\bf Emission Line Galaxies} \\
L2.3.1 & The average density with redshift $0.6 < z < 1.6$ shall be at least 1280 deg$^{-2}$. & The average density of {\tt ELG\_LOP} targets with redshift $0.6 < z < 1.6$ is 860 deg$^{-2}$. \\
 &  & The average density of {\tt ELG\_VLO} targets with redshift $0.6 < z < 1.6$ is 180 deg$^{-2}$. \\
L2.3.2 & The random redshift error shall be less than $\sigma_z = 0.0005 (1 + z)$. & The typical random redshift error is $\sigma_z = 0.000026 (1 + z)$. \\
L2.3.3 & The systematic in the mean redshift shall be less than $\Delta z = 0.0002 (1+z)$. & The typical systematic error in the mean redshift is $\sigma_z = 0.0000033 (1 + z)$. \\
L2.3.4 & The catastrophic redshift failures exceeding $1000\, \mathrm{km\,s}^{-1}$ shall be $<5$\%. & The rate of catastrophic redshift failures exceeding $1000\, \mathrm{km\,s}^{-1}$ is 0.2\%.\\
L2.3.5 & The redshift completeness shall be $>90$\% for each pointing averaged over all targets above the \oii\ flux limit.  & The redshift completeness over all fibers with targets is $\sim$70\% for the {\tt ELG\_LOP} and $\sim$94\% for the {\tt ELG\_VLO}.\\
L2.4 & {\bf Tracer Quasars ($0.9<z<2.1$)} \\
L2.4.1 & The average density with redshift $z < 2.1$ shall be at least 120 deg$^{-2}$. & The average density with redshift $z < 2.1$ is 144 deg$^{-2}$.\\
L2.4.2 & The random redshift error shall be less than $\sigma_z = 0.0025 (1 + z)$ (equivalent to $750\, \mathrm{km\,s}^{-1}$ rms). & The typical random redshift error is $\sigma_z = 0.00041 (1+z)$. \\
L2.4.3 & The systematic in the mean redshift shall be less than $\Delta z = 0.0004 (1+z)$. & The typical systematic error in the mean redshift is $\sigma_z = 0.000087 (1 + z)$.  \\
L2.4.4 & The catastrophic redshift failures exceeding $1000\, \mathrm{km\,s}^{-1}$ shall be $<5$\%. & The rate of catastrophic redshift failures exceeding $1000\, \mathrm{km\,s}^{-1}$ is 4.8\%.  \\
L2.4.5 & The redshift completeness shall be $>90$\% for each pointing averaged over all fibers with targets.  & The redshift completeness of confirmed quasars is 66\% (total completeness not recorded).\\
L2.5 & {\bf Ly-$\alpha$ Quasars} \\
L2.5.1 & The average density with redshift $z > 2.1$ and $r < 23.5$ mag shall be at least 50 deg$^{-2}$. & The average density with redshift $z > 2.1$ and $r < 23.0$ mag is 58.5 deg$^{-2}$.\\
L2.5.2 & The random redshift error shall be less than $\sigma_z = 0.0025 (1 + z)$ (equivalent to $750\, \mathrm{km\,s}^{-1}$ rms). & The typical random redshift error is $\sigma_z = 0.00027 (1+z)$.\\
L2.5.3 & The catastrophic redshift failures shall be $<2$\%. & The rate of catastrophic redshift failures exceeding $1000\, \mathrm{km\,s}^{-1}$ is 12.2\%. The rate of catastrophic redshift failures exceeding $3000\, \mathrm{km\,s}^{-1}$ is 1.8\%.\\
L2.5.4 & The $S/N$ per Angstrom (observer frame) shall be greater than 1 in the Ly$\alpha$ forest for $g = 23$ mag and scale with flux for brighter quasars. & To be determined. \\
\hline
\end{tabular}
\end{center}
\label{tab:level2}
\end{table*}

\begin{table*}[t]
\caption{MWS Spectroscopy, BGS Spectroscopy, Calibration, Fiber Assignment, and Target Selection}
\begin{center}
\begin{tabular}{l p{2.8in} | p{2.8in}}
\hline
\hline
\# & Requirement & Performance\\
\hline
 & {\bf Milky Way Survey} & The typical (median) radial velocity uncertainty is approximately  $\sigma_v = 1.3 \, \mathrm{km\,s}^{-1}$.\\
 & &  The rate of catastrophic redshift failures exceeding $20\, \mathrm{km\,s}^{-1}$ is approximately $0.6\%$.\\
 & {\bf Bright Galaxy Survey} & The average density of confirmed BGS Bright galaxies with redshift $0 < z < 0.4$ is 646 deg$^{-2}$.\\
 & & The typical random redshift error is $\sigma_z = 0.00003 (1+z)$.\\
 & & The typical systematic in the mean redshift is $\sigma_z = 0.000022 (1 + z)$. \\
 & &  The rate of catastrophic redshift failures exceeding $1000\, \mathrm{km\,s}^{-1}$ is approximately $0.5\%$.\\
 & & The redshift completeness is 99\% over all fibers that receive targets.\\
L2.6 & {\bf Spectrophotometric Calibration} \\
L2.6.1 & The Ly$\alpha$ QSO fractional flux calibration errors shall have power less than $1.2\,\mathrm{km\,s}^{-1}$ at $k \sim 0.001\,\mathrm{s\,km}^{-1}$. & To be determined.  \\
L2.7 & {\bf Fiber Completeness} \\
L2.7.1 & The fraction of targets that receive a fiber shall be at least 80\%. & The fraction of LRG targets that successfully acquire a spectrum is 89\%. \\
 & & The fraction of {\tt ELG\_LOP} targets that successfully acquire a spectrum is 69\%. \\
 & & The fraction of quasar targets that successfully acquire a spectrum is 99\%. \\
 & & The fraction of BGS Bright targets that successfully acquire a spectrum is 80\%. \\
 & & The fraction of MWS targets that successfully acquire a spectrum is 28\%. \\
L2.8 & {\bf Target Selection} \\
L2.8.1 & The LRG target density shall be 350 deg$^{-2}$, with at least 300 deg$^{-2}$ successfully measured. & The LRG target density is 624 deg$^{-2}$, with 533 deg$^{-2}$ successfully measured. \\
L2.8.2 & The ELG target density shall be 2400 deg$^{-2}$, with at least 1280 deg$^{-2}$ successfully measured.  & The {\tt ELG\_LOP} target density is 1941 deg$^{-2}$, with 938 deg$^{-2}$ successfully measured. \\
 &   & The {\tt ELG\_VLO} target density is 463 deg$^{-2}$, with  183 deg$^{-2}$ successfully measured. \\
L2.8.3 & The low-$z$ QSO target density shall be 170 deg$^{-2}$, with at least 120 deg$^{-2}$ successfully measured.  & The quasar target density is 311 deg$^{-2}$, with 144 deg$^{-2}$ successfully measured at $z<2.1$. \\
L2.8.4 & The Ly$\alpha$ QSO target density shall be 90 deg$^{-2}$, with at least 50 deg$^{-2}$ successfully measured.  & 58.5 deg$^{-2}$ quasars at $z>2.1$ are successfully measured from the overall quasar target sample. \\
& & The BGS Bright target density is 854 deg$^{-2}$, with 676 deg$^{-2}$ successfully measured.\\
& & The density of MWS targets (excluding the Faint samples) is 1637 deg$^{-2}$, with 458 deg$^{-2}$ successfully measured.\\\\
\hline
\end{tabular}
\end{center}
\label{tab:level2_cont}
\end{table*}

With effective exposure times of 1000 seconds and 180 seconds for dark time and bright time, respectively, survey simulations indicate that DESI will complete a 14,900 deg$^2$ footprint in five years.
From the central 14,000 deg$^2$ of that footprint, the target selection algorithms and fiber assignment efficiencies presented in Section~\ref{sec:ts}, and redshift performance reported in this section, we estimate the final statistics for the five year survey.
The redshift distributions for spectroscopically-confirmed, extragalactic targets is shown in Figure~\ref{fig:nz}. 
In total, we expect a final spectroscopic sample of 7.2 million unique stars (over all Bright tiles, including faint sources), 36.12 million unique galaxies, and 2.87 million unique quasars with reliable redshift estimates.
In Table~\ref{tab:level2} and Table~\ref{tab:level2_cont}, those results are compared to the science requirements crafted in 2014 that were used to inform the DESI instrument design \citep{desi-collaboration22a}.
Below we highlight some expected changes to the survey arising from the SV results.

Across the LRG, ELG, and quasar target samples, we expect to exceed the design requirements by a significant margin for random redshift errors.  
We also expect to exceed design requirements for catastrophic redshift failure rates for the galaxy samples by a large margin, while also surpassing the design requirement for quasars if the definition of a catastrophic failure is relaxed from 1,000 $\mathrm{km\,s}^{-1}$ to 3,000 $\mathrm{km\,s}^{-1}$.
Further improvements to the data reductions and redshift classification schemes may lead to better performance yet. 

The SV results led the DESI collaboration to reallocate fibers relative to the initial design to significantly increase the LRG and quasar sample sizes while keeping their combined observational cost almost unchanged.
It was originally assumed that LRG targets fainter than a $z$-band magnitude of 20 would require effective exposure times longer than 1000-seconds, leading to an average of 2000-seconds per target (two exposures) to achieve an adequate number of reliable redshift estimates.
Based on the high redshift success rates found in SV, the observational program was updated to only assign a single epoch of observation to all LRG targets, even those with magnitudes as faint as $z=20.6$.
With this modification, the target density was increased from 350 to 624 deg$^{-2}$.
The observational cost is 25 exposures deg$^{-2}$ lower than the original expectation, but now with 80\% more targets expected to be given a reliable redshift estimate.
The results from SV revealed that quasars could still be identified at a high confidence beyond the boundaries of the original selection.
By increasing the target density by 20\% to 311 deg$^{-2}$, we expect a 20\% increase in the number of Ly-$\alpha$ quasars and nearly the same fractional increase in the number of $z<2.1$ quasars.
With an average of 3.4 exposures for each high redshift quasar, the total cost of this program is 450 exposures deg$^{-2}$, only 30 more than the original design.  

While the LRG and quasar programs are expected to produce significantly larger clustering samples at effectively the same observational cost, the ELG program falls somewhat below what was originally planned in both target density and size of the clustering sample.
First, the assumed fiber assignment efficiencies of 80\% turned out to be slightly optimistic, as 69\% of the {\tt ELG\_LOP} sample is actually assigned a fiber.
Second, it proved challenging to identify the highest redshift galaxies from $grz$ imaging data beyond a certain density, so we reduced the target density of the prime sample from 2400 deg$^{-2}$ to 1941 deg$^{-2}$.
We provided the remaining targets from the lower priority {\tt ELG\_VLO} selection.
This split in the sample helped to preserve a higher fiber assignment efficiency for the targets most likely to produce a reliable redshift in the range not overlapping with the LRG sample.
A spectroscopic sample size of 400 deg$^{-2}$ was originally assumed over $1.1<z<1.6$ \citep{desi16a}, but as shown in Figure~\ref{fig:nz}, the {\tt ELG\_LOP} will exceed this expectation with a surface density of 450 deg$^{-2}$.
Even though the target density and total clustering sample sizes are lower than expected, the final selection produces a more efficient program in the redshift range that is most distinct from the other clustering samples.

Finally, even though the BGS and MWS samples did not drive the requirements for instrument design, the final samples still exceed expectations.
A BGS target sample of 700 deg$^{-2}$ was presented in the final survey design \citep{desi16a}, whereas 854 deg$^{-2}$ can be identified from a magnitude-limited sample and classified at high efficiency.
Likewise, the BGS Faint and MWS samples were not presented in detail, but combined, are expected to produce more than 10 million unique spectra over five years.

We repeated the study of total redshift efficiency and target redshift efficiency on the One-Percent Survey data.
The performance of the spectroscopic classification confirmed the redshift efficiencies reported here, thus validating the expected tracer counts for cosmology forecasts.
More detailed studies for each tracer such as differential efficiency rates as a function of magnitude can be found in the dedicated target selection and visual inspection papers.

\clearpage
\newpage

\section{Cosmological Forecasts}\label{sec:forecasts}
\begin{figure*}
\begin{center}
    \includegraphics[width=0.45\textwidth]{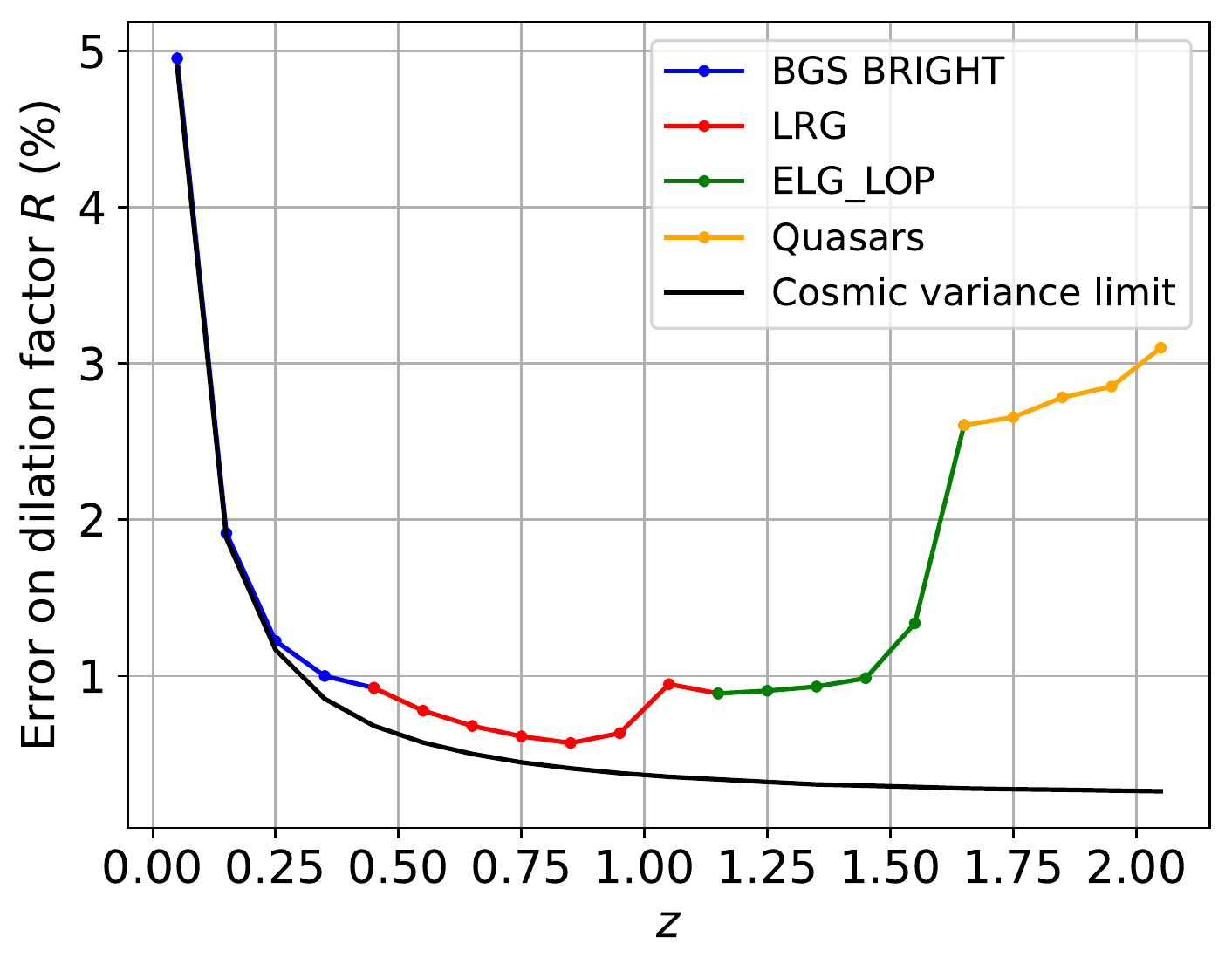}
    {\includegraphics[width=0.45\textwidth]{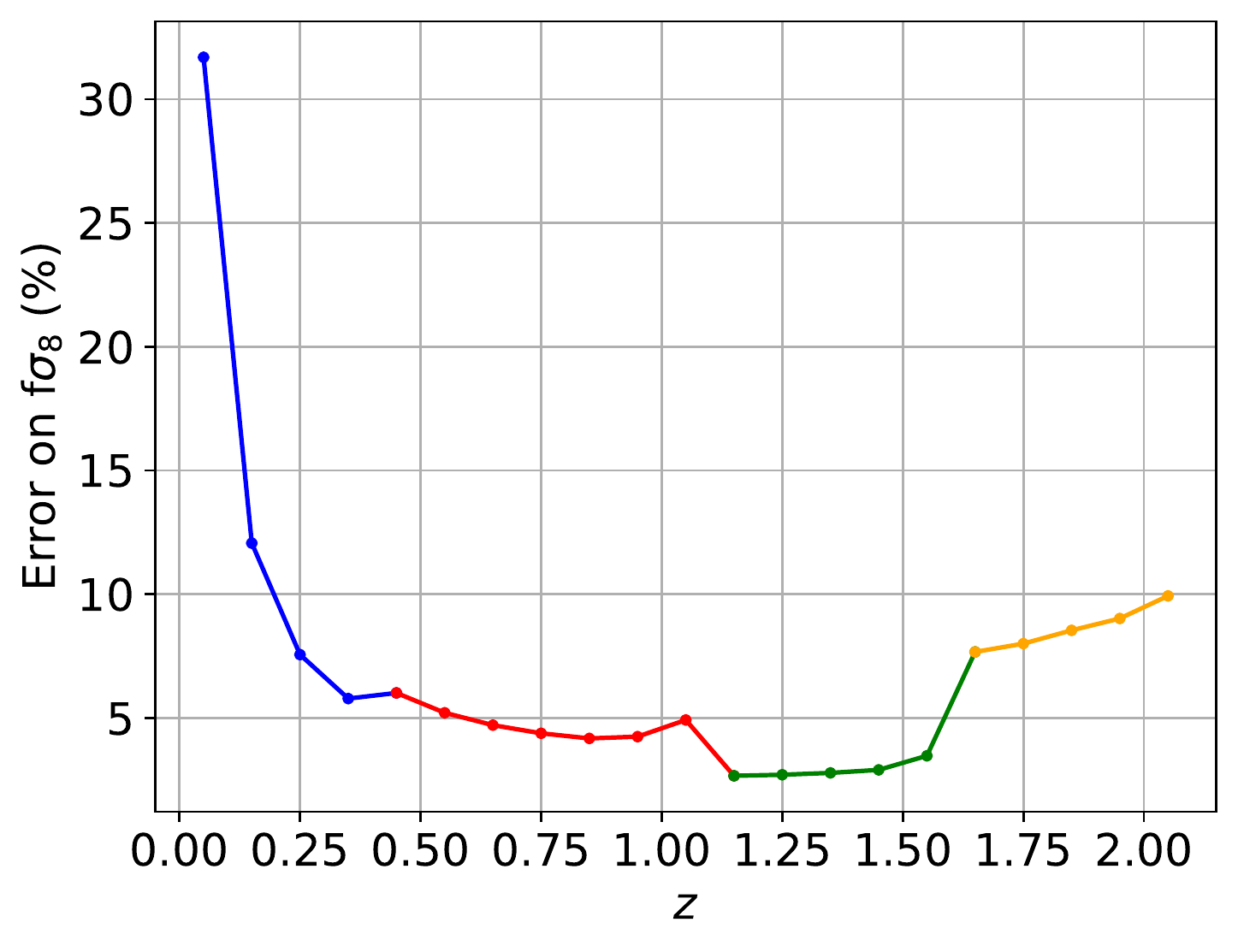}}
    \caption{{\bf Left:}  The forecasted precision on the dilation factor $R$ from DESI BAO measurements compared with the cosmic variance limit for a 14,000~deg$^2$ survey (black).  
    {\bf Right:}  The forecast precision on $f\sigma_8$ from DESI RSD measurements. In both panels, measurements from each tracer are represented by a different color and one tracer is assumed over a given redshift range. The redshift range excludes the Ly-$\alpha$ forest measurements for consistency in the illustration of BAO and RSD.
}
\label{fig:cosmicvariance}
\end{center}
\end{figure*}

Based on the derived target selection densities, fiber assignment efficiencies, and redshift efficiencies, we forecast the cosmological constraints for the 14,000 deg$^2$ DESI program.
We first demonstrate the statistical precision that we are expecting on the measurements of the distance scale through BAO and the growth of structure through RSD.
From these forecasts, we predict the precision expected on the cosmological parameters in various combinations of DESI BAO and RSD measurements both alone and with external data sets.

\subsection{BAO and RSD Forecasts}

The number density and redshift distribution over the 14,000 deg$^2$ footprint for each tracer used in the forecasts is shown in Table~\ref{tab:nz}.
These predicted number densities, along with an assumption on bias as described in the next paragraph, allow us to predict the sensitivity to $D_A(z)/r_d$, $H(z) r_d$, and $f\sigma_8$ in each redshift interval.
Here, $D_A$ is the angular diameter distance, $H$ is the Hubble parameter, $r_d$ is the sound horizon at the drag epoch, $f$ is the growth factor, and $\sigma_8$ is the amplitude of mass fluctuations in spheres of $8 \,h^{-1}$ Mpc; note that the quantity $f\sigma_8$ is essentially the amplitude of the velocity power spectrum and is readily probed by redshift-space distortion measurements. 
An additional parameter $R$ is often used to represent the precision resulting from the optimal combination of $D_A/r_d$ and $H r_d$ measurements. 
In redshift bins with multiple tracers, we use the densest tracer to keep our forecasts conservative.
The left panel of Figure~\ref{fig:cosmicvariance} shows the fractional uncertainty on the volume-averaged BAO measurement relative to the cosmic variance limit for a 14,000 deg$^2$ survey.
The right panel shows the fractional uncertainty on the growth of structure measurements from RSD.

\begin{table*}
\centering
\caption{
\label{tab:nz}
Cosmological tracers and forecasts of precision ($\times 100$\%) on BAO and RSD measurements.
Estimates of covariance between BAO and RSD parameters can be found with other supplemental data, as described immediately following the conclusion.
}
\begin{tabular}{l c c c c c c c }
\hline\hline
Redshift  & Surface & $nP_{k=0.14,\mu=0.6}$ & $V_{\rm eff}$ & $\frac{\sigma_{D_A/r_d}}{D_A/r_d}$ & $\frac{\sigma_{H r_d}}{H r_d}$ & $\frac{\sigma_R}{R}$\tablenotemark{\dag} & $\frac{\sigma_{f\sigma_8}}{f\sigma_8}$ \\
& Density (deg$^{-2}$) &  & ($h^{-3}$ $\text{Gpc}^3$)  & ($\%$) & ($\%$) & ($\%$) & ($\%$) \\
\hline
\multicolumn{8}{c}{\bf BGS BRIGHT}\\
\hline
$0.0 < z < 0.1$ & 101.1 & 338.54 & 0.04 & 6.65 & 13.92 &  4.95 & 31.64  \\
$0.1 < z < 0.2$ & 231.3 & 122.16 & 0.22 & 2.57 & 5.40 &  1.91 & 12.04 \\
$0.2 < z < 0.3$ & 216.9 & 47.11 & 0.54 & 1.64 & 3.41 & 1.21  & 7.54 \\
$0.3 < z < 0.4$ & 97.3 & 12.15 & 0.83 & 1.37 & 2.70 & 0.99  & 5.76 \\
\hline
\multicolumn{8}{c}{\bf LRG}\\
\hline
$0.4 < z < 0.5$ & 47.5 & 6.12 & 1.06 & 1.25 & 2.38 & 0.88  & 5.96 \\
$0.5 < z < 0.6$ & 65.6 & 6.35 & 1.42 & 1.05 & 1.99 & 0.74  & 5.16 \\
$0.6 < z < 0.7$ & 80.0 & 6.21 & 1.76 & 0.92 & 1.74 & 0.65  & 4.67 \\
$0.7 < z < 0.8$ & 93.2 & 6.08 & 2.07 & 0.84 & 1.56 & 0.59  & 4.34 \\
$0.8 < z < 0.9$ & 99.3 & 5.64 & 2.32 & 0.78 & 1.44 & 0.55  & 4.14 \\
$0.9 < z < 1.0$ & 63.7 & 3.23 & 2.09 & 0.87 & 1.52 & 0.59  & 4.19 \\
$1.0 < z < 1.1$ & 28.3 & 1.31 & 1.25 & 1.25 & 2.04 & 0.83  & 4.77 \\
\hline
\multicolumn{8}{c}{\bf  ELG\_LOP}\\
\hline
$1.1 < z < 1.2$ & 108.0 & 1.37 & 1.40 & 1.24 & 1.80 & 0.79  & 2.58 \\
$1.2 < z < 1.3$ & 103.6 & 1.23 & 1.35 & 1.26 & 1.80 & 0.80  & 2.62 \\
$1.3 < z < 1.4$ & 97.1 & 1.09 & 1.26 & 1.30 & 1.82 & 0.82  & 2.69 \\
$1.4 < z < 1.5$ & 87.7 & 0.93 & 1.13 & 1.37 & 1.89 & 0.87  & 2.80 \\
$1.5 < z < 1.6$ & 55.4 & 0.57 & 0.65 & 1.87 & 2.46 & 1.17  & 3.34 \\
\hline
\multicolumn{8}{c}{\bf Quasars}\\
\hline
$1.6 < z < 1.7$ & 12.1 & 0.22 & 0.17 & 3.39 & 4.76 & 2.16  & 7.30 \\
$1.7 < z < 1.8$ & 11.8 & 0.21 & 0.16 & 3.48 & 4.87 & 2.21  & 7.63 \\
$1.8 < z < 1.9$ & 11.1 & 0.19 & 0.14 & 3.67 & 5.14 & 2.34  & 8.17 \\
$1.9 < z < 2.0$ & 10.6 & 0.18 & 0.13 & 3.83 & 5.36 & 2.44  & 8.66 \\
$2.0 < z < 2.1$ & 9.5 & 0.16 & 0.10 & 4.22 & 5.90 & 2.69  & 9.58 \\
\hline
\multicolumn{8}{c}{\bf Ly-$\alpha$ auto-correlation and Quasar -- Ly-$\alpha$ cross-correlation\tablenotemark{\ddag}}\\
\hline
$2.1 < z < 2.2$ & 8.8 & & & 2.02 & 2.16 & 1.1 & \\
$2.2 < z < 2.3$ & 8.0 & & & 2.14 & 2.24 & 1.15 & \\
$2.3 < z < 2.4$ & 7.2 & & & 2.33 & 2.36 & 1.22 & \\
$2.4 < z < 2.5$ & 6.2 & & & 2.56 & 2.52 & 1.32 & \\
$2.5 < z < 2.6$ & 5.3 & & & 2.9 & 2.77 & 1.47 & \\
$2.6 < z < 2.7$ & 4.4 & & & 3.38 & 3.11 & 1.67 & \\
$2.7 < z < 2.8$ & 3.6 & & & 3.95 & 3.5 & 1.91 & \\
$2.8 < z < 2.9$ & 3.3 & & & 4.69 & 4.05 & 2.23 & \\
$2.9 < z < 3.0$ & 2.6 & & & 5.59 & 4.71 & 2.62 & \\
$3.0 < z < 3.1$ & 2.2 & & & 6.73 & 5.51 & 3.09 & \\
$3.1 < z < 3.2$ & 1.7 & & & 8.47 & 6.78 & 3.84 & \\
$3.2 < z < 3.3$ & 1.4 & & & 10.73 & 8.41 & 4.8 & \\
$3.3 < z < 3.4$ & 1.1 & & & 14.48 & 11.1 & 6.4 & \\
$3.4 < z < 3.5$ & 0.7 & & & 19.92 & 14.8 & 8.62 & \\
\hline
\end{tabular}
\tablenotetext{\dag}{The value $R$ represents the precision resulting from the optimal combination of $D_A(z)/r_d$ and $H(z) r_d$ measurements. 
In most cases, it can be considered a volume-averaged constraint on the BAO distance scale.}
\tablenotetext{\ddag}{Ly-$\alpha$ quasars will be used for both auto- and cross-correlation measurements.
Because the Ly-$\alpha$ forest is a continuous tracer, we do not compute volume density or effective volume.}
\end{table*}

For $z<0.4$, we forecast constraints using the BGS sample assuming that bias evolves as $b_{\text{BGS}}(z) = 1.34 / D(z)$, where $D(z)$ is the linear growth factor normalized by $D(z=0)=1$.
Over the redshift range $0.4<z<1.1$, we use the LRG sample assuming $b_{\text{LRG}}(z) = 1.7 / D(z)$.
For the ELG sample we use $b_{\text{ELG}}(z) = 0.84 / D(z)$ based on \cite{Mostek2013}, over the redshift range $1.1<z<1.6$.
We use quasars in two different ways. We consider them as discrete tracers of the matter density field to forecast constraints over the range $1.6<z<2.1$ assuming a bias of $b_{\text{QSO}}(z) = 1.2 / D(z)$, loosely based on \cite{Ross2009}.   
We also consider the line-of-sight absorption information to quasars in order to extract the clustering of the matter density field through the Ly-$\alpha$ forest.
We assume $b_{\text{QSO}}(z) = 1.1 / D(z)$ for redshifts $2.1<z<3.5$ when using quasars for measurements of the auto-correlation of the Ly-$\alpha$ forest and for the cross-correlation between the Ly-$\alpha$ forest and quasars \citep[e.g.][]{Font-Ribera14a}.

\begin{figure*}
\begin{center}
    \includegraphics[width=0.48\textwidth]{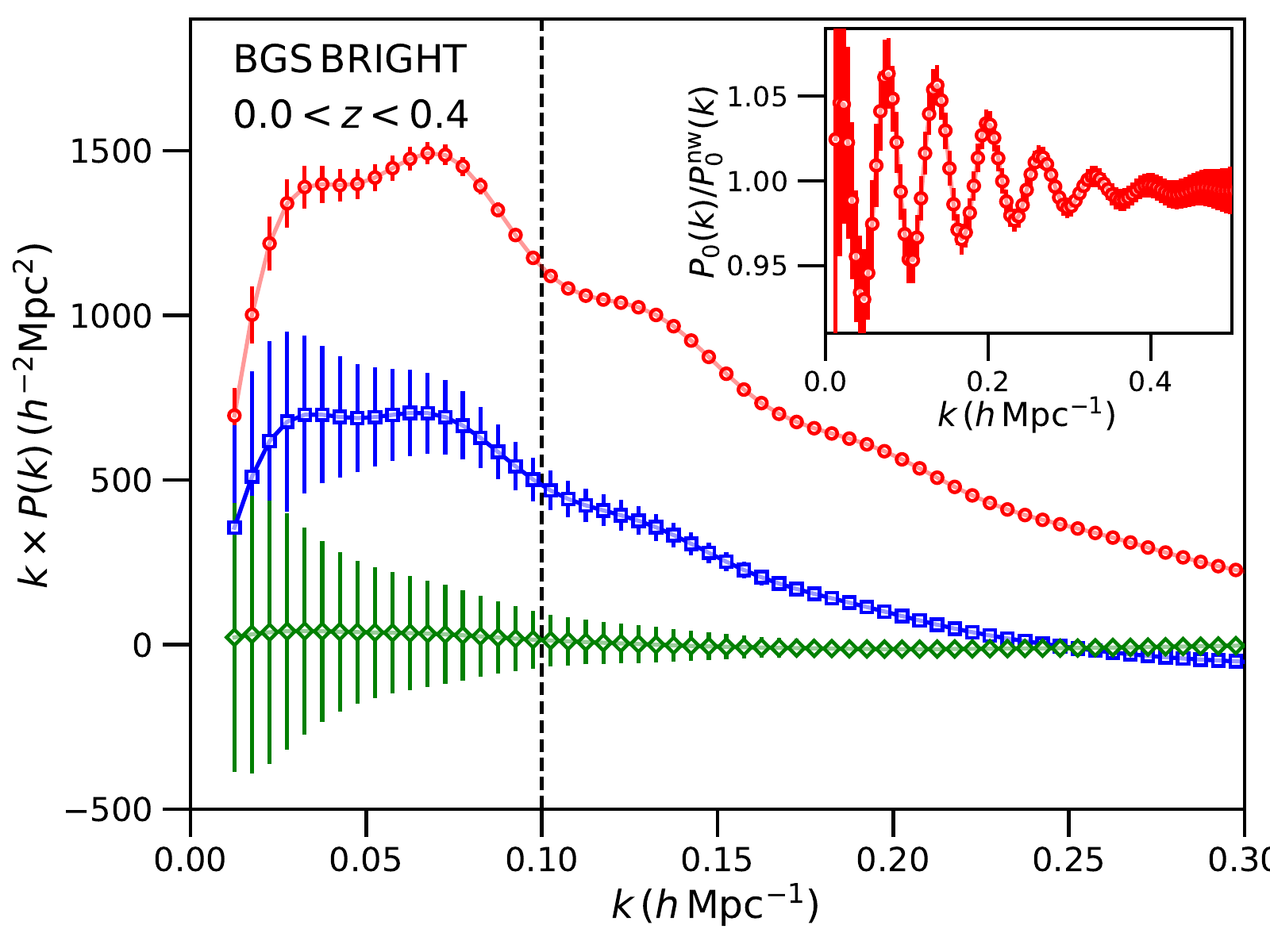}
    \includegraphics[width=0.48\textwidth]{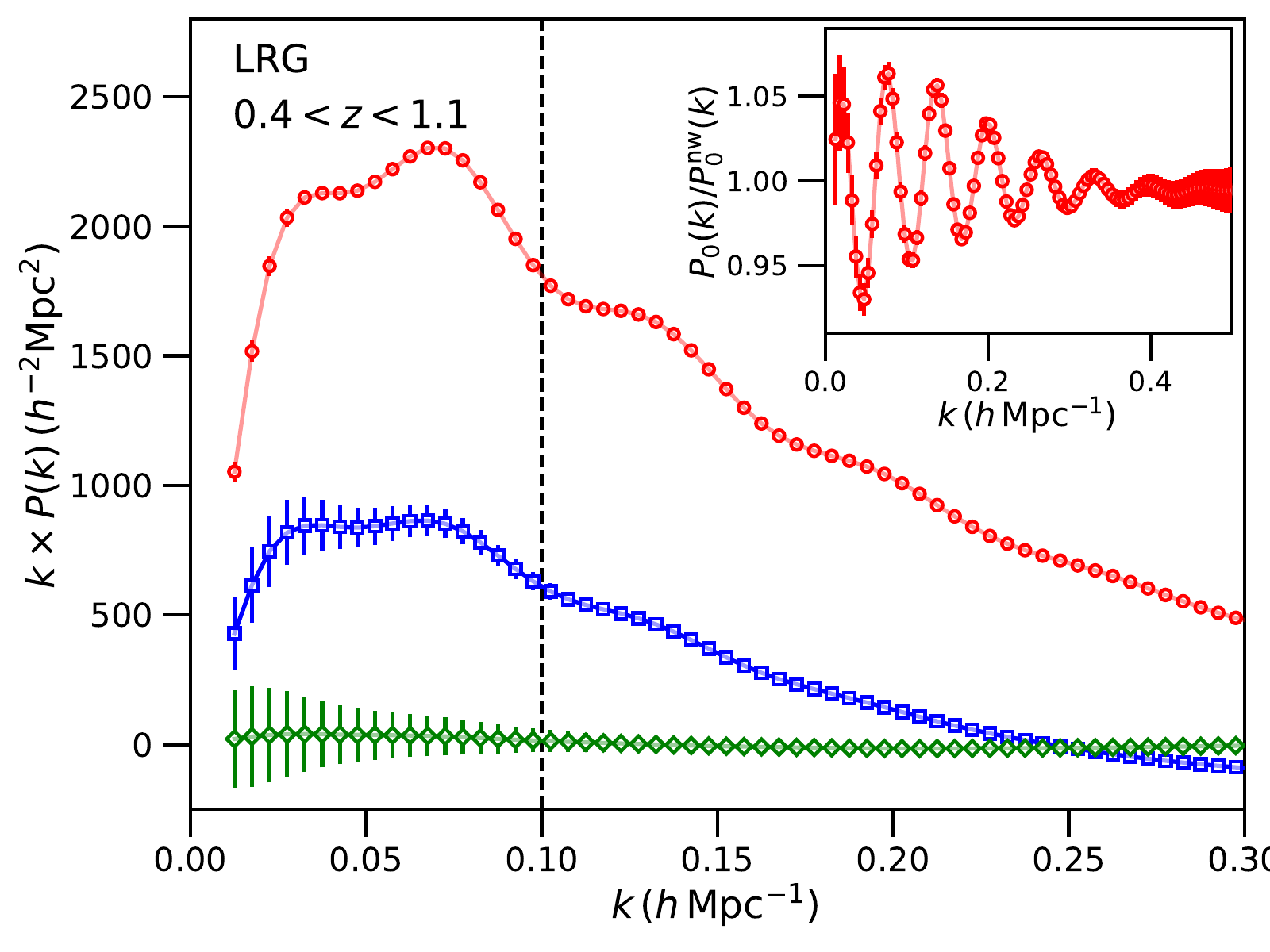}\\
    \includegraphics[width=0.48\textwidth]{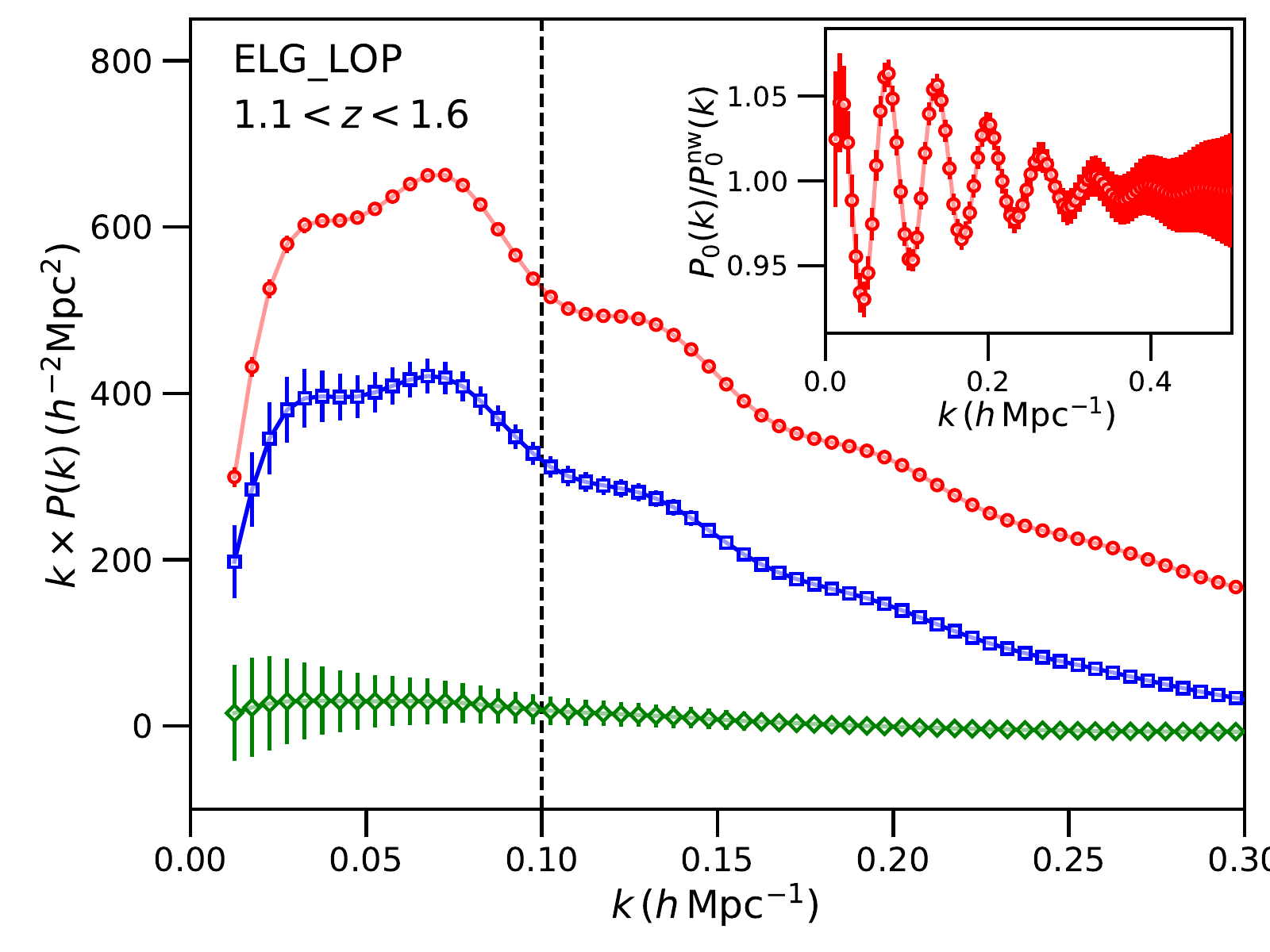}
    \includegraphics[width=0.48\textwidth]{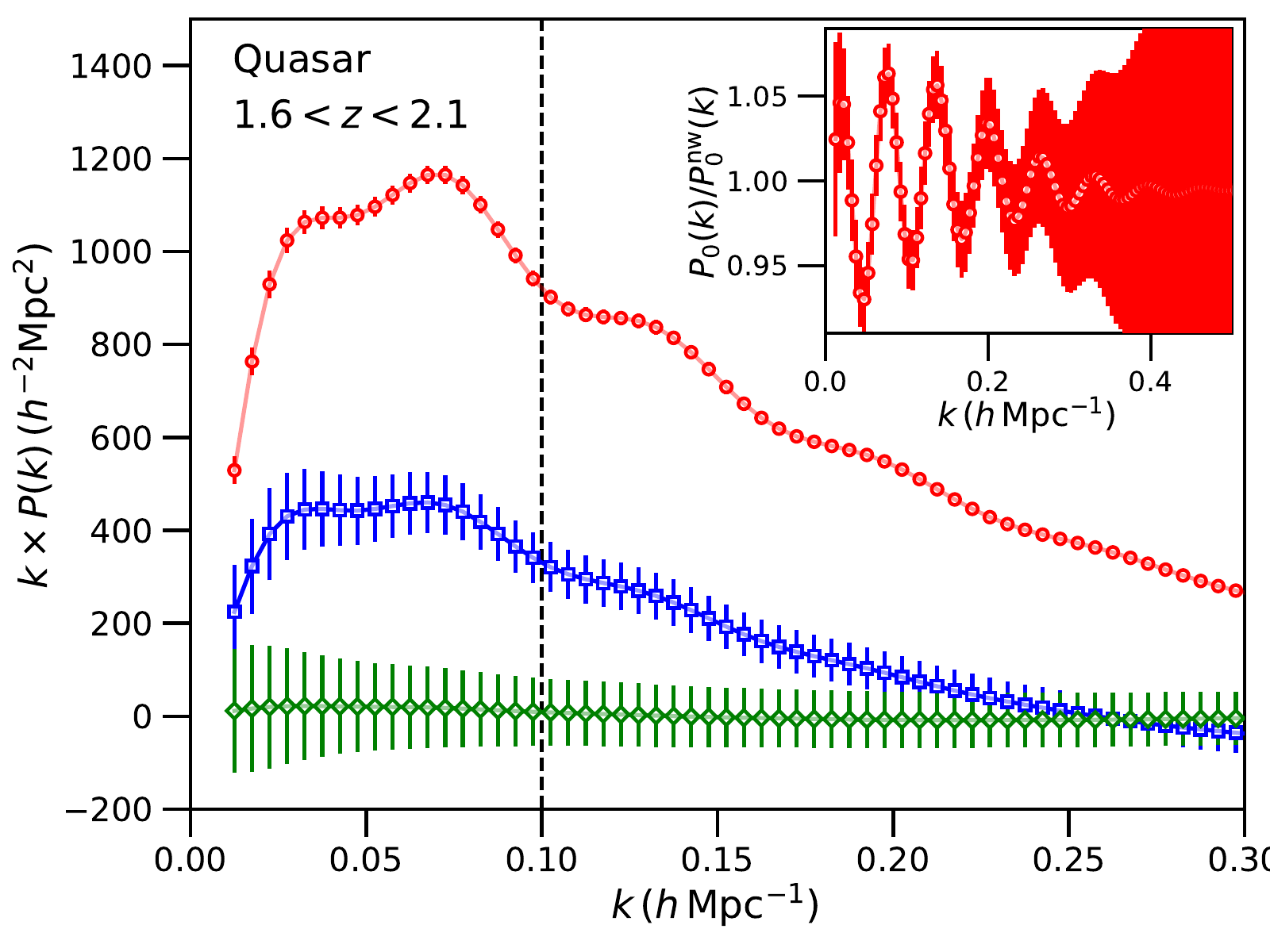}
    \caption{The assumed central values and predicted uncertainties as a function of wavenumber in the multipole expansion of the power spectrum for each of the discrete tracers in the DESI program.
    In each case, the red points represent the monopole, the blue points represent the quadrupole, and the green points represent the hexadecapole.
    The inset panels demonstrate the monopole of the predicted power spectrum relative to a featureless model in order to amplify the BAO feature.
}
\label{fig:powerspectra}
\end{center}
\end{figure*}

For the $z>2.1$ BAO projections involving the 
Ly-$\alpha$ forest, we use the
algorithm and code described in \cite{Font-Ribera2014}.
For the rest of our forecasts, we use the software {\texttt GoFish}\footnote{\url{https://github.com/ladosamushia/GoFish}}, a Fisher forecast tool for DESI galaxy clustering analysis.
The forecasts for BAO and RSD precision are based on the projected uncertainties in the clustering measurements. 
An illustration of the projected precision in the power spectrum for each tracer can be found in Figure~\ref{fig:powerspectra}.

In our Fisher forecast formalism, we include galaxy clustering information depending on the wavenumber interval of focus. 
From $0.10 < k < 0.50  \,h \, \text{Mpc}^{-1}$, we only use the information from the BAO feature in the power spectrum to constrain $D_A(z)/r_d$ and $H(z) r_d$. 
Over larger scales ($0.01 < k < 0.10\, h \,\text{Mpc}^{-1}$), in addition to the BAO feature information, we make use of the broadband galaxy power by using the power spectrum as a function of wavenumber and angle with respect to the line of sight to constrain $f\sigma_8$.
Because the forecasts rely primarily on scales that are much larger than the fiber patrol radius where pairs are less likely to be resolved (Figure~\ref{fig:pairwise}), we simply rely on the overall number densities and do not forecast uncertainties due to incompleteness on these scales.
Future cosmology analyses will introduce algorithms to recover information lost to fiber incompleteness \citep[e.g.][]{bianchi18,smith19,ross20a}.

To estimate the BAO constraints on $D_A(z)/r_d$ and $H(z) r_d$, we use the approach to isolate the BAO feature described in \citet{Hinton2020-Barry}. 
Additionally, to calculate the BAO uncertainties, we assume a degradation of the BAO damping scale following~\cite{Seo2007}, with a damping factor of the form 
\begin{equation} 
A(k,\mu,z) = \exp\left[-k^2\left( \frac{(1-\mu^2)\Sigma_\perp^2}{2} + \frac{\mu^2\Sigma_\parallel^2}{2} \right)\right].
\label{eq:damping_factor}
\end{equation}
Here the Lagrangian displacement distances are given by $\Sigma_\perp = 9.4 (\sigma_8(z)/0.9) \,h^{-1} \text{Mpc}$ and $\Sigma_\parallel = \Sigma_\perp (1+f(z))$, where both $\Sigma_\perp$ and $\Sigma_\parallel$ are multiplied by a factor $(\in [0.5, 1])$ to quantify the degradation of the standard BAO reconstruction due to shot noise, following~\cite{White2010}. 

When including the larger scales with RSD information, we assume
\begin{equation} 
P(k,\mu,z) = ( b(z) + f\mu^2 ) ^ 2 P_{\text{mass}}(k,z) A(k,\mu,z),
\label{eq:kaiser}
\end{equation}
where $A(k,\mu,z)$ is given by Equation~(\ref{eq:damping_factor}), $b$ is the linear bias parameter (which is marginalized over) and $P_{\text{mass}}(k, z)$ corresponds to the linear mass power spectrum.
The covariance matrix used for these Fisher
forecasts assumes the linear Kaiser model (i.e. Equation~(\ref{eq:kaiser}) with $A = 1$) and accounts for the shot noise.

It is worth noting that the forecast results obtained for $f \sigma_8$ should be taken with some caution as previous work on spectroscopic surveys showed that the Fisher forecasts for this parameter can be more optimistic compared to the results achieved from the data.
\citet{2021Foroozan} showed that the degradation was a result of degeneracies between the geometric parameters and optimistic assumptions about the scale where information from the linear regime can be extracted. 
Indeed, \citet{2021Foroozan} found that, when using linear theory, only using scales below $k = 0.08\,h^{-1} \mathrm{Mpc}$ in the forecasts resulted in good match between the forecast and measured $f\sigma_8$ uncertainties in past surveys. 
In our computation, the cut off scale is only slightly more optimistic (we use scales up to $k = 0.1\,h^{-1} \mathrm{Mpc}$ for the growth rate forecasts), but we account for the potential loss of the information due to nonlinear physics by applying the scale-dependent exponential damping from Equation~(\ref{eq:damping_factor}).

While not included in these results, \cite{cuceu21} also found a significant gain in information from the anisotropy in the full-shape of Ly-$\alpha$ forest correlations. Their forecasts show that adding this extra information would improve the BAO constraints on $D_A(z)$ and $H(z)$ by a factor of $1.5-1.8$. Therefore, such a measurement could result in significant further improvements of DESI constraints at high redshifts ($z > 2$).

\subsection{DESI as a Stage-IV Dark Energy Experiment}

\begin{table*}[htp]
\centering
\caption{
\label{tab:fom}
Aggregate precision on BAO/RSD measurements and forecast DETF Figure of Merit. We vary spatial curvature as well as dark energy parameters when deriving the FoM values. All cases include {\it Planck} CMB temperature and polarization data. All the numbers are for DESI, but we add in the last row the FoM for SDSS to demonstrate a factor of $\sim8.3$ improvement from DESI relative to a single Stage-III experiment paired with {\it Planck} CMB temperature and polarization data. 
}
\begin{tabular}{l c c c c c c c c }
\hline\hline
Redshift & Design  & Forecast & Design  & Forecast   & Forecast  & Design  & Forecast & Forecast \\
Range & $H(z)$  & $H(z)$  & $R(z)$  & $R(z)$ & $f \sigma_8$  & FoM & FoM (BAO only) & FoM (BAO+RSD) \\
\hline
$0<z<1.1$ & & 0.62\% & 0.28\% & 0.24\% & 1.56\% & &  &  \\
$1.1<z<1.9$ & & 0.82\% & 0.39\% & 0.37\% & 1.24\% & &  &  \\
$1.9<z<3.7$ & 1.05\% & 0.88\% & & 0.46\% &  & &  &  \\
$0<z<3.7$   & & 0.43\% & & 0.18\% & 0.95\% &  110 & 97 & 156.8 \\ 
\hline
$0<z<2.2$ & SDSS (Stage-III)    &  & & &  &  & & 24 \\ 
\hline
\hline
\end{tabular}
\end{table*}

As explained in Section~\ref{sec:srd}, we designed DESI with the goal to increase the DETF FoM by an order of magnitude beyond Stage-II results and thus qualify as a Stage-IV Dark Energy Experiment.
Implicit in the design was that DESI would achieve measurements of the BAO distance scale to a high precision over all redshifts accessible with the BGS, LRG, ELG, and quasar samples.
Those design requirements, and the aggregate precision expected from Table~\ref{tab:nz}, are found in Table~\ref{tab:fom}.
At redshift intervals $0<z<1.1$ and $1.1<z<1.9$, the forecasted precision on $R(z)$ is better than the primary science requirement that drove the DESI design.
In the highest redshift interval, accessible primarily through the Ly-$\alpha$ forest, the forecasted precision on $H(z)$ is better than the top-level requirement by almost 20\%.

These design requirements were expected to produce cosmological constraints with a DETF FoM of at least 110.
Using the predicted precision of the BAO and RSD measurements over all redshifts, we forecast the DETF FoM using the {\it Planck} measurements as the only external data set. 
We also vary the nuisance parameters used within the {\it Planck} {\texttt plik}, {\texttt commander} and {\texttt SimAll} likelihoods \citep[see][]{planckcosmo}.
We assume a cosmological model with time-varying dark energy equation of state parameters and spatial curvature as free parameters in addition to the standard $\Lambda$CDM parameters.
As shown in Table~\ref{tab:fom}, we expect to come close to reaching the design goal when using only the BAO data.
However, we expect to significantly exceed the Design FoM when adding the RSD measurements. 
The slight discrepancy between the Design FoM and the forecast FoM using only BAO arises from different assumptions about the CMB constraints.
At the time the final design report was completed, we had to rely on forecasts for {\it Planck} cosmological results, as opposed to the final {\it Planck} results that we use here.
For comparison, we also provide in the last row the FoM for SDSS.
The large improvement in DESI relative to the Stage-III SDSS program is as expected from a Stage-IV experiment when all {\it Planck} constraints are treated equally in the comparison.

\subsection{Predicted Cosmology Constraints}

For our predictions of the final cosmological constraints from DESI, we make use of the full suite of BAO and RSD forecasts found in Table~\ref{tab:nz}.
We use the publicly available code {\texttt Cobaya} \citep{2021-Cobaya} to fit the parameters to the data and infer their posterior distribution. 

The constraints are presented for an extension to the standard $\Lambda$CDM parameters where we vary the curvature density parameter, $\Omega_k$, and a time-evolving equation of state for dark energy, ($w_0$--$w_a$). 
This model is denoted $ow_0w_a$CDM as done in the DETF paper \citep{detf}. 
We also provide constraints for $w_0$ and $w_p$ since the latter is decorrelated from $w_a$.

\begin{figure}
\begin{center}
    \includegraphics[width=0.5\textwidth]{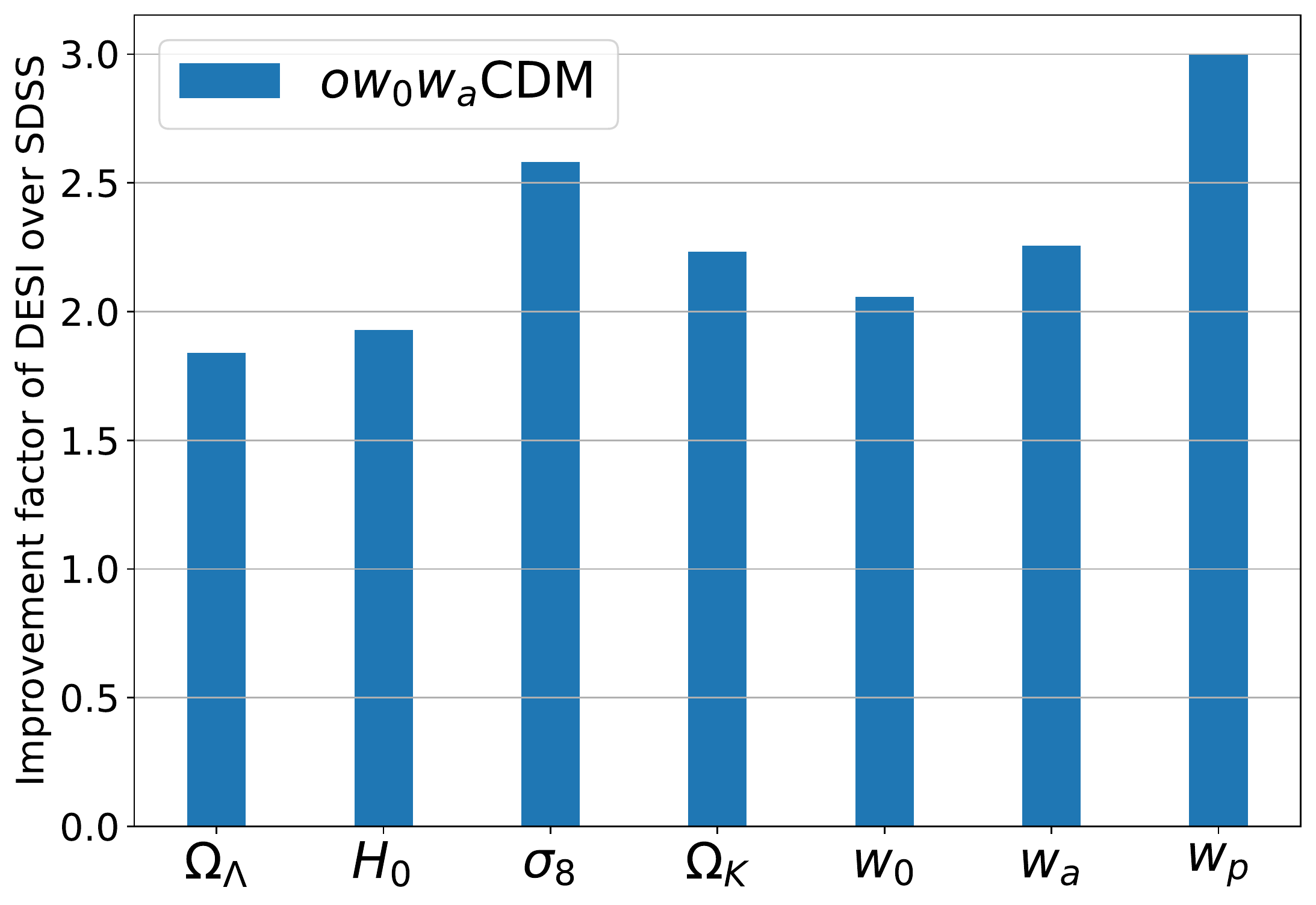}
    \caption{Relative improvement in 68\% confidence intervals over SDSS + {\it Planck} expected when constraining the standard $\Lambda$CDM parameters ($\Omega_{\Lambda}$, $H_0$, $\sigma_8$) along with the extension parameters ($\Omega_k$, $w_0$, $w_a$) using DESI + {\it Planck}.
}
\label{fig:advances}
\end{center}
\end{figure}

As a baseline to assess the advances we expect from DESI, we compute the constraints on the models using DESI and {\it Planck} data \citep{planckcosmo} and compare them with those from SDSS and {\it Planck}. 
As shown in 
Figure~\ref{fig:advances}, we expect significant gains across the full parameter space compared to SDSS. 
The largest of these gains appear in the projected constraints on $\sigma_8$ and $w_p$, demonstrating DESI's power to both constrain growth of structure and the equation of state for dark energy over a wide redshift range.
Relative to the SDSS and {\it Planck} results, we expect an improvement of a factor of $\sim 6.5$ in the DETF FOM for the $ow_0w_a$CDM model, as shown in Table~\ref{tab:fom}. 

\section{Conclusion}\label{sec:conclusion}
The Survey Validation data demonstrate that the target selection from legacy imaging data and spectroscopy from the DESI instrument on the Mayall Telescope will exceed initial expectations. 
In times when the survey speed ranges between $\frac{1}{6}$ and $\frac{2}{5}$, the 854 deg$^{-2}$ sample of BGS Bright targets will provide 647 deg$^{-2}$ spectroscopically-confirmed galaxies over the interval $0<z<0.4$.
In times with higher survey speeds, the LRG sample will produce 478 good redshifts deg$^{-2}$ over the redshift range $0.4<z<1.1$, the ELG sample 452 deg$^{-2}$ over $1.1<z<1.6$, and quasars at a density of 112 deg$^{-2}$ at higher redshifts, almost evenly split between direct tracers at $1.6<z<2.1$ and Lyman-$\alpha$ forest quasars at $2.1<z<3.5$.
The first year of observations are complete and progressing as expected from the survey simulations.

At the current rate of observing progress, DESI is expected to complete its full footprint covering 14,900 deg$^2$.
The central 14,000 deg$^2$ of this area is used to forecast the precision of cosmology measurements.
In these forecasts, we expect to reach a cumulative precision of 0.28\% on the isotropic BAO distance scale and 1.56\% on $f\sigma_8$ from the BGS and LRG samples at redshifts $z<1.1$.
Over the interval $1.1<z<1.9$, the ELG and quasar samples are expected to allow BAO measurements at 0.37\% precision and RSD measurements of $f\sigma_8$ to 1.24\% precision.
Finally, at the highest redshifts, we expect to reach a precision of 0.88\% on $H(z)r_d$ and 0.91\% precision on $D_A(z)/r_d$ using the BAO feature measured in the Ly-$\alpha$ forest relative to the sound horizon.
The Ly-$\alpha$ forest measurements will take advantage of both the Ly-$\alpha$ auto-correlation and the Ly-$\alpha$ -- quasar cross-correlation measurements.

In combining these measurements with information from {\it Planck} CMB measurements, these BAO and RSD measurements are forecast to provide dark energy constraints that correspond to a DETF Figure of Merit exceeding 150, thus qualifying DESI as a Stage-IV dark energy experiment.
When including additional information from lensing and Type~Ia supernovae, we expect an even more significant improvement in cosmological precision relative to Stage-II programs.

This paper is one of a series of results reporting the properties of the DESI MWS, BGS, LRG, ELG, and quasar target samples.  
These papers constitute the first key measurements from the DESI spectroscopic sample.
The first-year sample completed in June, 2022, and will provide the next series of key measurements from the DESI collaboration.
Just as the SV sample was used to confirm the one-point statistics for clustering studies, this first year sample will be used to test whether the DESI samples will meet expectations for BAO and RSD measurements.
The DESI collaboration has established five key projects toward this goal, all of which are well underway.

The first of these key projects is focused on the creation of the catalogs and the two-point statistics for each tracer.
Similar to those in BOSS \citep{reid16a} and eBOSS \citep{ross20a}, these catalogs will provide data samples corrected for observational systematic errors and a distribution of random positions and redshifts to convey the angular and radial coverage of the survey.
The two-point statistics from these catalogs will be presented in both configuration space and Fourier space, with a full characterization of the covariance between data points.
Work in support of this key project has already begun with studies of angular systematic errors \citep[e.g.][]{kitanidis20a,chaussidon22b}, clustering properties \citep[e.g.][]{zhou21a,zarrouk22a}, radial systematic errors, redshift classification, and fiber assignment corrections \citep[e.g.][]{bianchi18,smith19}.
Another key element to this effort will be the $N$-body simulations to test theoretical models and mock catalogs to approximate the program at a very large volume.
A series of numerical simulations have been compared at the halo level to assess the robustness of numerical simulation methods \citep{grove21}, while new techniques have been developed to suppress the effects of sample variance in these $N$-body simulations using approximate mocks \citep{ding22a}.
New high-fidelity mock galaxy catalogs have also been developed with mass resolution sufficient to resolve dark matter subhaloes for the BGS sample \citep{safonova21}, while various techniques have been developed to produce mock catalogs over volumes much larger than possible with $N$-body simulations \citep{balaguera22}.

The remaining key projects relate to the BAO measurements, RSD measurements, and cosmology constraints that will define DESI as a Stage-IV dark energy experiment.
The collaboration is now developing the BAO and RSD analyses for the BGS, LRG, ELG, and quasar samples on early data and on blinded catalogs.
Effort includes assessment of reconstruction methods, BAO and RSD fitting procedures, and systematic error calculation.
At higher redshift, quasars will be used to determine the BAO distance scale through auto-correlation in the Lyman-$\alpha$ forest and cross-correlation between the Lyman-$\alpha$ forest and quasars.
First results toward these Lyman-$\alpha$ forest studies include generation of mock catalogs \citep{Farr20a}, application of a convolutional neural network to characterize damped Lyman-$\alpha$ systems \citep{wang22}, a Ly$\alpha$ catalog \citep{perez23}, and a study on the effect of quasar redshift errors on Lyman-$\alpha$ forest correlation functions \citep{youles22,bault23,garcia23}.
Finally, the cosmological constraints from the first year measurements will be computed using many of the same tools as those used to make the forecasts in Section~\ref{sec:forecasts}.

We expect to obtain redshifts for 7.2 million unique stars, 36.12 million unique galaxies, and 2.87 million unique quasars over the main 14,000 deg$^2$ spectroscopic footprint.
The imaging data for these targets is already being used to constrain the BAO distance scale \citep[e.g.][]{sridhar20,zarrouk21a} and growth of structure through cross-correlations with the CMB \citep[e.g.][]{kitanidis21a,hang21a,white22}.
On smaller scales, the spectra will be used for direct constraints on neutrino mass through the one-dimensional Lyman-$\alpha$ forest power spectrum \citep[e.g.][]{karacayl22, Karacayli23, ravoux23}, while on the largest scales, the catalogs will be used to constrain non-Gaussianity in the primordial density field \citep[e.g.][]{mueller22}.
Beyond cosmology, DESI will provide spectroscopy to complement imaging from the Vera Rubin Observatory, provide new insights into galaxy evolution, and provide maps of the Milky Way and its neighbors \citep{dey23a} that can be used to infer its merger history, density profile, and other evolutionary characteristics.
When complete, the DESI program will offer the premier spectroscopic samples for cosmology and astrophysics.

\section*{Data Availability}

The Data Release 9 of the DESI Legacy Imaging Surveys is available at \url{https://www.legacysurvey.org/dr9/}. 

Documentation of DESI data access is maintained at \url{https://data.desi.lbl.gov/doc/access/}.

All data points used in published graphs are available in
\url{https://doi.org/10.5281/zenodo.10063934}.
\acknowledgements
{

This material is based upon work supported by the U.S. Department of Energy (DOE), Office of Science, Office of High-Energy Physics, under Contract No. DE–AC02–05CH11231, and by the National Energy Research Scientific Computing Center, a DOE Office of Science User Facility under the same contract. Additional support for DESI was provided by the U.S. National Science Foundation (NSF), Division of Astronomical Sciences under Contract No. AST-0950945 to the NSF’s National Optical-Infrared Astronomy Research Laboratory; the Science and Technology Facilities Council of the United Kingdom; the Gordon and Betty Moore Foundation; the Heising-Simons Foundation; the French Alternative Energies and Atomic Energy Commission (CEA); the National Council of Science and Technology of Mexico (CONACYT); the Ministry of Science and Innovation of Spain (MICINN), and by the DESI Member Institutions: \url{https://www.desi.lbl.gov/collaborating-institutions}. Any opinions, findings, and conclusions or recommendations expressed in this material are those of the author(s) and do not necessarily reflect the views of the U. S. National Science Foundation, the U. S. Department of Energy, or any of the listed funding agencies.

The authors are honored to be permitted to conduct scientific research on Iolkam Du’ag (Kitt Peak), a mountain with particular significance to the Tohono O’odham Nation.

}

\bibliographystyle{aasjournal}
\bibliography{references}

\end{document}